\documentclass[a4paper,11pt]{article}
\usepackage[utf8]{inputenc}
\usepackage[round,authoryear]{natbib}
\usepackage{amsmath,amsthm,amssymb}
\usepackage{dcolumn}
\usepackage{array}
\usepackage{booktabs}
\usepackage{caption}
\usepackage{subcaption}
\usepackage{rotating}
\usepackage{longtable}
\usepackage{fullpage}
\usepackage{comment}
\usepackage{multicol}
\usepackage{authblk}
\usepackage{etoolbox}
\makeatletter
\patchcmd{\@maketitle}{\vskip 2em}{\vskip 0.5em}{}{}
\makeatother

\usepackage{color,colortbl}
\usepackage{setspace}
\usepackage{hyperref}
\usepackage{placeins}
\usepackage[toc,page]{appendix}
\usepackage{adjustbox}
\usepackage{threeparttable}
\usepackage{tabularx}
\usepackage{multirow}
\usepackage{float}
\usepackage{graphicx}
\usepackage{tikz}
\usetikzlibrary{positioning,calc,fit,arrows.meta}

\hypersetup{pdfstartview={XYZ null null 1.00},bookmarksnumbered,hypertexnames=false}

\begin{document}
\doublespacing
\hyphenation{elas-tic-i-ties E-con-o-mies}
\renewcommand{\sectionautorefname}{Section}
\clubpenalty 9999
\widowpenalty 9999

\title{Electricity demand has not become more price-responsive despite ninety years of technological change\thanks{The data and code required to replicate all results are available at \url{https://meta-analysis.cz/electricity}. Kudela acknowledges support from the Grant Agency of Charles University (grant 120124); Irsova acknowledges support from the Czech Science Foundation (grant 24-11583S). We thank Carol Dahl for generously sharing her earlier data. Corresponding author: Zuzana Irsova, \href{mailto:zuzana.irsova@fsv.cuni.cz}{\texttt{zuzana.irsova@fsv.cuni.cz}}. All remaining errors are ours.}}

\author[1]{Peter Kudela}
\author[1,3,4]{Tomas Havranek}
\author[1,2,3]{Zuzana Irsova}
\author[1]{\protect\\Anna Kudelova}
\author[1]{Vojtech Sikl}
\affil[1]{Charles University, Prague}
\affil[2]{Anglo-American University, Prague}
\affil[3]{Meta-Research Innovation Center, Stanford}
\affil[4]{CEPR, London}

\date{\today}
\begin{singlespace}
\maketitle
\end{singlespace}

\vspace{-0.15cm}

\begin{singlespace}
\begin{abstract}
\noindent Energy planners have long assumed that electricity demand will grow more price-responsive as metering, automation, and storage spread, an assumption now embedded in decarbonization plans. We test it against the empirical record: 4,720 own-price elasticity estimates from 462 studies, with data spanning 1934--2024, ranked on a single ladder of identification quality from naive regressions to randomized experiments. Three findings emerge. First, the best-identified studies find \emph{smaller} responses than naive ones: the publication-bias-corrected short-run elasticity is about $-0.16$ (a 10\% rise in the electricity price cuts consumption by under 2\%), and only $-0.09$ among the best-identified studies, whose adjusted value is statistically indistinguishable from zero. Second, responsiveness grows with time to adjust, roughly doubling from $-0.16$ in the short run to $-0.38$ in the long run as the capital stock turns over, but this pattern has itself been stable for decades. Third, and most important, responsiveness shows no upward trend across nine decades of data; if anything, the most technology-rich settings, including time-of-use pricing, are the \emph{least} price-responsive in total consumption. Prices alone have not made total electricity consumption more responsive; broader demand flexibility will have to be engineered and paid for, through enabling technology, contracts, and program design.
\end{abstract}

\bigskip
\begin{tabular}{p{0.25\hsize}p{0.6\hsize}}
\textbf{Keywords:} & electricity demand, price elasticity, meta-analysis, identification, publication bias, demand flexibility\\
\textbf{JEL codes:} & Q41, Q48, C18\\
\end{tabular}
\end{singlespace}

\newpage

\section{Introduction}\label{sec:intro}

Surveying, in 1981, the fifteen residential time-of-use pricing experiments the U.S.\ Department of Energy had run since 1975, Dennis Aigner recorded the profession's expectation: with a full commitment to time-varying prices, ``appliance choices will be made with an eye to TOU response; new appliances will become widely available,'' and the pricing strategy ``surely must be even more desirable in the long-run'' \citep[p.~39]{aigner1985residential}. Four decades later, a recent working paper synthesizing time-based electricity rates opens with the same expectation, now attached to decarbonization: time-based rates ``may be a useful and efficient tool for encouraging demand-side flexibility'' in the transition to a high-renewable grid, and how the response will change ``with widespread adoption of emerging technologies'' remains, in the authors' words, an open prediction problem \citep{kahnlang2025slices}. Between these two statements, the assumption that price responsiveness of electricity demand is about to rise has underwritten smart-metering business cases \citep{faruqui2010household}, demand-response program design \citep{torriti2014time}, and the demand-side assumptions of energy-system models \citep{huntington2019industrializing}. It is among the more consequential quantitative assumptions in energy policy: if demand responds strongly to prices, decarbonized power systems need less storage and less firm capacity; if it does not, plans that rely on flexible demand will fall short.

The assumption is testable, and in the four decades since it was first recorded nobody has tested it head-on (\autoref{tab:premise} records it in the words of those who hold it). This paper brings the entire empirical record to bear on it: 4,720 own-price elasticity estimates from 462 studies, with underlying data spanning 1934--2024, to our knowledge the largest dataset assembled on the price elasticity of electricity demand, and one designed so that the question ``has responsiveness changed?'' can be separated from the two confounds that would otherwise contaminate the answer.

The first confound is identification. Reported elasticities depend on how a study confronts the simultaneity of price and quantity: tariff schedules respond to consumption, regulators respond to demand growth, and average prices constructed from bills divide revenue by the quantity being explained \citep{alberini2011response,ito2014}. A literature whose identification standards improve over time will exhibit spurious elasticity ``trends'' driven by changing methods rather than changing behavior. We therefore place every estimate on a single identification ladder, running from \emph{design-based} studies at the top (randomized experiments, mandated natural experiments, and difference-in-differences) down through instrumented specifications, panel fixed effects, and naive regressions, in the spirit of the design-quality tiers that \citet{kahnlang2025slices} apply to pricing pilots. The second confound is selective reporting. Statistically significant, correctly signed elasticities are easier to publish, which correlates estimates with their standard errors and inflates naive averages \citep{stanley2008meta,ioannidis2017power,brodeur2020methods}. If selection pressure has itself drifted over time, so has the visible literature, independently of consumers. We therefore run modern selection corrections within tiers and horizons, and we separate two clocks that previous surveys usually conflated: the vintage of a study's \emph{data} (do consumers behave differently?) and the study's \emph{publication} date (does the literature report differently?).

First, \emph{credibility shrinks the elasticity}. Moving up the ladder, the short-run elasticity falls in magnitude from $-0.37$ (raw naive mean) to $-0.09$ among design-based estimates (mean beyond bias). An elasticity of $-0.37$ means a 10\% rise in the electricity price cuts consumption by under 4\%; at $-0.09$ the same rise cuts it by under 1\%. Both fall well short of the 10\% price change, but the gap between them is large: the best-identified estimate is about a quarter the size of the naive one, shrinking a low response to almost none. Once observable study characteristics are held fixed, the design-based tier is statistically indistinguishable from zero ($+0.02$, $[-0.21,0.25]$) and significantly less elastic than the naive baseline ($p=0.006$). The instrumented tier, by contrast, is the one identification upgrade that remains as elastic as the naive literature ($-0.19$ adjusted, against $-0.24$ naive): instrumenting produces essentially none of the movement toward the experimental benchmark seen elsewhere on the ladder. Selective reporting differs sharply across the tiers: the observational literature shows pronounced funnel asymmetry (FAT $=-0.85$, $p<0.001$), while the design-based sample shows none ($p=0.97$). The corrected short-run consensus is small: bias-corrected estimates range from $-0.04$ to $-0.23$ across correction philosophies, with our preferred estimate at $-0.16$.

Second, \emph{responsiveness grows with adjustment time, not with calendar time}. The bias-corrected elasticity roughly doubles from the short run ($-0.16$) to the long run ($-0.38$), a gradient consistent with the within-study adjustment path that \citet{deryugina2020long} recover from a municipal-aggregation natural experiment, where the elasticity triples from $-0.09$ after six months to $-0.27$ after two years, and with the mechanism that \citet{costa2011electricity} identify in the housing stock: the long-run response operates through durable capital, so it arrives with the turnover of buildings and appliances, and it plateaus at the substitution possibilities the stock allows. The gradient itself is stable across eras of data: in every era since the mid-1970s the corrected long-run value has sat between $-0.29$ and $-0.47$. The long-run response has already arrived, at about $-0.38$, well short of $-1$, the unit-elastic level a fully flexible demand would reach.

Third, \emph{the elasticity shows no trend in the data vintage}. The short-run trend per decade of data is statistically indistinguishable from zero in every specification: the raw record, the bias-corrected record, and the fully adjusted record (both the data and publication clocks plus identification and composition moderators). The no-trend outcome is not just because the data are noisy: in the specifications precise enough to be informative, even the outer edge of the 95\% interval reaches only two-thirds to four-fifths of the way to a doubling of responsiveness over three decades, the canonical flexibility scenario. The point estimate in the fully adjusted specification actually leans toward \emph{less} elastic demand ($+0.017$ per decade, $p=0.53$). Estimates from time-of-use settings, the environments at the heart of the flexibility premise, are closer to zero than the rest of the literature (coefficient $+0.14$, $p=0.007$), and the design-based and time-of-use cells are consistent with the pilot-experiment literature, the $-0.02$ to $-0.10$ range of \citet{faruqui2010household} and the $-0.075$ average of \citet{kahnlang2025slices}.

This exercise is a descriptive cross-study audit of a literature, anchored by design-based benchmarks and within-study contrasts where they exist. No individual study spans ninety years or six identification tiers, so only a pooled reading of the whole record can ask whether the corrections the literature trusts reproduce what clean designs deliver, and whether the parameter planners extrapolate has ever moved. Our contribution is therefore a credibility audit of the assumption that the consensus elasticity is about to change, in the tradition of meta-research on selective reporting and the reliability of empirical economics \citep{ioannidis2017power,brodeur2020methods,dellavigna2022rcts,christensen2018transparency}.

Prior syntheses have examined the individual pieces but not the question itself. \citet{espey2004} include publication-year and data-vintage dummies in a 36-study sample and find drift; \citet{labandeira2017meta} report, in a single sentence, that a technical-progress trend ``is not significant at any level in any given specification''; \citet{zabaloy2022household} find significant data-vintage drifts of opposite signs for short- and long-run elasticities in Latin America; \citet{zhu2018meta} bin studies by period with no correction machinery; the surveys of \citet{dahl1993,dahl2011}, \citet{fatima2023price}, and \citet{marques2024latin} report static summaries. None makes the time path its central question, none runs it with bias correction, identification controls, and clustered inference, and none but \citet{espey2004} separates the data clock from the publication clock, the separation that, on our corpus, can account for the contradictory findings of these syntheses.

The paper proceeds as follows. \autoref{sec:data} describes the corpus and the three central variables. \autoref{sec:ladder} compares estimates across the identification ladder. \autoref{sec:selection} audits selective reporting along it. \autoref{sec:horizon} estimates the adjustment-horizon path. \autoref{sec:time} runs the calendar-time test and states the bound. \autoref{sec:policy} translates the results into the numbers planners and modelers should use. \autoref{sec:conclusion} concludes.

\section{Data}\label{sec:data}

\paragraph{Corpus.} We collect estimates of the own-price elasticity of electricity demand together with their standard errors from published studies and working papers. The final dataset contains 4,720 estimates from 462 studies (listed in \autoref{app:studies}), each reporting a usable measure of uncertainty. Our inclusion rule is deliberately simple: an estimate enters only if it reports a standard error, or a $t$-statistic or $p$-value from which one can be recovered; an estimate without any such measure cannot be corrected for publication bias and is excluded. A small number of estimates are methodologically borderline (notably: the construction-vintage elasticity of \citealp{costa2011electricity}; a separate, author-disavowed wrong-signed estimate; and an aggregate event-study elasticity); they carry a usable standard error and so meet our inclusion rule, and are individually immaterial at this sample size. Publication dates run from 1951 to 2026; the underlying data run from 1934 to 2024 (median data mid-year 1988). Each estimate carries 55 coded characteristics covering the data (country, sector, aggregation, frequency), the demand model (functional form, dynamics, controls), the price variable (average, marginal, time-of-use), the estimation method, and the publication outlet.

Our baseline sample places all comparable own-price elasticities on a common Marshallian (uncompensated) footing: the 3,324 estimates from 366 studies (1,647 short-run, 954 intermediate-run, 723 long-run) that carry a usable standard error and are either reported directly as Marshallian or converted from a compensated (Hicksian) estimate. The conversion applies the Slutsky relation $\varepsilon_M = \varepsilon_H - s\eta$ with electricity budget share $s=0.04$, a representative value for the electricity share of household spending in consumer expenditure surveys \citep{bls2024cex}, and each estimate's own income elasticity at the matching horizon.
 Because $s$ is small, the conversion shifts each estimate by roughly $0.01$ and leaves standard errors essentially unchanged (the delta-method adjustment is negligible), so the directly reported and the converted estimates can be read on a single scale.

Restricting to the 2,579 directly reported Marshallian estimates alone (the \emph{pure-Marshallian} sample; \autoref{app:robust}) leaves every headline result unchanged, as does varying $s$ between $0.02$ and $0.08$. A further 1,396 corpus estimates are held out of the baseline, either not convertible to the Marshallian footing or with an indirectly derived elasticity or standard error (\autoref{app:search}); \autoref{tab:heldout} shows they are statistically indistinguishable from it. Effects and standard errors are winsorized at 1\% throughout, and main-text inference clusters at the study level, with the model-averaging and full-battery appendices (\autoref{app:bma}, \autoref{app:pb}) additionally clustering two-way, by study and by underlying database. At the extremes of no winsorization and 5\%, the bias-corrected short-run estimate is $-0.1696$ and $-0.1683$, the long-run estimate $-0.4639$ and $-0.3572$, so the 1\% choice drives none of the headline conclusions.

\paragraph{The identification ladder.} The key variable is the \emph{price-identification strategy} of each estimate, coded on a six-tier ladder: (1) randomized experiments and randomized encouragement designs; (2) mandated natural experiments with exogenous price variation; (3) difference-in-differences designs; (4) instrumented specifications (IV, 2SLS/3SLS, GMM); (5) panel fixed effects and structural demand systems without instrumentation; (6) naive time-series or cross-section regressions with no attempt to address price endogeneity. Coding rules and study-level examples appear in \autoref{app:coding}. Because tiers 1--3 are thin in this literature (136, 10, and 26 estimates respectively) we group them as \emph{design-based} throughout the main text; 878 estimates whose strategy cannot be classified from the primary study are excluded from ladder analyses and retained elsewhere. In the coded corpus, sixteen studies report estimates on more than one tier.

\paragraph{Adjustment horizon and the two clocks.} Each estimate is classified as short-run (within roughly a year), intermediate-run, or long-run, using the primary study's own labels where given and the model structure otherwise (static single-equation estimates on levels with cointegration pre-tests are long-run; dynamic models contribute both a short- and a long-run estimate). We treat the horizon as an object of interest in its own right: \autoref{sec:horizon} estimates the elasticity as a function of it.

Finally, every estimate carries two dates: the mid-year of the data used (the \emph{data clock}; 23 estimates with no reported data years are assigned publication year minus three) and the publication year (the \emph{publication clock}). Only changes in consumer behavior can shift the first; the second also shifts with publication practices, methods fashion, and selection pressure. Most previous surveys used one or the other; \autoref{sec:time} uses both at once.

\begin{figure}[t]
\centering
\caption{Nine decades of estimates, and no visible drift}
\label{fig:corpus}
\begin{threeparttable}
\includegraphics[width=.95\textwidth]{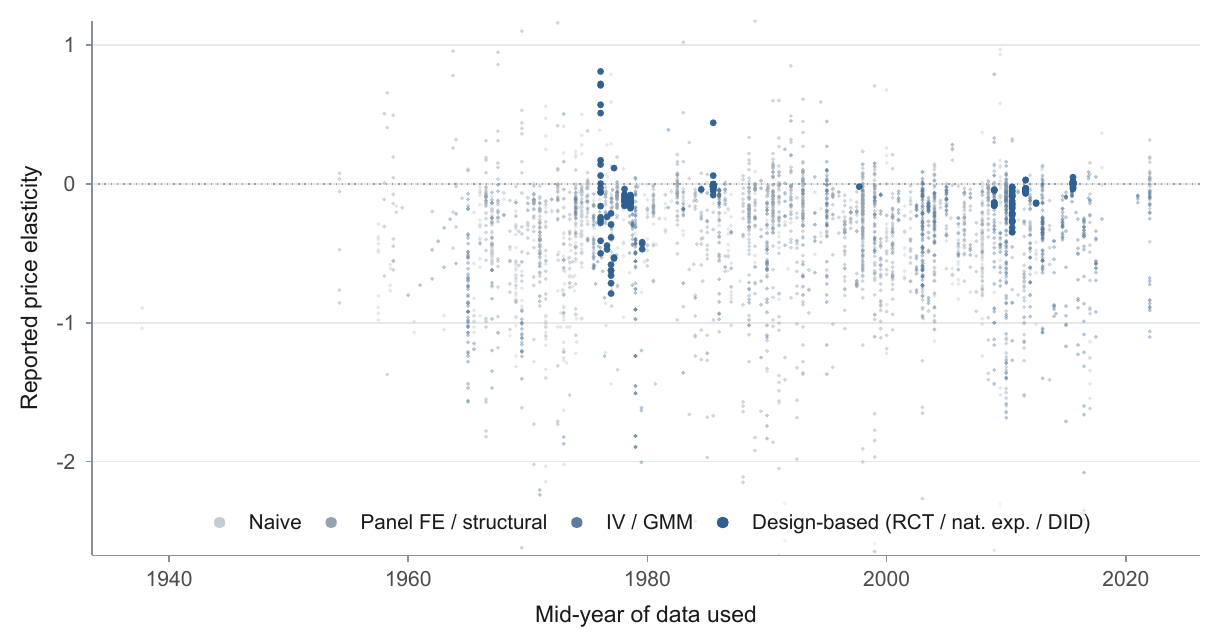}
\begin{tablenotes}[flush]
\footnotesize
\item \textit{Notes:} Reported own-price elasticities against the mid-year of the underlying data, colored by identification tier; the 3,843 tier-classified estimates with usable standard errors are shown, the vertical axis truncated at $[-2.5, 1]$ so that 44 outlying estimates fall outside the frame; these outliers are omitted from the figure only and kept in every analysis (unclassified estimates are likewise omitted from the plot but not from non-ladder analyses). Two features organize the paper: the raw record shows no visible drift over nine decades, and the design-based literature (dark blue) arrives late and thin (8 studies contribute short-run design-based estimates), so composition must be held fixed before any trend is read as behavior.
\end{tablenotes}
\end{threeparttable}
\end{figure}

\section{The identification ladder}\label{sec:ladder}

Does the elasticity survive scrutiny of its identification? \autoref{tab:ladder} lines the short-run literature up along the ladder. The unadjusted column reports the raw mean within each tier; the adjusted column reports the conditional mean from a single precision-controlling meta-regression: elasticity on its standard error (absorbing small-study effects), tier indicators, and a common set of study characteristics (sector, price measurement, data structure, region, demand controls, functional form), evaluated at a zero standard error with other characteristics at sample means, with study-clustered 95\% intervals. This is a descriptive cross-study association: which design a study uses is not randomly assigned. We let the within-study evidence and the design-based benchmark carry the causal weight; the design-based benchmark itself rests on the cross-study comparison.

\begin{table}[t]
\centering
\caption{Well-identified estimates have smaller short-run response}
\label{tab:ladder}
\begin{tabular}{lccccc}
\toprule
Tier & $N$ & Studies & Unadjusted & Adjusted & 95\% CI \\ \midrule
Design-based (RCT/nat.~exp./DID) & 81 & 8 & $-0.091$ & $0.022$ & [-0.21, 0.25] \\
Instrumented (IV/GMM) & 549 & 52 & $-0.304$ & $-0.195$ & [-0.35, -0.04] \\
Panel FE / structural & 655 & 98 & $-0.209$ & $-0.126$ & [-0.28, 0.03] \\
Naive & 316 & 70 & $-0.365$ & $-0.238$ & [-0.37, -0.11] \\
\bottomrule
\end{tabular}
\par\smallskip
\begin{minipage}{\linewidth}
\footnotesize\textit{Notes:} The table presents the short-run price elasticity by identification tier. Headline sample (Marshallian-equivalent, usable SE), short run; 1,601 estimates from 220 studies enter the adjusted regression (tier-classified subset). Unadjusted is the raw mean within the tier. Adjusted is the conditional mean from one meta-regression of the elasticity on its standard error, tier indicators (naive omitted), and study characteristics, evaluated at SE $=0$ and characteristics at sample means; 95\% study-clustered intervals. Relative to naive, the design-based shift is $+0.260$ ($p=0.006$), the fixed-effects shift $+0.112$ ($p=0.003$), the IV shift $+0.043$ ($p=0.32$). RCT, randomized controlled trial; nat.~exp., natural experiment; DID, difference-in-differences; IV, instrumental variables; GMM, generalized method of moments; FE, fixed effects.
\end{minipage}
\end{table}

The pattern is a gradient with one exception. Raw means fall in magnitude from $-0.37$ (naive) to $-0.09$ (design-based); adjusted, the design-based tier sits at $+0.02$ with an interval of $[-0.21, 0.25]$, and is significantly less elastic than the naive baseline. That interval is wide: we read the design-based cell as consistent with a small elasticity while unable to exclude a moderate one. In the menu (\autoref{tab:menu}), $-0.09$ is the raw and PET value and $+0.02$ the composition-adjusted prediction. The exception is instrumenting: the IV tier is the only identification upgrade that stays as elastic as the naive literature after adjustment ($-0.19$ against $-0.24$ naive), statistically indistinguishable from naive and overshooting the design-based benchmark. Within the sixteen studies that report estimates on more than one tier, the same picture appears in miniature: the within-study contrast between the cleaner and the lower tier averages $-0.08$ (cleaner estimates \emph{more} elastic, driven by IV-versus-naive pairs), but it is small relative to its dispersion ($t=-0.79$) and positive in half the studies. The within-study evidence is underpowered, so we report it for completeness without leaning on it (\autoref{fig:withinstudy} in \autoref{app:robust} plots every contrast). None of these within-study contrasts involves a design-based estimate, so the within-study evidence bears on the IV and fixed-effects ranking rather than on the design-based benchmark, and the argument depends on the cross-study ladder and the funnel evidence of the next section.

Two readings of the IV overshoot are possible, and the data cannot separate them here: instruments may correct attenuation from classical average-price measurement error \citep{alberini2011response}, legitimately raising magnitudes; or weak instruments and specification search may inflate them. \citet{ito2014} shows the average-price coefficient is itself the behaviorally relevant object rather than an attenuated one, which cuts against the attenuation-correction reading. What the ladder does show is the direction that matters for the flexibility debate: \emph{no identification upgrade makes demand look more price-responsive than the naive literature already claims}: the best designs point toward less responsiveness.

Price measurement provides a second check on the ladder. Average-price estimates carry classical division bias and inflated standard errors; marginal-price estimates are cleaner but scarcer. Bias-corrected short-run elasticities are $-0.15$ (marginal price) and $-0.17$ (average price): the correction for measurement regime moves the consensus by two hundredths, far less than the flexibility premise requires.

\section{Selective reporting along the ladder}\label{sec:selection}

A meta-analysis that ignores selective reporting conflates the literature's publication preferences with the underlying parameters \citep{stanley2008meta,ioannidis2017power}. The concern is concrete here: intuitive negative elasticities and statistically significant elasticities are easier to publish, which leaves a correlation between estimates and standard errors that funnel-based estimators detect and net out.

\begin{table}[t]
\centering
\begin{threeparttable}
\caption{Publication bias inflates the elasticities across tiers and horizons}
\label{tab:battery}
\small
\begin{tabular*}{\textwidth}{@{\extracolsep{\fill}}lccccccc@{}}
\toprule
Sample & $N$ & Studies & Mean & PW & PET & PEESE & WAAP \\ \midrule
Long-run elasticities (full sample) & 723 & 151 & $-0.583$ & $-0.378$ & $-0.377$ & $-0.498$ & $-0.376$ \\
Intermediate elasticities (full sample) & 954 & 108 & $-0.440$ & $-0.149$ & $-0.331$ & $-0.416$ & $-0.148$ \\
Short-run elasticities (full sample) & 1,647 & 226 & $-0.267$ & $-0.114$ & $-0.163$ & $-0.231$ & $-0.111$ \\
\midrule
Short-run, design-based & 81 & 8 & $-0.091$ & $-0.078$ & $-0.095$ & $-0.114$ & $-0.080$ \\
Short-run, instrumented (IV) & 549 & 52 & $-0.304$ & $-0.104$ & $-0.203$ & $-0.276$ & $-0.102$ \\
Short-run, panel FE / structural & 655 & 98 & $-0.209$ & $-0.150$ & $-0.088$ & $-0.164$ & $-0.151$ \\
Short-run, naive & 316 & 70 & $-0.365$ & $-0.108$ & $-0.245$ & $-0.318$ & $-0.106$ \\
Short-run, marginal price & 401 & 48 & $-0.301$ & $-0.054$ & $-0.148$ & $-0.230$ & $-0.049$ \\
Short-run, average price & 1,030 & 166 & $-0.264$ & $-0.107$ & $-0.168$ & $-0.236$ & $-0.105$ \\
Short-run, preferred only & 426 & 191 & $-0.302$ & $-0.159$ & $-0.182$ & $-0.271$ & $-0.158$ \\
\bottomrule
\end{tabular*}
\par\smallskip
\begin{minipage}{\textwidth}
\footnotesize\textit{Notes:} The table presents the publication-bias battery by horizon and tier. Headline sample; the full-sample rows pool all estimates at each horizon, below which the short-run sample is split by tier and by price regime. Mean is the raw mean; PW the precision-weighted mean; PET and PEESE the funnel-asymmetry-corrected estimates \citep{Stanley2005,stanley2008meta}; WAAP the weighted average of adequately powered estimates \citep{ioannidis2017power}. Study-clustered inference. FAT $p$-values for the asymmetry test: short-run all $<0.001$; design-based $0.97$; marginal-price subsample $0.001$. ``Preferred only'' uses the estimates flagged as preferred by the original studies (426 estimates from 191 studies). The sample-size-weighted corrector on the short-run full sample (1,647 estimates with a usable weight) gives $-0.04$ (se $0.032$), the range's low end.
\end{minipage}
\end{threeparttable}
\end{table}

\autoref{tab:battery} reports the full battery of tests. First, selection is real and it inflates: the short-run funnel-asymmetry test rejects strongly (FAT $=-0.85$, $p<0.001$), and every corrector pulls the raw mean of $-0.27$ toward zero, to $-0.16$ (PET), $-0.23$ (PEESE), $-0.11$ (WAAP), $-0.10$ (top decile by precision), $-0.04$ (se $0.032$; 1,647 estimates; sample-size weighted); a fuller battery in \autoref{app:pb}, adding the MAIVE estimator of \citet{irsova2023spurious}, $p$-uniform* \citep{van2021correcting}, and a selection model \citep{andrews2019identification}, stays in the same range. We report the range without adjudicating among correction philosophies; the corrected short-run consensus lies between $-0.04$ and $-0.23$, and nowhere near the raw mean. Second, it is the \emph{observational} literature that shows selection bias: the design-based sample shows no funnel asymmetry (FAT $p=0.97$), though with 81 estimates from 8 studies the test is under-powered, so selection can be neither detected nor excluded there; we rest the design-based benchmark on design credibility rather than on this null FAT. The slope itself is small ($\gamma=0.029$) and stays insignificant under few-cluster-robust inference (wild cluster bootstrap $p=0.918$ Rademacher, $p=0.941$ Webb; CR3-Satterthwaite $p=0.991$), so the null is not an artifact of the asymptotic approximation with only 8 clusters. The cell's bias-corrected mean is, in the same way, indistinguishable from zero under the wild-cluster bootstrap ($-0.09$, $p=0.07$, 95\% CI $[-0.23,0.04]$).\footnote{The other thin-cluster cell the analysis leans on, the most recent long-run data decade (\autoref{fig:decades}, 2010s onward, 9 studies), is robust in the opposite direction: its bias-corrected mean, if anything more elastic than the $-0.29$ to $-0.47$ era band, stays firmly negative under the same bootstrap ($p=0.002$, 95\% CI $[-0.72,-0.38]$), so the strength of the recent long-run response is not a few-cluster artifact.} Leave-one-study-out sharpens this caveat rather than resolving it: dropping any one of the 8 studies moves the FAT slope as low as $-1.96$ and the $p$-value as high as $0.99$, so the design-based null is a property of the full 8-study cell, not a result that survives study-by-study scrutiny. Third, the asymmetry is not an artifact of constructed standard errors or average-price measurement error: it survives, essentially undiminished, in the marginal-price subsample (FAT $=-0.97$, $p=0.001$), which is the diagnostic that separates genuine selection from errors-in-variables mechanics. A caliper test points the same way from a different angle: there is no excess mass of estimates just above the conventional $|t|=1.96$ significance threshold, even in the observational subsample (share above within a $\pm0.2$ window $=0.51$, $p=0.87$; full headline sample $0.53$, $p=0.39$), so the selective reporting this literature displays acts on the magnitude of estimates (a small-study effect) rather than on crossing the significance threshold. The two tests are complementary rather than competing: a literature filtered on the sign and size of estimates, not on their $p$-values, is precisely one that shows funnel asymmetry without threshold bunching.

Because the correctors themselves rely on the standard errors that the selection story calls into question, we triangulate with weights that do not use them: weighting by sample size yields a corrected short-run elasticity of $-0.04$ (se $0.032$; 1,647 estimates), statistically indistinguishable from zero and the least elastic value in the battery. Restricting to the estimates the original studies flag as preferred (426 estimates from 191 studies) leaves the picture unchanged (PET $-0.18$), as does the long run on the same restriction (251 estimates from 137 studies; PET $-0.36$), and the calendar-time trend on the preferred-only short-run sample stays indistinguishable from zero ($-0.056$ per decade, $p=0.084$), if anything more negative than the pooled headline. All of these checks converge: \emph{the bias-corrected short-run price elasticity of electricity demand is small}, and every standard correction moves it closer to zero.

\section{Responsiveness grows with adjustment time}\label{sec:horizon}

If short-run responsiveness is small, the flexibility case shifts to the long run: given time, capital adjusts. The record supports this mechanism and quantifies it.

\begin{figure}[t]
\centering
\caption{The bias-corrected elasticity grows with the adjustment horizon, tracking the quasi-experimental benchmark}
\label{fig:horizon}
\includegraphics[width=.8\textwidth]{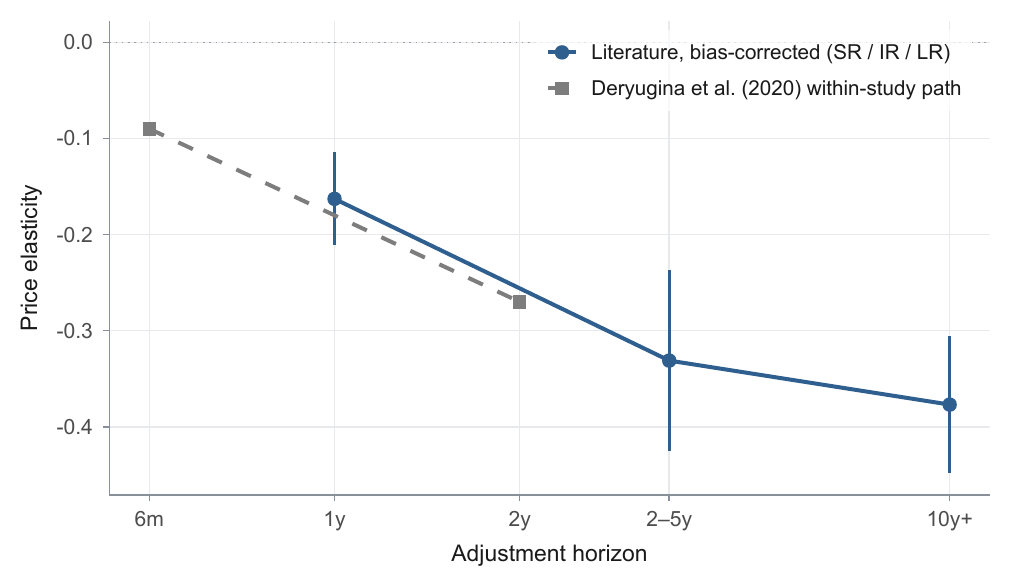}
\par\smallskip
\begin{minipage}{\textwidth}\footnotesize\textit{Notes:} Bias-corrected (PET) elasticity by horizon on the headline sample, with study-clustered 95\% intervals: short run $-0.163$ (se $0.025$; 1,647 estimates, 226 studies), intermediate run $-0.331$ (se $0.048$; 954/108), long run $-0.377$ (se $0.037$; 723/151). Dashed: the within-study adjustment path of \citet{deryugina2020long}, $-0.09$ at six months to $-0.27$ at two years. Horizon placement of the literature bins is notational (SR $\approx$ within a year; IR 2--5 years; LR beyond).\end{minipage}
\end{figure}

\autoref{fig:horizon} plots the bias-corrected elasticity by horizon: $-0.16$ in the short run, $-0.33$ in the intermediate run, $-0.38$ in the long run: the response roughly doubles as adjustment time accumulates and then plateaus. This is not an artifact of which studies happen to populate which horizon bin: among the 100 studies that report both a short- and a long-run estimate, the same-study contrast goes from $-0.238$ to $-0.593$, a within-study ratio of means of $2.492$ (median study-level ratio $2.571$), with 86.0\% of studies showing a more elastic long run and a paired $t$ of $-8.68$. A within-study fixed-effects regression on the 1,241 estimates from the 112 studies reporting more than one horizon, with study fixed effects absorbing every cross-study difference in composition, puts the same steepening at $-0.12$ (se $0.065$) from short to intermediate run and $-0.22$ (se $0.028$) from short to long run: the pooled gradient is not a compositional artifact of which studies populate which bin.

Two studies help interpret the adjustment profile. Within a single natural experiment, \citet{deryugina2020long} track the same set of over 250 municipally-aggregated communities in a difference-in-differences design and find the elasticity tripling from $-0.09$ at six months to $-0.27$ after two years: the literature-wide gradient is directionally consistent with what this dynamic design delivers, a rare point of agreement between this literature and design-based work. Because this study also supplies the short-run design-based cell, it grounds the horizon path but is not independent corroboration of both findings. And the plateau has a mechanism: \citet{costa2011electricity} show that electricity consumption is embodied in the housing stock at construction (an implied elasticity of about $-0.22$ at construction time), so the long-run response works through the turnover of buildings and appliances and stops at the substitution possibilities the stock allows. Descriptive dynamics in the literature point the same way: the median adjustment speed among dynamic models implies half of the long-run response within roughly a year and a half and ninety percent within five and a half years \citep{dahl2011}. \autoref{fig:metairf} in \autoref{app:robust} draws this adjustment path as a single continuous curve, pinned to the corpus's own short- and long-run values and consistent with these speeds.

The policy-relevant corollary follows directly: \emph{the long run is already in the data}. A planner invoking ``the long run'' is invoking a central estimate of $-0.38$, not the unit-elastic $-1$ that fully flexible demand would imply; we note that this long-run value comes from the observational correctors (the design-based long-run cell is a single study, too thin to report), and that the within-study long-run projection of \citet{deryugina2020long} carries a 95\% interval reaching $-1$, so $-0.38$ is a central estimate rather than a design-audited bound. The gradient itself is stable across eras: within every data-vintage era from the 1970s to the 2010s, the long-run corrected elasticity sits between $-0.29$ and $-0.47$ (\autoref{tab:horizonera}). In the pooled record, adjustment time roughly doubles responsiveness (era-by-era the gradient varies with the short-run dips of the 1980s but shows no tendency to steepen toward the present).

\begin{table}[t]
\centering
\begin{threeparttable}
\caption{The horizon gradient is stable across eras of data}
\label{tab:horizonera}
\small
\begin{tabular*}{0.7\textwidth}{@{\extracolsep{\fill}}lcccc@{}}
\toprule
Horizon & $\leq$1975 & 1976--90 & 1991--05 & $\geq$2006 \\
\midrule
Short run & $-0.179$ & $-0.090$ & $-0.158$ & $-0.223$ \\
 & $(0.038)$ & $(0.041)$ & $(0.034)$ & $(0.043)$ \\
Intermediate run & $-0.418$ & $-0.256$ & $-0.258$ & $-0.241$ \\
 & $(0.080)$ & $(0.081)$ & $(0.060)$ & $(0.112)$ \\
Long run & -- & $-0.429$ & $-0.287$ & $-0.472$ \\
 &  & $(0.057)$ & $(0.046)$ & $(0.055)$ \\
\bottomrule
\end{tabular*}
\par\smallskip
\begin{minipage}{0.7\textwidth}
\footnotesize\textit{Notes:} Bias-corrected (PET) elasticity within horizon $\times$ era cells, headline sample, study-clustered standard errors in parentheses. The long-run cell before 1976 has 51 estimates from 12 studies but is imprecisely estimated (a wide interval) and is suppressed. The intermediate run varies non-monotonically across eras; the short- and long-run values in the most recent era are not statistically distinguishable from the earliest era \emph{shown} (for the long run this baseline is 1976--90, because the $\leq$1975 long-run cell is suppressed as imprecise), and no horizon exhibits a monotone rise toward the present. These era comparisons are imprecise and should be read as a limit on power rather than as evidence of exact stability.
\end{minipage}
\end{threeparttable}
\end{table}

\section{The calendar-time test}\label{sec:time}

A small measured elasticity could still be reconciled with the flexibility premise in three ways. The true response could be larger, hidden by weak identification. Responsiveness could have grown over time, with selective reporting concealing the rise. Or the extra response could still be to come, arriving only in the long run. Sections~\ref{sec:ladder}--\ref{sec:horizon} closed off all three. What remains is the premise's core empirical claim, that responsiveness has been rising as technology diffuses, and it is directly testable on the data clock.

\begin{table}[t]
\centering
\begin{threeparttable}
\caption{Two clocks, and neither shows a trend}
\label{tab:twoclocks}
\small
\begin{tabular*}{0.95\textwidth}{@{\extracolsep{\fill}}lcc@{}}
\toprule
 & Data clock & Publication clock \\
\midrule
\multicolumn{3}{@{}l}{\emph{Short run}} \\
\hspace{0.5cm}Data clock only & $-0.018$ ($0.013$) & -- \\
\hspace{0.5cm}Two clocks & $-0.033$ ($0.031$) & $0.019$ ($0.033$) \\
\hspace{0.5cm}Two clocks + composition & $0.017$ ($0.027$) & $-0.030$ ($0.028$) \\
\hspace{1.0cm}$+$ country fixed effects & $0.044$ ($0.033$) & $-0.045$ ($0.033$) \\
\addlinespace
\multicolumn{3}{@{}l}{\emph{Intermediate run}} \\
\hspace{0.5cm}Data clock only & $0.012$ ($0.028$) & -- \\
\hspace{0.5cm}Two clocks & $0.000$ ($0.077$) & $0.012$ ($0.065$) \\
\hspace{0.5cm}Two clocks + composition & $0.043$ ($0.064$) & $0.030$ ($0.057$) \\
\hspace{1.0cm}$+$ country fixed effects & $0.053$ ($0.065$) & $-0.003$ ($0.060$) \\
\addlinespace
\multicolumn{3}{@{}l}{\emph{Long run}} \\
\hspace{0.5cm}Data clock only & $0.011$ ($0.036$) & -- \\
\hspace{0.5cm}Two clocks & $-0.092$ ($0.048$) & $0.135$ ($0.047$) \\
\hspace{0.5cm}Two clocks + composition & $-0.019$ ($0.053$) & $0.073$ ($0.050$) \\
\hspace{1.0cm}$+$ country fixed effects & $0.076$ ($0.056$) & $0.007$ ($0.058$) \\
\bottomrule
\end{tabular*}
\par\smallskip
\begin{minipage}{0.95\textwidth}
\footnotesize\textit{Notes:} Each cell is a coefficient per decade from a study-clustered meta-regression that always controls for the standard error. The \emph{Data clock} column is the trend in the elasticity per decade of the underlying \emph{data} (do consumers behave differently over time?); the \emph{Publication clock} column is the trend per decade of \emph{publication} year (does the literature report differently over time?). The four rows are nested specifications: \emph{Data clock only} includes just the data clock; \emph{Two clocks} adds the publication clock, so the two are separated; \emph{Two clocks $+$ composition} further adds the identification tier and study characteristics of \autoref{tab:ladder}, that is, the changing \emph{composition} of the literature, leaving any residual data-vintage trend. The final row within each block, \emph{$+$ country fixed effects}, adds a full set of country dummies to that specification, so the data-vintage trend is identified only from within-country variation and cannot reflect the changing country mix of the literature; it stays insignificant at every horizon and, if anything, turns more positive (short run $+0.044$, $p=0.19$; long run $+0.076$, $p=0.17$), further from the negative slope a rising-flexibility world would require. Unlike \autoref{tab:ladder}, which drops the estimates whose strategy could not be classified (leaving 1,601 in the short run), the composition-adjusted rows retain them in the naive reference category, so every row here is estimated on the full sample. Short-run sample: 1,647 estimates, 226 studies, data span 63 years; intermediate run: 954 estimates, 108 studies; long-run: 723 estimates, 151 studies, 54 years. A negative coefficient means the literature moves toward more elastic (more negative) values. No linear data-clock coefficient is significant at 5\% in any specification. Quadratic specifications yield a significant squared term in the long run ($p<0.001$) but not the short run (squared $p=0.06$); we therefore label the long-run trend inconclusive rather than flat. The short-run quadratic does carry a significant linear term ($-0.030$ per decade, $p=0.03$), but every linear short-run specification above is insignificant, consistent with the non-monotone mid-decade swings of \autoref{fig:decades} rather than a trend.
\end{minipage}
\end{threeparttable}
\end{table}

\begin{figure}[t]
\centering
\caption{Bias-corrected elasticity by decade of data}
\label{fig:decades}
\begin{threeparttable}
\includegraphics[width=.95\textwidth]{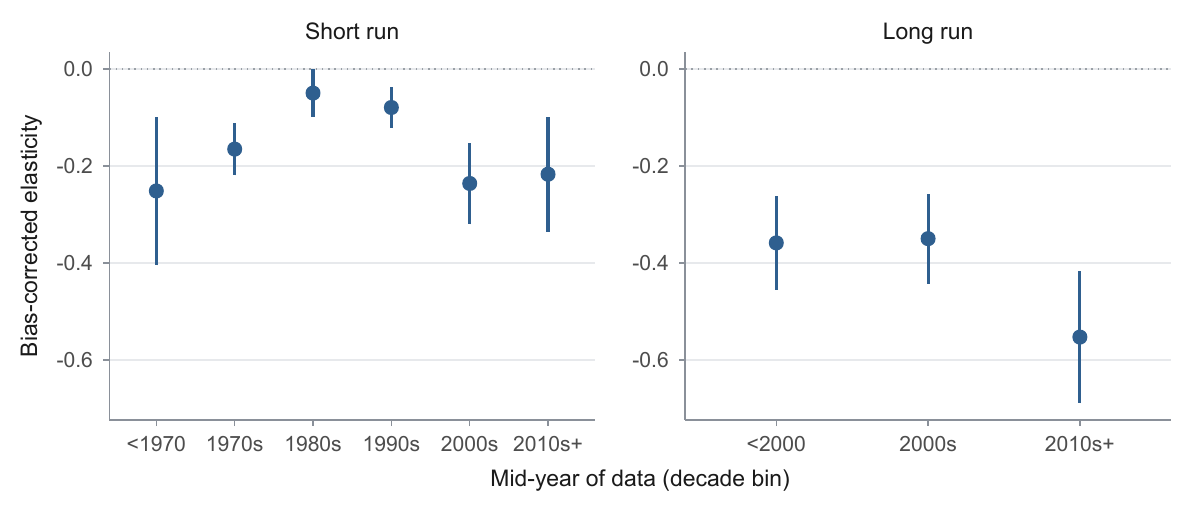}
\begin{tablenotes}[flush]
\footnotesize
\item \textit{Notes:} PET-corrected elasticity within decade bins of the data mid-year, headline sample, 95\% study-clustered intervals. Short run: $-0.251$ ($<$1970), $-0.165$ (1970s), $-0.050$ (1980s), $-0.079$ (1990s), $-0.236$ (2000s), $-0.217$ (2010s+). The 2010s+ value is statistically indistinguishable from the 1970s value; the series swings across the middle decades (the 1980s and 2000s intervals do not overlap). Long run: $-0.36$, $-0.35$, $-0.55$ across coarse bins, the last on 9 studies.
\end{tablenotes}
\end{threeparttable}
\end{figure}

\autoref{tab:twoclocks} reports the trend under progressively stricter specifications; \autoref{fig:decades} shows the decade-by-decade profile. The answer is the same throughout. In the raw record, the short-run trend is $-0.010$ per decade of data ($p=0.43$). Controlling for precision, $-0.018$ ($p=0.17$). Separating the two clocks leaves neither significant: the data clock is $-0.033$ ($p=0.27$) and the publication clock $+0.019$ ($p=0.57$); jointly, the two calendar coefficients are indistinguishable from zero as well (Wald $\chi^2_2=2.18$, $p=0.34$), so the modest apparent drift that earlier surveys read as ``elasticities changing over time'' \citep{espey2004,zabaloy2022household} splits into two pieces, each indistinguishable from zero, rather than a behavioral trend in consumers. Holding identification and composition fixed, the data-clock point estimate turns positive: $+0.017$ per decade ($p=0.53$),\footnote{Entering the controls one block at a time (data clock alone, then the publication clock, then the identification tier, then study composition) confirms the point estimate: $+0.017$ ($p=0.53$) at the full, composition-adjusted specification, identical to the headline.} if anything, a drift toward \emph{less} price-responsiveness as the credibility revolution is netted out. The sign convention matters here: a rising-flexibility world requires this coefficient to be robustly \emph{negative}. The pooled coefficient could still mask a rise concentrated in the settings the premise is built on, so we interact the data clock with sector and region and re-estimate it within each: none is robustly negative. The residential interaction is not distinguishable from the pooled trend ($-0.021$ per decade, $p=0.30$), and the composition-adjusted data-clock slope is negative but insignificant in Europe ($-0.067$, $p=0.18$), in post-2000 data ($-0.037$, $p=0.60$), and in the residential-and-post-2000 intersection the premise is specifically about ($-0.079$, $p=0.35$); the one subsample where the slope is significant, the United States ($+0.099$, $p=0.002$), carries the wrong sign for a rising-flexibility world. We detect no rise in the settings that matter most to the premise.

\paragraph{The bound.} Statistical insignificance alone would say little; what matters is the width of the interval. Take the canonical flexibility scenario: short-run responsiveness roughly doubles over three decades, from $-0.16$ to $-0.33$, an increase of about $0.16$ in magnitude. Specification by specification, the edge of the 95\% interval most favorable to that scenario allows a cumulative three-decade increase of $0.10$ (raw record, 64\% of the required movement), $0.13$ (precision-controlled bivariate, 80\%), and $0.11$ (two clocks plus composition, 66\%). The one exception is instructive: splitting the two clocks \emph{without} composition controls inflates the data-clock standard error enough (the two clocks are collinear) that its interval alone could accommodate a doubling, which is why the composition-adjusted specification is the informative one. In every specification precise enough to bound change, the record does more than fail to show the assumed rise: it bounds any concealed rise below what the scenario needs.

The formal version of that bound is an equivalence test. The doubling scenario requires responsiveness to rise by $0.05428$ per decade, a full doubling spread evenly across three decades. The test asks, one-sided, whether the data are still consistent with a rise that large. In all three specifications they are not: it rejects a doubling-sized rise at the 5\% level ($p=0.0002$ raw, $p=0.0026$ precision-controlled, $p=0.004$ two clocks plus composition). So the exclusion rests on a test, not on eyeballing where an interval ends. How large a rise could each specification have caught in the first place? At 80\% power, the two precise specifications would detect a rise of just $65$--$67\%$ of a doubling, so even a partial rise would not have slipped past them. The composition-adjusted specification is noisier: measured against zero, it could only detect a rise larger than a full doubling ($138\%$). It still rejects the doubling anyway, because the equivalence test measures the estimate against the rise margin, not against zero, and the estimate is positive, pointing the opposite way and so sitting comfortably clear of that margin. The long-run record, being noisier (a per-decade interval of $\pm 0.10$ around zero and a significant quadratic), cannot support a symmetric claim, and we label it inconclusive at current power rather than flat.

\paragraph{The new-regime objection.} The record's data end in 2024, and the premise is about the future: perhaps smart meters, automation, and time-varying rates change the regime after the sample. We make no claim about that post-sample future. But the settings the premise is built on already appear in the record. Where they do, demand is no more price-responsive than elsewhere. The objection can be addressed inside the data, because the record already contains the tariff instrument central to the premise, time-of-use pricing. The record covers the pricing instrument rather than the full enabling-technology regime (mass smart metering, automation, and storage) that the premise projects for the post-sample future, and it samples that regime thinly. We therefore read the result below as suggestive of the technology-rich corner rather than dispositive for it.

\begin{figure}[t]
\centering
\caption{The technology-rich cells are the least elastic}
\label{fig:newregime}
\begin{threeparttable}
\includegraphics[width=.85\textwidth]{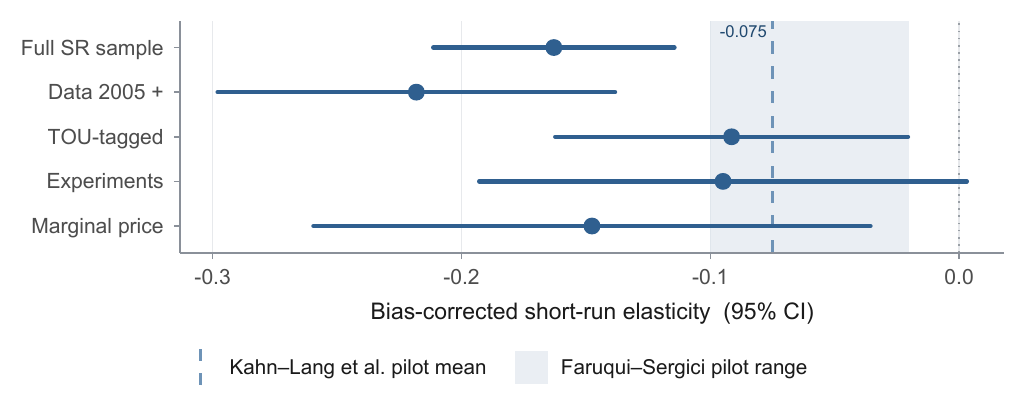}
\begin{tablenotes}[flush]
\footnotesize
\item \textit{Notes:} PET-corrected short-run elasticity with 95\% study-clustered intervals, by cell of the headline sample ($n$: full 1,647; post-2005 499; TOU 210; experiments 81; marginal price 401). Shaded band: the own-price elasticity range from dynamic-pricing pilots surveyed by \citet{faruqui2010household}; dashed line: the $-0.075$ pilot average of \citet{kahnlang2025slices}.
\end{tablenotes}
\end{threeparttable}
\end{figure}

\autoref{fig:newregime} compares corrected short-run elasticities across cells: the full sample ($-0.16$), data from 2005 onward ($-0.22$), marginal-price studies ($-0.15$), time-of-use settings ($-0.09$), and design-based studies ($-0.09$; the field experiments of \citealp{jessoe2014knowledge} and \citealp{ito2018moral}, and the natural experiment of \citealp{deryugina2020long}); the broader pilot literature concurs, with the $-0.02$ to $-0.10$ range of \citet{faruqui2010household} and the $-0.075$ average that \citet{kahnlang2025slices} extract from post-2000 U.S.\ pilots. That same pilot literature reports substitution elasticities of $0.07$ to $0.40$ and peak own-price values as large as $-0.79$. Those describe intertemporal load-shifting, which is outside the own-price level object we measure. In the meta-regression, the time-of-use indicator enters at $+0.14$ ($p=0.007$): conditional on precision, estimates from the most technologically enabled settings are closer to zero than the literature at large. The model-averaged counterpart is smaller and less firmly selected ($+0.07$, PIP $=0.73$; \autoref{tab:bmasr}), so we read the time-of-use signal as directionally consistent but less firmly identified than the design-based tier. Nothing in the technology-rich corner of the record hints at a rise in the own-price \emph{level} elasticity we measure. The metric is narrow by construction: an own-price level elasticity does not capture intraday load-shifting or intertemporal substitution, which net out in total consumption, so these cells bound the level response but are silent on engineered demand-response flexibility. What does move pilot responses is hardware and program design (enabling technology adds over twenty percentage points to peak reductions in \citet{kahnlang2025slices}), which supports our central conclusion: flexibility comes from engineering, and prices alone do not deliver it.

Aigner's conjecture \citep{aigner1985residential} can now be answered. Households replaced their appliances several times over, metering and displays arrived at scale, and automated response began its rollout. Four further decades of estimates accumulated, and the predicted growth in responsiveness never materialized. The conservation effect the early experiments measured turned out to be, as Aigner himself suspected it might, the full extent of the response.

\section{What should planners and models use}\label{sec:policy}

\begin{table}[t]
\centering
\begin{threeparttable}
\caption{A menu of corrected consensus elasticities for planning and modeling}
\label{tab:menu}
\small
\begin{tabular*}{0.95\textwidth}{@{\extracolsep{\fill}}lcc@{}}
\toprule
Object & Preferred & Range across correctors \\
\midrule
Short run, full sample & $-0.16$ & $-0.04$ to $-0.23$ \\
Short run, design-based & $-0.09$ & $-0.08$ to $-0.11$ \\
Short run, time-of-use settings & $-0.09$ & -- \\
Intermediate run & $-0.33$ & $-0.15$ to $-0.42$ \\
Long run & $-0.38$ & $-0.38$ to $-0.50$ \\
\addlinespace
Trend allowance, per decade & $0.00$ & $95\%$ interval $(-0.04, +0.07)$ adj. \\
 & & $(-0.04, +0.01)$ bivariate \\
\bottomrule
\end{tabular*}
\par\smallskip
\begin{minipage}{0.95\textwidth}
\footnotesize\textit{Notes:} Bias-corrected values on the headline sample (ranges across PET, PEESE, WAAP, precision- and sample-size weighting where computed). The trend allowance row reports the 95\% interval on the short-run data clock in the composition-adjusted specification and, for comparison, the precision-controlled bivariate one; the unrestricted two-clock split is too collinear to be informative (data and publication clocks correlate at $r=0.915$, VIF $=6.12$, inflating the data-clock standard error $2.35\times$; see \autoref{sec:time}).
\end{minipage}
\end{threeparttable}
\end{table}

Energy-system models rarely document their demand elasticities; where visible, household own-price values of $-0.2$ to $-0.6$ in the long run are common \citep{huntington2019industrializing}. \autoref{tab:menu} states what the audited record supports. The menu implies three planning rules. First, \emph{match the elasticity to the horizon}: $-0.16$ for operations within a year, $-0.38$ for capital-turnover horizons. The upper end of the modeled range (beyond $-0.5$) has little support in the corrected record. These menu values are residential-dominated; the long-run response is roughly half as large in industrial and commercial settings (\autoref{tab:sumlr}), so a whole-system planner should apply a demand-weighted blend smaller in magnitude than $-0.38$. Second, \emph{any assumed growth in responsiveness is a technology assumption rather than a demand assumption}: the historical record caps organic growth at roughly $0.04$ per decade in magnitude in every specification precise enough to bound change, so a scenario in which the effective elasticity doubles must specify the hardware, contracts, and automation that deliver the difference, and should price them, because the same pilot literature that finds small elasticities finds large effects of enabling technology. Third, \emph{treat headline literature means as upper bounds}: the raw mean of $-0.27$ reflects selective reporting; every correction philosophy and every identification upgrade moves the number toward zero. For guidance below the aggregate, \autoref{tab:bpcountry} reports a best-practice elasticity for each country with at least five headline estimates, the value an ideally-designed study would recover there; the cross-country spread is modest, and no country's profile supports a large or rising response.

\section{Conclusion}\label{sec:conclusion}

We asked whether the price responsiveness of electricity demand has risen, as four decades of energy planning has assumed it would. Across 4,720 estimates from 462 studies spanning nine decades of data, the answer is no. The corrected short-run elasticity is small ($-0.16$ corrected, $-0.09$ in design-based studies, indistinguishable from zero once composition is held fixed), and nearly every step up the identification ladder, every publication-bias correction, and every technology-rich subsample moves it closer to zero; the one exception is the post-2005 data cell ($-0.22$), which is more elastic than the full sample yet still sits inside the corrected range and shows no upward trend. Responsiveness does grow with adjustment time, doubling toward $-0.38$ as the durable stock turns over, exactly as the quasi-experimental adjustment path and the embodied-capital mechanism predict; but that gradient has been in the data since the 1970s and has not steepened. And on the clock that matters (the vintage of the data, separated from the vintage of the publications), the elasticity shows no trend: in the specifications precise enough to bound it, the record caps any concealed three-decade rise at two-thirds to four-fifths of what a doubling scenario assumes, and the point estimate leans the other way.

We do not claim that responsiveness cannot rise. Automation that responds to prices without human attention, storage that arbitrages tariffs, and contracts that delegate curtailment are all technologies that create price responsiveness, and the pilot record shows they work. Our claim is about the assumption that does the planning work today: demand has never yet become more price-elastic merely because pricing became more sophisticated, and a half-century of evidence limits how much such a change could have gone undetected. Prices alone have not made total electricity consumption more responsive, and we find no evidence that this is changing; flexibility in the broader sense will depend on enabling technology, contracts, and program design.

Three limitations bound these conclusions. First, holding identification and composition fixed moves the short-run data clock from $-0.033$ to $+0.017$. Because design quality is itself correlated with data vintage, the composition controls could in principle absorb a genuine behavioral trend, though the raw record independently shows no trend, which mitigates this concern. Second, the corrected values are read at a zero standard error, an out-of-support extrapolation of the funnel. Third, the design-based benchmark rests on eight short-run studies.

\section*{Use of artificial intelligence}
Claude Opus 4.8 by Anthropic and GPT~5.6 Sol by OpenAI assisted in cross-checking the findings and editing the text, following the guidance of \citet{Cook2026b} and \citet{Cook2026} on the use of AI in meta-analysis. All results were produced by running the analysis code on the dataset and can be reproduced from the public replication package. The authors are responsible for all of the paper's content.

\newpage
\begin{singlespace}
\footnotesize                  
\setlength{\bibsep}{4pt}       
\begin{multicols}{2}
\bibliographystyle{electricity}
\bibliography{electricity}

@article{ackah2014demand,
  title={On the demand dynamics of electricity in {Ghana}: do exogenous non-economic variables count?},
  author={Ackah, Ishmael and Frank, ADU and Takyi, Richard Opoku},
  journal={International Journal of Energy Economics and Policy},
  volume={4},
  number={2},
  pages={149--153},
  year={2014},
  publisher={{I}lhan {O}ZT{U}RK}
}

@TechReport{Acton1976,
  author =      {Acton, J. P. and B.M. Mitchell and R.S. Mowill},
  title =       {{Residential Demand for Electricity in Los Angeles: An Econometric Study of Disaggregated Data}},
  institution = {Rand Corporation, Santa Monica, California},
  year =        {1976},
  type =        {Report},
  number =      {R-1899-NSF}
}

@Article{Acton1980,
  author =  {Acton, J. P. and B. M. Mitchell and R. Sohlberg},
  title =   {{Estimating Residential Electricity Demand Under Declining-Block Tariffs: An Econometric Study Using Micro-Data}},
  journal = {Applied Economics},
  year =    {1980},
  volume =  {12},
  number =  {2},
  pages =   {145-161}
}

@TechReport{ADNDE1981,
  author =      {ADNDE},
  title =       {{Forecasts of Energy Demand and Supply: Primary and Secondary Fuels, {Australia}: 1980-81 to 1089-90}},
  institution = {Australian Department of National Development and Energy, Australian Governemt Publishing Service, Canberra, Australia},
  year =        {1981},
  type =        {Technical Report}
}

@incollection{aigner1985residential,
  title={The residential electricity time-of-use pricing experiments: what have we learned?},
  author={Aigner, Dennis},
  booktitle={Social experimentation},
  pages={11--54},
  year={1985},
  publisher={University of Chicago Press}
}

@article{page2021prisma,
  title={The {PRISMA} 2020 statement: an updated guideline for reporting systematic reviews},
  author={Page, Matthew J. and McKenzie, Joanne E. and Bossuyt, Patrick M. and Boutron, Isabelle and Hoffmann, Tammy C. and Mulrow, Cynthia D. and others},
  journal={BMJ},
  volume={372},
  pages={n71},
  year={2021},
  publisher={BMJ Publishing Group}
}

@TechReport{Akmal2001,
  author =      {Akmal, M. and D.I. Stern},
  title =       {{Residential Energy Demand in {Australia}: An Application of Dynamic OLS}},
  institution = {Centre for Resource and Environmental Studies, Australian National University, Canberra, Australia},
  year =        {2001},
  type =        {Working Papers in Ecological Economics},
  number =      {104/2001}
}

@article{al2018estimating,
  title={Estimating the determinants of electricity consumption in {Jordan}},
  author={Al-Bajjali, Saif Kayed and Shamayleh, Adel Yacoub},
  journal={Energy},
  volume={147},
  pages={1311--1320},
  year={2018},
  publisher={Elsevier}
}

@article{al2018exploring,
  title={Exploring drivers of sectoral electricity demand in {Indonesia}},
  author={Al Irsyad, Muhammad and Nepal, Rabindra and Halog, Anthony},
  journal={Energy Sources, Part B: Economics, Planning, and Policy},
  volume={13},
  number={9-10},
  pages={383--391},
  year={2018},
  publisher={Taylor \& Francis}
}

@article{alberini2011residential,
  title={Residential consumption of gas and electricity in the {US}: The role of prices and income},
  author={Alberini, Anna and Gans, Will and Velez-Lopez, Daniel},
  journal={Energy Economics},
  volume={33},
  number={5},
  pages={870--881},
  year={2011},
  publisher={Elsevier}
}

@article{alberini2011response,
  title={Response of residential electricity demand to price: The effect of measurement error},
  author={Alberini, Anna and Filippini, Massimo},
  journal={Energy Economics},
  volume={33},
  number={5},
  pages={889--895},
  year={2011},
  publisher={Elsevier}
}

@article{alberini2019response,
  title={Response to extreme energy price changes: Evidence from {Ukraine}},
  author={Alberini, Anna and Khymych, Olha and {S}casny, Milan},
  journal={The Energy Journal},
  volume={40},
  number={1},
  pages={189--212},
  year={2019},
  publisher={SAGE Publications Sage CA: Los Angeles, CA}
}

@Article{Al-Faris2002,
  author =  {Al-Faris, A. R. F.},
  title =   {{The Demand for Electricity in the {GCC} Countries}},
  journal = {Energy Policy},
  year =    {2002},
  volume =  {30},
  number =  {2},
  pages =   {117-135}
}

@Article{Al-Sahlawi1999,
  author =  {Al-Sahlawi, M. A.},
  title =   {{Electricity Planning with Demand Estimation and Forecasting in {Saudi Arabia}}},
  journal = {Energy Studies Review},
  year =    {1999},
  volume =  {9},
  number =  {1},
  pages =   {82-88}
}

@article{alter2011empirical,
  title={An empirical analysis of electricity demand in {Pakistan}},
  author={Alter, Noel and Syed, Shabib Haider},
  journal={International Journal of Energy Economics and Policy},
  volume={1},
  number={4},
  pages={116--139},
  year={2011},
  publisher={{I}lhan {O}ZT{U}RK}
}

@article{amarawickrama2008electricity,
  title={Electricity demand for {Sri Lanka}: a time series analysis},
  author={Amarawickrama, Himanshu A and Hunt, Lester C},
  journal={Energy},
  volume={33},
  number={5},
  pages={724--739},
  year={2008},
  publisher={Elsevier}
}

@article{amusa2009aggregate,
  title={Aggregate demand for electricity in {South Africa}: An analysis using the bounds testing approach to cointegration},
  author={Amusa, Hammed and Amusa, Kafayat and Mabugu, Ramos},
  journal={Energy Policy},
  volume={37},
  number={10},
  pages={4167--4175},
  year={2009},
  publisher={Elsevier}
}

@TechReport{Anderson1971,
  author =      {Anderson, K. P.},
  title =       {{Towards Econometric Estimation of Industrial Energy Demand: An Experimental Application to the Primary Metal Industry}},
  institution = {Rand Corporation, Santa Monica, California},
  year =        {1971},
  type =        {Report},
  number =      {R-719-NSF}
}

@Article{Anderson1973a,
  author =  {Anderson, K. P.},
  title =   {{Residential Demand for Electricity: Econometric Estimates for {California} and the {United States}}},
  journal = {The Journal of Business},
  year =    {1973a},
  volume =  {46},
  number =  {4},
  pages =   {526-553}
}

@TechReport{Anderson1973b,
  author =      {Anderson, K. P.},
  title =       {{Residential Energy Use: An Econometric Analysis}},
  institution = {Rand Corporation, Santa Monica, California},
  year =        {1973b},
  type =        {Report},
  number =      {R-1297-NSF}
}

@TechReport{Anderson1974,
  author =      {Anderson, K. P.},
  title =       {{The Price Elasticity of Residential Energy Use}},
  institution = {Rand Corporation, Santa Monica, California},
  year =        {1974},
  type =        {Report},
  number =      {P-5180}
}

@article{andrews2019identification,
  title={Identification of and correction for publication bias},
  author={Andrews, Isaiah and Kasy, Maximilian},
  journal={American Economic Review},
  volume={109},
  number={8},
  pages={2766--2794},
  year={2019},
  publisher={American Economic Association 2014 Broadway, Suite 305, Nashville, TN 37203}
}

@Article{Andrikopoulos1989,
  author =  {Andrikopoulos, A. A. and J. A. Brox and C.C. Paraskevopoulos},
  title =   {{Interfuel and Interfactor Substitution in {Ontario} Manufacturing, 1962-1982}},
  journal = {Applied Economics},
  year =    {1989},
  volume =  {21},
  number =  {12},
  pages =   {1667-1681}
}

@Article{Ang1992,
  author =  {Ang, B.W. and T.N. Goh and X.Q. Liu},
  title =   {{Residential Electricity Demand in {Singapore}}},
  journal = {Energy},
  year =    {1992},
  volume =  {17},
  number =  {1},
  pages =   {37-46}
}

@Article{Apte1983,
  author =  {Apte, P.G.},
  title =   {{Substitution Among Energy and Non-Energy Inputs in Selected {Indian} Manufacturing Industries: An Econometric Analysis}},
  journal = {Indian Economic Journal},
  year =    {1983},
  volume =  {31},
  number =  {2},
  pages =   {71-94}
}

@Article{Archibald1982,
  author =  {Archibald, R.B. and D.H. Finifter and C.E. Moody Jr.},
  title =   {{Seasonal Variation in Residential Electricity Demand: Evidence from Survey Data}},
  journal = {Applied Economics},
  year =    {1982},
  volume =  {14},
  number =  {1},
  pages =   {167-181}
}

@Article{Arisoy2014,
  author =  {Arisoy, I. and I. Ozturk},
  title =   {{Estimating Industrial and Residential Electricity Demand in {Turkey}: A Time Varying Parameter Approach}},
  journal = {Energy},
  year =    {2014},
  volume =  {66},
  number =  {1},
  pages =   {959-964}
}

@article{aroonruengsawat2012impact,
  title={The impact of state level building codes on residential electricity consumption},
  author={Aroonruengsawat, Anin and Auffhammer, Maximilian and Sanstad, Alan H},
  journal={The Energy Journal},
  volume={33},
  number={1},
  pages={31--52},
  year={2012},
  publisher={SAGE Publications Sage CA: Los Angeles, CA}
}

@Article{Arsenault1995,
  author =  {Arsenault, E. and J.T. Bernard and C.W. Carr and E. Genest-Laplante},
  title =   {{A Total Energy Demand Model of Quebec: Forecasting Properties}},
  journal = {Energy Economics},
  year =    {1995},
  volume =  {17},
  number =  {2},
  pages =   {163-171}
}

@TechReport{Asadoorian2006,
  author =      {Asadoorian, M. O. and R. S. Eckaus and C. A. Schlosser},
  title =       {{Modeling Climate Feedbacks to Energy Demand: The Case of {China}}},
  institution = {MIT Joint Program on the Science and Policy of Global Change, Cambridge, Massachusetts},
  year =        {2006},
  type =        {Report},
  number =      {135/2006}
}

@article{asadoorian2008modeling,
  title={Modeling climate feedbacks to electricity demand: The case of {China}},
  author={Asadoorian, Malcolm O and Eckaus, Richard S and Schlosser, C Adam},
  journal={Energy Economics},
  volume={30},
  number={4},
  pages={1577--1602},
  year={2008},
  publisher={Elsevier}
}

@article{aslam2023untangling,
  title={Untangling electricity demand elasticities: Insights from heterogeneous household groups in {Pakistan}},
  author={Aslam, Misbah and Ahmad, Eatzaz},
  journal={Energy},
  volume={282},
  pages={128827},
  year={2023},
  publisher={Elsevier}
}

@Article{Atakhanova2005,
  author =  {Atakhanova, Z. and P. Howie},
  title =   {{Electricity Demand in {Kazakhstan}}},
  journal = {Energy Policy},
  year =    {2005},
  volume =  {35},
  number =  {7},
  pages =   {3729-3743}
}

@article{athukorala2010estimating,
  title={Estimating short and long-term residential demand for electricity: New evidence from {Sri Lanka}},
  author={Athukorala, PPA Wasantha and Wilson, Clevo},
  journal={Energy Economics},
  volume={32},
  pages={S34--S40},
  year={2010},
  publisher={Elsevier}
}

@article{athukorala2019household,
  title={Household demand for electricity: The role of market distortions and prices in competition policy},
  author={Athukorala, Wasantha and Wilson, Clevo and Managi, Shunsuke and Karunarathna, Muditha},
  journal={Energy Policy},
  volume={134},
  pages={110932},
  year={2019},
  publisher={Elsevier}
}

@TechReport{Atkinson1979a,
  author =      {Atkinson, S. F.},
  title =       {{A Comparative Analysis of Consumer Response to Time-of-Use Electricity Pricing: Arizona and Wisconsin}},
  institution = {Electric Power Research Institute, Palo Alto, California},
  year =        {1979a},
  type =        {EPRI Report},
  number =      {EA-1304},
  booktitle =   {Paper presented at EPRI Workshop on "Modeling and Analysis of Electricity Demand by Time-of-Day" on June 12-14, 1978, San Diego, California},
  editor =      {D. Aigner}
}

@Article{Atkinson1979b,
  author =  {Atkinson, S. F.},
  title =   {{Responsiveness to Time-of-Day Electricity Pricing: First Empirical Results}},
  journal = {Journal of Econometrics},
  year =    {1979b},
  volume =  {9},
  number =  {1-2},
  pages =   {79-103}
}

@article{auray2019price,
  title={Price elasticity of electricity demand in {France}},
  author={Auray, Stephane and Caponi, Vincenzo and Ravel, Benoit},
  journal={Economie et Statistique},
  volume={513},
  number={1},
  pages={91--103},
  year={2019},
  publisher={Persee-Portail des revues scientifiques en SHS}
}

@article{azevedo2011residential,
  title={Residential and regional electricity consumption in the {US} and {EU}: How much will higher prices reduce CO2 emissions?},
  author={Azevedo, Ines M Lima and Morgan, M Granger and Lave, Lester},
  journal={The Electricity Journal},
  volume={24},
  number={1},
  pages={21--29},
  year={2011},
  publisher={Elsevier}
}

@inproceedings{babatunde2009demand,
  title={The demand for residential electricity in {Nigeria}: A bound testing approach},
  author={Babatunde, M Adetunji and Shuaibu, M Isa},
  booktitle={Proceedings of 2nd International Workshop on Empirical Methods in Energy Economics},
  year={2009}
}

@Article{Badri1992,
  author =  {Badri, M. A.},
  title =   {{Analysis of Demand for Electricity in the {United States}}},
  journal = {Energy},
  year =    {1992},
  volume =  {17},
  number =  {7},
  pages =   {725-733}
}

@Article{Balabanoff1994,
  author =  {Balabanoff, S.},
  title =   {{The Dynamics of Energy Demand in {Latin America}}},
  journal = {OPEC Review},
  year =    {1994},
  volume =  {18},
  number =  {4},
  pages =   {467-488}
}

@TechReport{Banda2007,
  author =      {Banda, S. H. and L. E. B. Verdugo},
  title =       {{Translog Cost Functions: An Application for {Mexican} Manufacturing}},
  institution = {Banco de Mexico, Mexico City, Mexico},
  year =        {2007},
  type =        {Working paper},
  number =      {8/2007}
}

@Article{Barnes1981,
  author =  {Barnes, R. and R. Gillingham and R. Hagemann},
  title =   {{The Short Run Residential Demand for Electricity}},
  journal = {Review of Economics and Statistics},
  year =    {1981},
  volume =  {63},
  number =  {4},
  pages =   {541-552}
}

@TechReport{Basu1976,
  author =      {Basu, D. R.},
  title =       {{Demand Systems for Energy Commodities, U.K. 1948 - 1972}},
  institution = {University of Birmingham, Birmingham, United Kingdom},
  year =        {1976},
  type =        {Working paper},
  booktitle =   {{Future Energy Policies for the UK: An Optimal Control Approach}},
  chapter =     {{Demand Systems for Energy Commodities, U.K. 1948 - 1972}},
  editor =      {Basu, D. R.},
  pages =       {88-103}
}

@Book{Baughman1979,
  author    = {Baughman, M. L. and Joskow, P. L. and Kamat, D. P.},
  publisher = {MIT Press: Cambridge, Massachusetts},
  title     = {{Electric Power in the {United States}: Models and Policy Analysis}},
  year      = {1979},
}

@Article{Beenstock1999,
  author =  {Beenstock, M. and E. Goldin and D. Nabot},
  title =   {{The Demand for Electricity in {Israel}}},
  journal = {Energy Economics},
  year =    {1999},
  volume =  {21},
  number =  {2},
  pages =   {168-183}
}

@Article{Beierlein1981,
  author =  {Beierlein, J. G. and J. W. Dunn and J. C. Jr. Mcconnon},
  title =   {{The Demand for Electricity and Natural Gas in the Northeastern {United States}}},
  journal = {Review of Economics and Statistics},
  year =    {1981},
  volume =  {63},
  number =  {3},
  pages =   {403-408}
}

@article{bekhet2011assessing,
  title={Assessing the Elasticities of Electricity Consumption for rural and urban areas in {Malaysia}: A Non-linear approach},
  author={Bekhet, Hussain Ali and Othman, NS},
  journal={International Journal of Economics and Finance},
  volume={3},
  number={1},
  pages={208--217},
  year={2011}
}

@Article{Belanger1990,
  author =  {Belanger, D. and J.-T. Bernard and R. Dubois},
  title =   {{Demand for Non-Energy Petroleum Products: The Case for Quebec}},
  journal = {Energy Economics},
  year =    {1990},
  volume =  {12},
  number =  {3},
  pages =   {177-184}
}

@Article{Bernard1987,
  author =  {Bernard, J.-T. and M. Lemieux and S. Thivierge},
  title =   {{Residential Energy Demand: An Integrated Two-Level Approach}},
  journal = {Energy Economics},
  year =    {1987},
  volume =  {9},
  number =  {3},
  pages =   {139-144}
}

@Article{Bernard1996,
  author =  {Bernard, J.-T. and D. Bolduc and D. Belanger},
  title =   {{Quebec Residential Electricity Demand: A Microeconometric Approach}},
  journal = {Canadian Journal of Economics},
  year =    {1996},
  volume =  {29},
  number =  {1},
  pages =   {92-114}
}

@article{bernard2011pseudo,
  title={A pseudo-panel data model of household electricity demand},
  author={Bernard, Jean-Thomas and Bolduc, Denis and Yameogo, Nadege-Desiree},
  journal={Resource and Energy Economics},
  volume={33},
  number={1},
  pages={315--325},
  year={2011},
  publisher={Elsevier}
}

@InProceedings{Berndt1980,
  author =    {Berndt, E. R. and G. May and G. C. Watkins},
  title =     {{An Econometric Model of Alberta Electricity Demand}},
  booktitle = {{Energy Policy Modeling: United States and Canadian Experiences}},
  year =      {1980},
  editor =    {Ziemba, W. T. and S. L. Schwartz and E. Koenigsberg},
  pages =     {103-116},
  publisher = {Martinus Nijhoff Publishing: Leiden}
}

@Article{Berndt1984,
  author =  {Berndt, E. R. and R. Samaniego},
  title =   {{Residential Electricity Demand in {Mexico}: A Model Distinguishing Access from Consumption}},
  journal = {Land Economics},
  year =    {1984},
  volume =  {60},
  number =  {3},
  pages =   {268-269}
}

@TechReport{Bernstein2006,
  author =      {Bernstein, M. A. and J. Griffin},
  title =       {{Regional Differences in the Price-Elasticity of Demand for Energy}},
  institution = {Rand Corporation, Santa Monica, California},
  year =        {2006},
  type =        {Subcontract Report prepared for the National Renewable Energy Laboratory},
  number =      {SR-620-39512}
}

@Article{Betancourt1981,
  author =  {Betancourt, R. R.},
  title =   {{An Econometric Analysis of Peak Electricity Demand in the Short-Run}},
  journal = {Energy Economics},
  year =    {1981},
  volume =  {3},
  number =  {1},
  pages =   {14-29}
}

@article{bianco2009electricity,
  title={Electricity consumption forecasting in {Italy} using linear regression models},
  author={Bianco, Vincenzo and Manca, Oronzio and Nardini, Sergio},
  journal={Energy},
  volume={34},
  number={9},
  pages={1413--1421},
  year={2009},
  publisher={Elsevier}
}

@article{bianco2010analysis,
  title={Analysis and forecasting of nonresidential electricity consumption in {Romania}},
  author={Bianco, Vincenzo and Manca, Oronzio and Nardini, Sergio and Minea, Alina A},
  journal={Applied Energy},
  volume={87},
  number={11},
  pages={3584--3590},
  year={2010},
  publisher={Elsevier}
}

@TechReport{Bigano2006,
  author =      {Bigano, A. and F. Bosello and G. Marano},
  title =       {{Energy Demand and Temperature: A Dynamic Panel Analysis}},
  institution = {Fondazione Eni Enrico Mattei, Milano, Italy},
  year =        {2006},
  type =        {FEEM Working Paper},
  number =      {112/2006}
}

@TechReport{Bjerkholt1983,
  author =      {Bjerkholt, O. and J. Rinde},
  title =       {{Consumption Demand in the MSG Model}},
  institution = {Central Bureau of Statistics, Oslo, Norway},
  year =        {1983},
  type =        {SamfunnsOkonomiske studier},
  number =      {53}
}

@Article{Bjorner2001,
  author =  {Bjorner, T. B. and M. Togeby and H. H. Jensen},
  title =   {{Industrial Companies' Demand for Electricity: Evidence from a Micropanel}},
  journal = {Energy Economics},
  year =    {2001},
  volume =  {23},
  number =  {5},
  pages =   {595-617}
}

@article{blasch2017explaining,
  title={Explaining electricity demand and the role of energy and investment literacy on end-use efficiency of {Swiss} households},
  author={Blasch, Julia and Boogen, Nina and Filippini, Massimo and Kumar, Nilkanth},
  journal={Energy Economics},
  volume={68},
  pages={89--102},
  year={2017},
  publisher={Elsevier}
}

@article{blazquez2013residential,
  title={Residential electricity demand in {Spain}: new empirical evidence using aggregate data},
  author={Blazquez, Leticia and Boogen, Nina and Filippini, Massimo},
  journal={Energy Economics},
  volume={36},
  pages={648--657},
  year={2013},
  publisher={Elsevier}
}

@Article{Blundell1999,
  author =  {Blundell, R. and J. M. Robin},
  title =   {{Estimation in Large and Disaggregated Demand Systems: An Estimator for Conditionally Linear Systems}},
  journal = {Journal of Applied Econometrics},
  year =    {1999},
  volume =  {14},
  number =  {3},
  pages =   {209-232}
}

@article{bom2019kinked,
  title={A kinked meta-regression model for publication bias correction},
  author={Bom, Pedro RD and Rachinger, Heiko},
  journal={Research Synthesis Methods},
  volume={10},
  number={4},
  pages={497--514},
  year={2019},
  publisher={Wiley Online Library}
}

@article{boogen2017dynamic,
  title={Dynamic models of residential electricity demand: Evidence from {Switzerland}},
  author={Boogen, Nina and Datta, Souvik and Filippini, Massimo},
  journal={Energy Strategy Reviews},
  volume={18},
  pages={85--92},
  year={2017},
  publisher={Elsevier}
}

@TechReport{borenstein2009electricity,
  author      = {Borenstein, Severin},
  title       = {{To What Electricity Price Do Consumers Respond? Residential Demand Elasticity Under Increasing-Block Pricing}},
  institution = {Center for the Study of Energy Markets, Haas School of Business, University of California, Berkeley},
  type        = {Working Paper},
  number      = {CSEM WP 195},
  year        = {2009}
}

@Article{Bose1999,
  author =  {Bose, R. K. and M. Shukla},
  title =   {{Elasticities of Electricity Demand in {India}}},
  journal = {Energy Policy},
  year =    {1999},
  volume =  {27},
  number =  {3},
  pages =   {137-146}
}

@Article{Botero1990,
  author =  {Botero, J. and Castano E. and C. E. Velez},
  title =   {{Modelo Economico De Demanda De Energia Electrica En La Industria Colombiana}},
  journal = {Lecturas de Economia},
  year =    {1990},
  volume =  {32-33},
  pages =   {97-124}
}

@Article{Branch1993,
  author =  {Branch, R.},
  title =   {{Short Run Income Elasticity of Demand for Residential Electricity Using Consumer Expenditure Survey Data}},
  journal = {The Energy Journal},
  year =    {1993},
  volume =  {14},
  number =  {4},
  pages =   {111-121}
}

@Article{Brenton1997,
  author =  {Brenton, P.},
  title =   {{Estimates of the Demand for Energy Using Cross-Country Consumption Data}},
  journal = {Applied Economics},
  year =    {1997},
  volume =  {29},
  number =  {7},
  pages =   {851-859}
}

@article{burke2018electricity,
  title={Electricity subsidy reform in {Indonesia}: Demand-side effects on electricity use},
  author={Burke, Paul J and Kurniawati, Sandra},
  journal={Energy Policy},
  volume={116},
  pages={410--421},
  year={2018},
  publisher={Elsevier}
}

@article{burke2018price,
  title={The price elasticity of electricity demand in the {United States}: A three-dimensional analysis},
  author={Burke, Paul J and Abayasekara, Ashani},
  journal={The Energy Journal},
  volume={39},
  number={2},
  pages={123--146},
  year={2018},
  publisher={SAGE Publications Sage CA: Los Angeles, CA}
}

@article{byrne2021experimental,
  title={An experimental study of monthly electricity demand (in) elasticity},
  author={Byrne, David P and Nauze, Andrea La and Martin, Leslie A},
  journal={The Energy Journal},
  volume={42},
  number={2},
  pages={205--222},
  year={2021},
  publisher={SAGE Publications Sage CA: Los Angeles, CA}
}

@article{campbell2018price,
  title={Price and income elasticities of electricity demand: Evidence from {Jamaica}},
  author={Campbell, Alrick},
  journal={Energy Economics},
  volume={69},
  pages={19--32},
  year={2018},
  publisher={Elsevier}
}

@article{cao2019chinese,
  title={{Chinese} residential electricity consumption: Estimation and forecast using micro-data},
  author={Cao, Jing and Ho, Mun Sing and Li, Yating and Newell, Richard G and Pizer, William A},
  journal={Resource and Energy Economics},
  volume={56},
  pages={6--27},
  year={2019},
  publisher={Elsevier}
}

@article{cao2023experiment,
  title={An experiment in own-price elasticity estimation for non-residential electricity demand in the {US}},
  author={Cao, Kang Hua and Qi, HS and Li, Raymond and Woo, Chi-Keung and Tishler, Asher and Zarnikau, Jay},
  journal={Utilities Policy},
  volume={81},
  pages={101489},
  year={2023},
  publisher={Elsevier}
}

@Article{Cargill1971,
  author =  {Cargill, T.E. and R.A. Meyer},
  title =   {{Estimating the Demand for Electricity by Time of Day}},
  journal = {Applied Economics},
  year =    {1971},
  volume =  {3},
  number =  {4},
  pages =   {233-246}
}

@Article{Carlevaro1983,
  author =  {Carlevaro, F. and C. Spierer},
  title =   {{Dynamic Energy Demand Models with Latent Equipment}},
  journal = {European Economic Review},
  year =    {1983},
  volume =  {23},
  number =  {2},
  pages =   {161-194}
}

@Article{Cavoulacos1983,
  author =  {Cavoulacos, P. and M. Caramanis},
  title =   {{Energy and Other Factor Input Demand in {Greek} Manufacturing, 1963-1975}},
  journal = {Greek Economic Review},
  year =    {1983},
  volume =  {5},
  number =  {2},
  pages =   {158-181}
}

@article{cebula2012us,
  title={{US} residential electricity consumption: the effect of states' pursuit of energy efficiency policies},
  author={Cebula, Richard J},
  journal={Applied Economics Letters},
  volume={19},
  number={15},
  pages={1499--1503},
  year={2012},
  publisher={Taylor \& Francis}
}

@Article{Chang1981a,
  author =  {Chang, H. S. and W. S. Chern},
  title =   {{Specification, Estimation, and Forecasts of Industrial Demand and Price of Electricity}},
  journal = {Energy Systems and Policy},
  year =    {1981a},
  volume =  {5},
  number =  {3},
  pages =   {219-242}
}

@Article{Chang1981b,
  author =  {Chang, H. S. and W. S. Chern},
  title =   {{A Study on the Demand for Electricity and the Variation in the Price Elasticities for Manufacturing Industries}},
  journal = {Journal of Economics and Business},
  year =    {1981b},
  volume =  {33},
  number =  {2},
  pages =   {122-131}
}

@Article{Chang1991,
  author =  {Chang, H. S. and Y. Hsing},
  title =   {{The Demand for Residential Electricity: New Evidence on Time-Varying Elasticities}},
  journal = {Applied Economics},
  year =    {1991},
  volume =  {23},
  number =  {7},
  pages =   {1251-1256}
}

@Article{Chaudhary1999,
  author =  {Chaudhary, M. A. and E. Ahmad and A. A. Burki and M. A. Khan},
  title =   {{Industrial Sector Input Demand Responsiveness and Policy Interventions}},
  journal = {Pakistan Development Review},
  year =    {1999},
  volume =  {38},
  number =  {4},
  pages =   {1083-1100}
}

@article{chaudhry2010panel,
  title={A Panel Data Analysis of Electricity Demand in {Pakistan}.},
  author={Chaudhry, Azam Amjad},
  journal={Lahore Journal of Economics},
  volume={15},
  year={2010}
}

@InProceedings{Chern1975,
  author =    {Chern, W. S.},
  title =     {{Estimating Industrial Demand for Electricity: Methodology and Empirical Evidence}},
  booktitle = {{Energy: Mathematics and Models}},
  year =      {1975},
  series =    {Proceedings of the "Siam Institute for Mathematics and Society Conference" held on July 7-11, 1975, At Alta, Utah},
  pages =     {103-120}
}

@InProceedings{Chern1978,
  author =    {Chern, W. S.},
  title =     {{Aggregate Demand for Energy in the {United States}}},
  booktitle = {{Econometric Studies in Energy Demand and Supply}},
  year =      {1978},
  editor =    {Maddala, G. S. and W. S. Chern and G. S. Gill},
  pages =     {5-41},
  publisher = {Praeger Publishers: New York}
}

@Article{Chern1988,
  author =  {Chern, W. S. and E. Bouis},
  title =   {{Structural Changes in Residential Electricity Demand}},
  journal = {Energy Economics},
  year =    {1988},
  volume =  {10},
  number =  {3},
  pages =   {213-222}
}

@Article{Chishti1993,
  author =  {Chishti, S.},
  title =   {{Recursively Bootstrapped Probability Distribution of Electricity Demand Forecast in {Pakistan}}},
  journal = {The Journal of Energy and Development},
  year =    {1993},
  volume =  {18},
  number =  {2},
  pages =   {223-231}
}

@PhdThesis{Choi2002,
  author = {Choi, J.},
  title =  {{Short-Run and Long-Run Elasticities of Electricity Demand in the Public Sector: A Case Study of the U.S. Navy Bases}},
  school = {Department of Economics, George Washington University, Washington, D.C.},
  year =   {2002},
  type =   {Ph.D. thesis}
}

@Article{Christodoulakis1997,
  author =  {Christodoulakis, N. M. and S. C. Kalyvitis},
  title =   {{The Demand for Energy in {Greece}: Assessing the Effects of the Community Support Framework 1994-1999}},
  journal = {Energy Economics},
  year =    {1997},
  volume =  {19},
  number =  {4},
  pages =   {393-416}
}

@Article{Christopoulos2000,
  author =  {Christopoulos, D.},
  title =   {{The Demand for Energy in {Greek} Manufacturing}},
  journal = {Energy Economics},
  year =    {2000},
  volume =  {22},
  number =  {5},
  pages =   {569-586}
}

@Article{Chung1981,
  author =  {Chung, C. and D. Aigner},
  title =   {{Industrial and Commercial Demand for Electricity by Time-of-Day: A {California} Case Study}},
  journal = {The Energy Journal},
  year =    {1981},
  volume =  {2},
  number =  {3},
  pages =   {91-110}
}

@article{cialani2018household,
  title={Household and industrial electricity demand in {Europe}},
  author={Cialani, Catia and Mortazavi, Reza},
  journal={Energy Policy},
  volume={122},
  pages={592--600},
  year={2018},
  publisher={Elsevier}
}

@article{cicchetti1975alternative,
  title={Alternative price measures and the residential demand for electricity: A specification analysis},
  author={Cicchetti, Charles J and Smith, V Kerry},
  journal={Regional Science and Urban Economics},
  volume={5},
  number={4},
  pages={503--516},
  year={1975},
  publisher={Elsevier}
}

@TechReport{CISEPA1998,
  author =      {CISEPA},
  title =       {{Proyeccion del Consumo Mensual de Energia Electrica, Junio 1997-Diciembre 2000}},
  institution = {Centro de Investigaciones Sociologicas, Economicas, Politicas y Antropologicas, Lima, Peru},
  year =        {1998},
  type =        {Informe Final: Consultoria para Comision de Tarifas de Energia}
}

@Article{Considine2000,
  author =  {Considine, T.},
  title =   {{The Impacts of Weather Variations on Energy Demand and Carbon Emissions}},
  journal = {Resource and Energy Economics},
  year =    {2000},
  volume =  {22},
  number =  {4},
  pages =   {295-312}
}

@TechReport{Coughlin1995,
  author =      {Coughlin, R. M.},
  title =       {{The Estimation of Residential Price Elasticities for {New England} During A Period of Increasing Demand-Side Management}},
  institution = {Electric Power Research Institute, Palo Alto, California},
  year =        {1995},
  type =        {EPRI Technical report},
  number =      {TR-105012}
}

@article{cuddington2015estimating,
  title={Estimating short and long-run demand elasticities: a primer with energy-sector applications},
  author={Cuddington, John T and Dagher, Leila},
  journal={The Energy Journal},
  volume={36},
  number={1},
  pages={185--210},
  year={2015},
  publisher={SAGE Publications Sage CA: Los Angeles, CA}
}

@PhdThesis{Dahan1996,
  author = {Dahan, A. A.},
  title =  {{Energy Consumption in {Yemen}: Economics and Policy 1970-1990}},
  school = {Department of Mining and Geology, University of Arizona, Arizona},
  year =   {1996},
  type =   {Ph.D. thesis}
}

@TechReport{dahl1993,
  author      = {Dahl, Carol A.},
  title       = {{A Survey of Energy Demand Elasticities in Support of the Development of the NEMS}},
  institution = {University Library of Munich, Germany},
  type        = {MPRA Paper},
  number      = {13962},
  year        = {1993}
}

@inproceedings{dahl2011,
  title={A global survey of electricity demand elasticities},
  author={Dahl, Carol},
  booktitle={Institutions, Efficiency and Evolving Energy Technologies, 34th IAEE International Conference, June 19-23, 2011},
  year={2011},
  organization={International Association for Energy Economics}
}

@article{davis2008durable,
  title={Durable goods and residential demand for energy and water: evidence from a field trial},
  author={Davis, Lucas W},
  journal={The RAND Journal of Economics},
  volume={39},
  number={2},
  pages={530--546},
  year={2008},
  publisher={Wiley Online Library}
}

@TechReport{DeCian2007,
  author =      {De Cian, E. and E. Lanzi and R. Roson},
  title =       {{The Impact of Temperature Change on Energy: A Dynamic Panel Analysis}},
  institution = {Fondazione Eni Enrico Mattei, Milano, Italy},
  year =        {2007},
  type =        {FEEM Working Paper},
  number =      {46/2007}
}

@Article{Delfino1995,
  author =  {Delfino, J. A.},
  title =   {{La Demanda Industrial de Energia en Argentia: Una Estimacion Integral por Etapas}},
  journal = {Economica},
  year =    {1995},
  volume =  {41},
  number =  {2},
  pages =   {125-149}
}

@Article{Denton1999,
  author =  {Denton, F.T. and D.C. Mountain and B.G. Spencer},
  title =   {{Energy Use in the Commercial Sector: Estimated Intensities and Costs for {Canada} Based on {US} Survey Data}},
  journal = {Energy Studies Review},
  year =    {1999},
  volume =  {9},
  number =  {1},
  pages =   {24-46}
}

@Article{Denton2003,
  author =  {Denton, F. T. and D. C. Mountain and B. G. Spencer},
  title =   {{Energy Demand with Declining Rate Schedules: An Econometric Model for the U.S. Commercial Sector}},
  journal = {Land Economics},
  year =    {2003},
  volume =  {79},
  number =  {1},
  pages =   {86-105}
}

@Article{Dergiades2008,
  author =  {Dergiades, T. and L. Tsoulfidis},
  title =   {{Estimating Residential Demand for Electricity in the {United States}, 1965--2006}},
  journal = {Energy Economics},
  year =    {2008},
  volume =  {30},
  number =  {5},
  pages =   {2722-2730}
}

@article{deryugina2020long,
  title={The long-run dynamics of electricity demand: Evidence from municipal aggregation},
  author={Deryugina, Tatyana and MacKay, Alexander and Reif, Julian},
  journal={American Economic Journal: Applied Economics},
  volume={12},
  number={1},
  pages={86--114},
  year={2020},
  publisher={American Economic Association 2014 Broadway, Suite 305, Nashville, TN 37203-2425}
}

@Article{DeVita2006,
  author =  {De Vita, G. and K. Endresen and L.C. Hunt},
  title =   {{An Empirical Analysis of Energy Demand in {Namibia}}},
  journal = {Energy Policy},
  year =    {2006},
  volume =  {34},
  number =  {18},
  pages =   {3447--3463}
}

@Article{Diabli1998,
  author =  {Diabli, A.},
  title =   {{The Demand for Electric Energy in {Saudi Arabia}: An Empirical Investigation}},
  journal = {OPEC Review},
  year =    {1998},
  volume =  {22},
  number =  {1},
  pages =   {13--29}
}

@article{dilaver2011industrial,
  title={Industrial electricity demand for {Turkey}: a structural time series analysis},
  author={Dilaver, Zafer and Hunt, Lester C},
  journal={Energy Economics},
  volume={33},
  number={3},
  pages={426--436},
  year={2011},
  publisher={Elsevier}
}

@article{dilaver2011turkish,
  title={{Turkish} aggregate electricity demand: An outlook to 2020},
  author={Dilaver, Zafer and Hunt, Lester C},
  journal={Energy},
  volume={36},
  number={11},
  pages={6686--6696},
  year={2011},
  publisher={Elsevier}
}

@Book{Dobozi1988,
  title =     {{An Empirical Estimation of the Price Responsiveness of the Hungarian Economy: The Case of Energy Demand}},
  publisher = {Hungarian Scientific Council for the World Economy: Budapest},
  year =      {1988},
  author =    {Dobozi, I.}
}

@Article{Dodgson1990,
  author =  {Dodgson, J. S. and R. Millward and R. Ward},
  title =   {{The Decline in Residential Electricity Consumption in {England} and {Wales}}},
  journal = {Applied Economics},
  year =    {1990},
  volume =  {22},
  number =  {1},
  pages =   {56-68}
}

@Article{Donatos1989,
  author =  {Donatos, G. S. and G. J. Mergos},
  title =   {{Energy Demand in {Greece}: The Impact of Two Energy Crisis}},
  journal = {Energy Economics},
  year =    {1989},
  volume =  {11},
  number =  {2},
  pages =   {147-152}
}

@Article{Donatos1991,
  author =  {Donatos, G. S. and G. J. Mergos},
  title =   {{Residential Demand for Electricity: The Case of {Greece}}},
  journal = {Energy Economics},
  year =    {1991},
  volume =  {31},
  number =  {1},
  pages =   {41-47}
}

@article{dong2018price,
  author  = {Dong, Jinwoo and Kim, Young-Duk},
  title   = {{Price Elasticity of Electricity Demand with Temperature Effect in {South Korea}: Empirical Evidence}},
  journal = {Global Conference on Business and Finance Proceedings},
  volume  = {13},
  number  = {1},
  pages   = {219--225},
  year    = {2018}
}

@inproceedings{dong2020estimating,
  title={Estimating the price elasticity of electricity of urban residential consumers in eastern {China}},
  author={Dong, Zhen and Liu, Zhe and Liu, Jing and Li, Lihua and Zhao, Jingqian},
  booktitle={2020 Asia Energy and Electrical Engineering Symposium (AEEES)},
  pages={881--885},
  year={2020},
  organization={IEEE}
}

@Article{Donnelly1984,
  author =  {Donnelly, W. A.},
  title =   {{Residential Electricity Demand Modeling in the {Australian} Capital Territory: Preliminary Results}},
  journal = {The Energy Journal},
  year =    {1984},
  volume =  {5},
  number =  {2},
  pages =   {119-131}
}

@TechReport{Donnelly1984a,
  author =      {Donnelly, W. A. and M. Diesendorf},
  title =       {{Note on an Econometric Analysis of Peak Electricity Demand in the Short-Run}},
  institution = {Centre for Resource and Environmental Studies, Australian National University, Canberra, Australia},
  year =        {1984},
  type =        {CRES Working Paper},
  number =      {18/1984}
}

@Article{Donnelly1984b,
  author =  {Donnelly, W. A. and H. D. W. Saddler},
  title =   {{The Retail Demand for Electricity in {Tasmania}}},
  journal = {Australian Economic Papers},
  year =    {1984},
  volume =  {23},
  number =  {42},
  pages =   {52-60}
}

@InProceedings{Donnelly1985,
  author =    {Donnelly, W. A.},
  title =     {{Electricity Demand Modelling}},
  booktitle = {{New Mathematical Advances in Economic Dynamics}},
  year =      {1985},
  editor =    {Batten, D. F. and P. F. Lesse},
  pages =     {179-195},
  publisher = {New York University Press: New York}
}

@Book{Donnelly1987,
  title =     {{The Econometrics of Energy Demand: A Survey of Applications}},
  publisher = {Praeger Publishers: New York},
  year =      {1987},
  author =    {Donnelly, W. A.}
}

@Article{Douthitt1989,
  author =  {Douthitt, R. A.},
  title =   {{An Economic Analysis of the Demand for Residential Space Heating in {Canada}}},
  journal = {Energy},
  year =    {1989},
  volume =  {14},
  number =  {4},
  pages =   {187-197}
}

@Book{Dubin1985,
  title =     {{Consumer Durable Choice and the Demand for Electricity}},
  publisher = {North-Holland Publishing Company: New York},
  year =      {1985},
  author =    {Dubin, J. A.}
}

@Article{Duncan1976,
  author =  {Duncan, R. C. and H. P. Binswanger},
  title =   {{Energy Sources: Substitutability and Biases in {Australia}}},
  journal = {Australian Economic Papers},
  year =    {1976},
  volume =  {15},
  number =  {27},
  pages =   {289-301}
}

@Article{Dunstan1988,
  author =  {Dunstan, R. H. and R. H. Schmidt},
  title =   {{Structural Changes in Residential Energy Demand}},
  journal = {Energy Economics},
  year =    {1988},
  volume =  {10},
  number =  {3},
  pages =   {206-212}
}

@article{durmaz2020estimation,
  title={Estimation of residential electricity demand in {Hong Kong} under electricity charge subsidies},
  author={Durmaz, Tunc and Pommeret, Aude and Tastan, Huseyin},
  journal={Energy Economics},
  volume={88},
  pages={104742},
  year={2020},
  publisher={Elsevier}
}

@article{ekpo2011dynamics,
  title={The dynamics of electricity demand and consumption in {Nigeria}: application of the bounds testing approach},
  author={Ekpo, Udo N and Chuku, Chuku A and Effiong, Ekpeno L},
  journal={Current Research Journal of Economic Theory},
  volume={3},
  number={2},
  pages={43--52},
  year={2011},
  publisher={Maxwell Science Publishing}
}

@Article{Eltony1993,
  author =  {Eltony, M. N. and Y. H. Mohammad},
  title =   {{The Structure of Demand for Electricity in the Gulf Cooperation Council Countries}},
  journal = {The Journal of Energy and Development},
  year =    {1993},
  volume =  {18},
  number =  {2},
  pages =   {213-221}
}

@Article{Eltony1995,
  author =  {Eltony, M. N.},
  title =   {{The Sectoral Demand for Electricity in {Kuwait}}},
  journal = {OPEC Review},
  year =    {1995},
  volume =  {19},
  number =  {1},
  pages =   {37-44}
}

@Article{Eltony1996,
  author =  {Eltony, M. N. and A. Hoque},
  title =   {{A Cointegrating Relationship in the Demand for Energy: The Case of Electricity in {Kuwait}}},
  journal = {The Journal of Energy and Development},
  year =    {1996},
  volume =  {21},
  number =  {2},
  pages =   {293-302}
}

@Article{Eltony1999,
  author =  {Eltony, M. N. and M. Hajeeh},
  title =   {{Electricity Demand by the Commercial Sector in {Kuwait}: An Econometric Analysis}},
  journal = {OPEC Review},
  year =    {1999},
  volume =  {23},
  number =  {1},
  pages =   {23-33}
}

@Article{Eltony2004,
  author =  {Eltony, M. N.},
  title =   {{A Model for Forecasting and Planning: The Case for Energy Demand in {Kuwait}}},
  journal = {The Journal of Energy and Development},
  year =    {2004},
  volume =  {30},
  number =  {1},
  pages =   {91-108}
}

@Article{Eltony2006,
  author =  {Eltony, M. N.},
  title =   {{Industrial Energy Policy: A Case Study of Demand in {Kuwait}}},
  journal = {OPEC Review},
  year =    {2006},
  volume =  {30},
  number =  {2},
  pages =   {85-103}
}

@Article{Eltony2007,
  author =  {Eltony, M. N. and A. Al-Awadhi},
  title =   {{The Commercial Sector Demand for Energy in {Kuwait}}},
  journal = {OPEC Review},
  year =    {2007},
  volume =  {31},
  number =  {1},
  pages =   {17-26}
}

@InProceedings{Erickson1973,
  author =    {Erickson, E. W. and R. M. Spann and R. Ciliano},
  title =     {{Substitution and Usage in Energy Demand: An Econometric Estimation of Long-Run and Short-Run Effects}},
  booktitle = {{Energy Modeling: Art, Science, Practice}},
  year =      {1973},
  editor =    {Searl, M. F.},
  pages =     {190-212},
  publisher = {Resources for the Future, Inc.: Washington D.C.}
}

@TechReport{Eskeland1994,
  author =      {Eskeland, G. S. and E. Jimenez and L. Liu},
  title =       {{Energy Pricing and Air Pollution: Econometric Evidence from Manufacturing in {Chile} and {Indonesia}}},
  institution = {World Bank, Washington, D.C.},
  year =        {1994},
  type =        {Policy Research Working Paper},
  number =      {1323/1994}
}

@article{eskeland2010electricity,
  title={Electricity demand in a changing climate},
  author={Eskeland, Gunnar S and Mideksa, Torben K},
  journal={Mitigation and Adaptation Strategies for Global Change},
  volume={15},
  pages={877--897},
  year={2010},
  publisher={Springer}
}

@article{espey2004,
  title={Turning on the lights: A meta-analysis of residential electricity demand elasticities},
  author={Espey, James A and Espey, Molly},
  journal={Journal of Agricultural and Applied Economics},
  volume={36},
  number={1},
  pages={65--81},
  year={2004},
  publisher={Cambridge University Press}
}

@article{fan2019impacts,
  title={Impacts of climate change on electricity demand in {China}: An empirical estimation based on panel data},
  author={Fan, Jing-Li and Hu, Jia-Wei and Zhang, Xian},
  journal={Energy},
  volume={170},
  pages={880--888},
  year={2019},
  publisher={Elsevier}
}

@Article{Fatai2003,
  author =  {Fatai, K. and L. Oxley and F.G. Scrimgeour},
  title =   {{Modeling and Forecasting the Demand for Electricity in {New Zealand}: A Comparison of Alternative Approaches}},
  journal = {The Energy Journal},
  year =    {2003},
  volume =  {24},
  number =  {1},
  pages =   {75-103}
}

@MastersThesis{fatima2023price,
  author = {Fatima, Zainab},
  title  = {{Price \& Income Elasticity of Residential Electricity Demand in {Asia}: A Meta-Analysis}},
  school = {Charles University, Prague},
  year   = {2023}
}

@article{fell2014new,
  title={A new look at residential electricity demand using household expenditure data},
  author={Fell, Harrison and Li, Shanjun and Paul, Anthony},
  journal={International Journal of Industrial Organization},
  volume={33},
  pages={37--47},
  year={2014},
  publisher={Elsevier}
}

@Article{Filippini1995a,
  author =  {Filippini, M.},
  title =   {{Swiss Residential Demand for Electricity by Time-of-Use}},
  journal = {Resource and Energy Economics},
  year =    {1995a},
  volume =  {17},
  number =  {3},
  pages =   {281-290}
}

@Article{Filippini1995b,
  author =  {Filippini, M.},
  title =   {{Electricity Demand by Time-of-Use: An Application of the Household AIDS Model}},
  journal = {Energy Economics},
  year =    {1995b},
  volume =  {17},
  number =  {3},
  pages =   {197-204}
}

@Article{Filippini1999,
  author =  {Filippini, M.},
  title =   {{Swiss Residential Demand for Electricity}},
  journal = {Applied Economics Letters},
  year =    {1999},
  volume =  {6},
  number =  {8},
  pages =   {533-538}
}

@Article{Filippini2004,
  author =  {Filippini, M. and S. Pachauri},
  title =   {{Elasticities of Electricity Demand in Urban {Indian} Households}},
  journal = {Energy Policy},
  year =    {2004},
  volume =  {32},
  number =  {3},
  pages =   {429-436}
}

@Article{Filippini2011,
  author =  {Filippini, M.},
  title =   {{Short- and Long-Run Time-Of-Use Price Elasticities in {Swiss} Residential Electricity Demand}},
  journal = {Energy Policy},
  year =    {2011},
  volume =  {39},
  number =  {10},
  pages =   {5811--5817}
}

@article{filippini2018habits,
  title={Habits and rational behaviour in residential electricity demand},
  author={Filippini, Massimo and Hirl, Bettina and Masiero, Giuliano},
  journal={Resource and Energy Economics},
  volume={52},
  pages={137--152},
  year={2018},
  publisher={Elsevier}
}

@Book{Fisher1962,
  title =     {{A Study in Econometrics: The demand for electricity in the {United States}}},
  publisher = {North-Holland Publishing Company: Amsterdam},
  year =      {1962},
  author =    {Fisher, F. M. and G. S. Kaysen}
}

@Article{Fouquet1995,
  author =  {Fouquet, R.},
  title =   {{The Impact of VAT Introduction on {UK} Residential Energy Demand an Investigation Using the Cointegration Approach}},
  journal = {Energy Economics},
  year =    {1995},
  volume =  {17},
  number =  {3},
  pages =   {237-247}
}

@article{frondel2019,
  title={Heterogeneity in {German} residential electricity consumption: a quantile regression approach},
  author={Frondel, Manuel and Sommer, Stephan and Vance, Colin},
  journal={Energy Policy},
  volume={131},
  pages={370--379},
  year={2019},
  publisher={Elsevier}
}

@article{furukawa2019publication,
  title={Publication bias under aggregation frictions: Theory, evidence, and a new correction method},
  author={Furukawa, Chishio},
  journal={Evidence, and a New Correction Method (March 29, 2019)},
  year={2019}
}

@article{gam2012electricity,
  title={Electricity demand in Tunisia},
  author={Gam, Imen and Rejeb, Jaleleddine Ben},
  journal={Energy Policy},
  volume={45},
  pages={714--720},
  year={2012},
  publisher={Elsevier}
}

@TechReport{Garbacz1983a,
  author =      {Garbacz, C.},
  title =       {{Residential Energy Demand: A National Micro-Based Model}},
  institution = {Presented at the "American Economic Association Meeting" on Dec. 28-30, 1983, San Francisco, California},
  year =        {1983a}
}

@Article{Garbacz1983b,
  author =  {Garbacz, C.},
  title =   {{Electricity Demand and the Elasticity of Intra-Marginal Price}},
  journal = {Applied Economics},
  year =    {1983b},
  volume =  {15},
  number =  {5},
  pages =   {699-701}
}

@Article{Garbacz1983c,
  author =  {Garbacz, C.},
  title =   {{A Model of Residential Demand for Electricity Using National Household Sample}},
  journal = {Energy Economics},
  year =    {1983c},
  volume =  {5},
  number =  {2},
  pages =   {124-128}
}

@Article{Garbacz1984a,
  author =  {Garbacz, C.},
  title =   {{A National Micro-Data Based Model of Residential Electricity Demand: New Evidence on Seasonal Variation}},
  journal = {Southern Economic Journal},
  year =    {1984a},
  volume =  {51},
  number =  {1},
  pages =   {235-249}
}

@TechReport{Garbacz1984b,
  author =      {Garbacz, C.},
  title =       {{Regional Residential Electricity Demand}},
  institution = {Department of Economics, Missouri University of Science and Technology, Rolla, Missouri},
  year =        {1984b},
  type =        {Technical report}
}

@Article{Garbacz1984c,
  author =  {Garbacz, C.},
  title =   {{Residential Electricity Demand: A Suggested Appliance Stock Equation}},
  journal = {The Energy Journal},
  year =    {1984c},
  volume =  {5},
  number =  {2},
  pages =   {150-154}
}

@Article{Garbacz1986,
  author =  {Garbacz, C.},
  title =   {{Seasonal and Regional Residential Electricity Demand: A Micro-Based National Model}},
  journal = {The Energy Journal},
  year =    {1986},
  volume =  {7},
  number =  {2},
  pages =   {121-134}
}

@Article{Garcia-Cerrutti2000,
  author =  {Garcia-Cerrutti, L. M.},
  title =   {{Estimating Elasticities of Residential Energy Demand from Panel County Data Using Dynamic Random Variables Models with Heteroskedastic and Correlated Error Terms}},
  journal = {Energy Economics},
  year =    {2000},
  volume =  {22},
  number =  {4},
  pages =   {355-366}
}

@article{gautam2018estimating,
  title={Estimating sectoral demands for electricity using the pooled mean group method},
  author={Gautam, Tej K and Paudel, Krishna P},
  journal={Applied Energy},
  volume={231},
  pages={54--67},
  year={2018},
  publisher={Elsevier}
}

@incollection{george2010dilution,
  title={Dilution priors: Compensating for model space redundancy},
  author={George, Edward I},
  booktitle={Borrowing Strength: Theory Powering Applications--A Festschrift for Lawrence D. Brown},
  volume={6},
  pages={158--166},
  year={2010},
  publisher={Institute of Mathematical Statistics}
}

@InProceedings{Gill1978,
  author =    {Gill, S. G. and G. S. Maddala},
  title =     {{Residential Demand for Electricity in the TVA Area}},
  booktitle = {{Econometric Studies in Energy Demand and Supply}},
  year =      {1978},
  editor =    {Maddala, G. S. and W. S. Chern and G. S. Gill},
  pages =     {44-59},
  publisher = {Praeger Publishers: New York}
}

@Article{Glakpe1985,
  author =  {Glakpe, E. and R. Fazzolare},
  title =   {{Economic Demand Analysis for Electricity in West {Africa}}},
  journal = {The Energy Journal},
  year =    {1985},
  volume =  {6},
  number =  {1},
  pages =   {137-144}
}

@Book{Gollnick1975,
  title =     {{Dynamic Structure of Household Expenditures in the Federal Republic of {Germany}: Analysis and Projections 1955-1969/1971 and 1975/1977}},
  publisher = {North-Holland Publishing Company: Amsterdam},
  year =      {1975},
  author =    {Gollnick, H. G. L.}
}

@Article{Green1986,
  author =  {Green, R. D. and A. G. Salley and G. Grass and A. S. Osei},
  title =   {{The Demand for Heating Fuels: A Disaggregated Modeling Approach}},
  journal = {Atlantic Economic Journal},
  year =    {1986},
  volume =  {14},
  number =  {4},
  pages =   {1-14}
}

@Article{Gundimeda2008,
  author =  {Gundimeda, H. and G. Kohlin},
  title =   {{Fuel Demand Elasticities for Energy and Environmental Policies: {Indian} Sample Survey Evidence}},
  journal = {Energy Economics},
  year =    {2008},
  volume =  {30},
  number =  {2},
  pages =   {517-546}
}

@TechReport{Guo1994,
  author =      {Guo, C. and J. R. Tybout},
  title =       {{How Relative Prices Affect Fuel Use Patterns in Manufacturing: Plant Level Evidence from {Chile}}},
  institution = {World Bank, Washington, D.C.},
  year =        {1994},
  type =        {Policy Research Working Paper},
  number =      {1297/1994}
}

@article{halicioglu2007residential,
  title={Residential electricity demand dynamics in {Turkey}},
  author={Halicioglu, Ferda},
  journal={Energy Economics},
  volume={29},
  number={2},
  pages={199--210},
  year={2007},
  publisher={Elsevier}
}

@Article{Hall1986,
  author =  {Hall, B.},
  title =   {{Major {OECD} Country Industrial Sector Interfuel Substitution Estimates, 1960-79}},
  journal = {Energy Economics},
  year =    {1986},
  volume =  {9},
  number =  {2},
  pages =   {74-89}
}

@Article{Halvorsen1975,
  author =  {Halvorsen, R.},
  title =   {{Residential Demand for Electric Energy}},
  journal = {Review of Economics and Statistics},
  year =    {1975},
  volume =  {57},
  number =  {1},
  pages =   {12-18}
}

@Article{Halvorsen1976,
  author =  {Halvorsen, R.},
  title =   {{Demand for Electric Energy in the {United States}}},
  journal = {Southern Economic Journal},
  year =    {1976},
  volume =  {42},
  number =  {4},
  pages =   {610-625}
}

@Article{Halvorsen1977,
  author =  {Halvorsen, R.},
  title =   {{Energy Substitution in U. S. Manufacturing}},
  journal = {Review of Economics and Statistics},
  year =    {1977},
  volume =  {59},
  number =  {4},
  pages =   {381-388}
}

@InProceedings{Halvorsen1979,
  author =    {Halvorsen, R. and J. Ford},
  title =     {{Substitution Among Energy, Capital, and Labor Inputs in U.S. Manufacturing}},
  booktitle = {{The Structure of Energy Markets: Advances in the Economics of Energy and Resources}},
  year =      {1979},
  editor =    {Pindyck, R. S.},
  volume =    {1},
  pages =     {51-75},
  publisher = {JAI Press Inc.: Greenwich, Connecticut}
}

@Article{Halvorsen2001,
  author =  {Halvorsen, B. and B.M. Larsen},
  title =   {{The Flexibility of Household Electricity Demand Over Time}},
  journal = {Resource and Energy Economics},
  year =    {2001},
  volume =  {23},
  number =  {1},
  pages =   {1-18}
}

@Article{Hartman1981,
  author =  {Hartman, R. S. and A. Werth},
  title =   {{Short-Run Residential Demand for Fuels: A Disaggregated Approach}},
  journal = {Land Economics},
  year =    {1981},
  volume =  {57},
  number =  {2},
  pages =   {195-212}
}

@article{hasanov2016modeling,
  title={Modeling and forecasting electricity demand in {Azerbaijan} using cointegration techniques},
  author={Hasanov, Fakhri J and Hunt, Lester C and Mikayilov, Ceyhun I},
  journal={Energies},
  volume={9},
  number={12},
  pages={1045},
  year={2016},
  publisher={MDPI}
}

@Article{Hausman1979,
  author =  {Hausman, J.},
  title =   {{Individual Discount Rates and the Purchase and Utilization of Energy-Using Durables}},
  journal = {The Bell Journal of Economics},
  year =    {1979},
  volume =  {10},
  number =  {1},
  pages =   {197-212}
}

@Article{Hawkins1975,
  author =  {Hawkins, R. G.},
  title =   {{The Demand for Electricity: A Cross-Section Study of {New South Wales} and the {Australian} Capital Territory}},
  journal = {Economic Record},
  year =    {1975},
  volume =  {51},
  number =  {133},
  pages =   {1-18}
}

@Article{Hawkins1977,
  author =  {Hawkins, R. G.},
  title =   {{Factor Demands and the Production Function in Selected {Australian} Manufacturing Industries}},
  journal = {Australian Economic Papers},
  year =    {1977},
  volume =  {16},
  number =  {28},
  pages =   {97-111}
}

@Article{Hawkins1978,
  author =  {Hawkins, R. G.},
  title =   {{A Vintage Model of the Demand for Energy and Employment in {Australian} Manufacturing Industry}},
  journal = {The Review of Economic Studies},
  year =    {1978},
  volume =  {45},
  number =  {3},
  pages =   {479-494}
}

@InProceedings{He2004,
  author    = {He, L. and D. Lambert},
  title     = {{Chinese Industrial Energy Demand}},
  booktitle = {{Proceedings of the 24th United States Association for Energy Economics and International Association for Energy Economics North American Conference}},
  address   = {Washington, D.C.},
  year      = {2004}
}

@Article{Henderson1983,
  author =  {Henderson, J. S.},
  title =   {{The Economics of Electricity Demand Charges}},
  journal = {The Energy Journal},
  year =    {1983},
  volume =  {4},
  number =  {Special Issue},
  pages =   {127-140}
}

@Article{Henriksson2014,
  author =  {Henriksson, E. and P. Soderholm and L. Warell},
  title =   {{Industrial Electricity Demand and Energy Efficiency Policy: The Case of the {Swedish} Mining Industry}},
  journal = {Energy Efficiency},
  year =    {2014},
  volume =  {7},
  number =  {3},
  pages =   {477-491}
}

@Article{Henson1984,
  author =  {Henson, S. E.},
  title =   {{Electricity Demand Estimates Under Increasing-Block Rates}},
  journal = {Southern Economic Journal},
  year =    {1984},
  volume =  {51},
  number =  {1},
  pages =   {147-156}
}

@Article{Herriges1994,
  author =  {Herriges, J. A. and K. K. King},
  title =   {{Residential Demand for Electricity Under Inverted Block Rates: Evidence from A Controlled Experiment}},
  journal = {Journal of Business and Economic Statistics},
  year =    {1994},
  volume =  {12},
  number =  {4},
  pages =   {419-430}
}

@Article{Hesse1986,
  author =  {Hesse, D. M. and H. Tarkka},
  title =   {{The Demand for Capital, Labor, and Energy in {European} Manufacturing Industry Before and After the Oil Price Shocks}},
  journal = {Scandinavian Journal of Economics},
  year =    {1986},
  volume =  {88},
  number =  {3},
  pages =   {529-546}
}

@TechReport{Hieronymus1976,
  author =      {Hieronymus, W. H.},
  title =       {{Long-Range Forecasting Properties of State-of-The-Art Models of Demand for Electric Energy}},
  institution = {Charles River Associates, Inc., Cambridge, Massachusetts},
  year =        {1976},
  type =        {Final Report prepared for the Electric Power Research Institute}
}

@Article{Hill1983,
  author =  {Hill, D. H. and D. A. Ott and L. D. Taylor and J. M. Walker},
  title =   {{Incentive Payments in Time-of-Day Electricity Pricing Experiments: The Arizona Experience}},
  journal = {The Review of Economics and Statistics},
  year =    {1983},
  volume =  {65},
  number =  {1},
  pages =   {59-65}
}

@TechReport{hirth2022very,
  author      = {Hirth, Lion and Khanna, Tarun and Ruhnau, Oliver},
  title       = {{The (very) short-term price elasticity of {German} electricity demand}},
  institution = {ZBW -- Leibniz Information Centre for Economics, Kiel and Hamburg},
  year        = {2022}
}

@Article{Hogan1989,
  author =  {Hogan, W. W.},
  title =   {{A Dynamic Putty---Semi-Putty Model of Aggregate Energy Demand}},
  journal = {Energy Economics},
  year =    {1989},
  volume =  {11},
  number =  {1},
  pages =   {53-69}
}

@Article{Holtedahl2004,
  author =  {Holtedahl, P. and F. L. Joutz},
  title =   {{Residential Electricity Demand in {Taiwan}}},
  journal = {Energy Economics},
  year =    {2004},
  volume =  {26},
  number =  {2},
  pages =   {201-224}
}

@article{hondroyiannis2004estimating,
  title={Estimating residential demand for electricity in {Greece}},
  author={Hondroyiannis, George},
  journal={Energy Economics},
  volume={26},
  number={3},
  pages={319--334},
  year={2004},
  publisher={Elsevier}
}

@Article{Horowitz2007,
  author =  {Horowitz, M. J.},
  title =   {{Changes in Electricity Demand in the {United States} from the 1970s to 2003}},
  journal = {The Energy Journal},
  year =    {2007},
  volume =  {28},
  number =  {3},
  pages =   {93-119}
}

@Article{Houston1982,
  author =  {Houston, D. A.},
  title =   {{Revenue Effects from Changes in A Declining Block Pricing Structure}},
  journal = {Land Economics},
  year =    {1982},
  volume =  {58},
  number =  {3},
  pages =   {351-336}
}

@Article{Houthakker1951,
  author =  {Houthakker, H. S.},
  title =   {{Some Calculations of Electricity Consumption in Great {Britain}}},
  journal = {Journal of the Royal Statistical Society: Series A},
  year =    {1951},
  volume =  {114},
  number =  {3},
  pages =   {351-371}
}

@Article{Houthakker1974,
  author =  {Houthakker, H. and P. K. Verleger, Jr. and D. Sheehan},
  title =   {{Dynamic Demand Analysis for Gasoline and Residential Electricity}},
  journal = {American Journal of Agricultural Economics},
  year =    {1974},
  volume =  {56},
  number =  {2},
  pages =   {412-418}
}

@Article{Houthakker1980,
  author =  {Houthakker, H. S.},
  title =   {{Residential Electricity Revisited}},
  journal = {The Energy Journal},
  year =    {1980},
  volume =  {1},
  number =  {1},
  pages =   {29-41}
}

@Article{Hsiao1989,
  author =  {Hsiao, C. and D. C. Mountain and M.W. L. Chan and K. Y. Tsui},
  title =   {{Modelling {Ontario} Regional Electricity System Demand Using A Mixed Fixed and Random Coefficients Approach}},
  journal = {Regional Science and Urban Economics},
  year =    {1989},
  volume =  {19},
  number =  {4},
  pages =   {565-587}
}

@Article{Hsiao1994,
  author =  {Hsiao, C. and D. C. Mountain},
  title =   {{A Framework for Regional Modeling and Impact Analysis: An Analysis of Demand for Electricity by Large Municipalities in {Ontario}, {Canada}}},
  journal = {Journal of Regional Science},
  year =    {1994},
  volume =  {34},
  number =  {3},
  pages =   {361-385}
}

@InProceedings{Hsueh1986,
  author =    {Hsueh, L.-M. and J. L. Gerner},
  title =     {{A Model of Home Heating and the Calculation of Rates of Return on Housing Thermal Improvements Investment}},
  booktitle = {{Energy Decisions for the Future: Challenges and Opportunities}},
  year =      {1986},
  editor =    {Miyata, M. and K. Matsui},
  series =    {Proceeding of the "IAEE 8th Annual International Conference," Tokyo, Japan},
  pages =     {423-442}
}

@Article{Hughes-Cromwick1985,
  author =  {Hughes-Cromwick, E. L.},
  title =   {{Nairobi Households and Their Energy Use: Economic Analysis of Consumption Patterns}},
  journal = {Energy Economics},
  year =    {1985},
  volume =  {7},
  number =  {4},
  pages =   {265-278}
}

@article{hung2015dynamic,
  title={Dynamic demand for residential electricity in {Taiwan} under seasonality and increasing-block pricing},
  author={Hung, Ming-Feng and Huang, Tai-Hsin},
  journal={Energy Economics},
  volume={48},
  pages={168--177},
  year={2015},
  publisher={Elsevier}
}

@TechReport{Huntington1982,
  author =      {Huntington, H. G. and E. Soffer},
  title =       {{Demand for Energy in the Commercial Sector}},
  institution = {Data Resources, Inc., Lexington, Massachusetts},
  year =        {1982},
  type =        {Final Report to Electric Power Research Institute},
  number =      {2330/1982}
}

@InProceedings{Hyndman1980,
  author =    {Hyndman, R. and Y. Kotowitz and F. Mathewson},
  title =     {{The Residential Demand for Electric Energy and Natural Gas in {Canada}}},
  booktitle = {{Energy Policy Modeling: United States and Canadian Experiences, Specialized Energy Policy Models I}},
  year =      {1980},
  editor =    {Ziemba, W.T. and S. L. Schwartz and E. Koenigsberg},
  pages =     {86-102},
  publisher = {Martinus Nijhoff Publishing: Leiden}
}

@Book{IEEJ1986,
  title =     {{Petroleum Demand Econometric Study for {Japan}}},
  publisher = {The Institute of Energy Economics of Japan: Tokyo},
  year =      {1986},
  author =    {IEEJ}
}

@Article{Ilmakunnas1989,
  author =  {Ilmakunnas, P. and H. Torma},
  title =   {{Structural Change in Factor Substitution in {Finnish} Manufacturing}},
  journal = {Scandinavian Journal of Economics},
  year =    {1989},
  volume =  {91},
  number =  {4},
  pages =   {705-721}
}

@article{inglesi2010aggregate,
  title={Aggregate electricity demand in {South Africa}: Conditional forecasts to 2030},
  author={Inglesi, Roula},
  journal={Applied Energy},
  volume={87},
  number={1},
  pages={197--204},
  year={2010},
  publisher={Elsevier}
}

@Article{Inglesi-Lotz2011,
  author =  {Inglesi-Lotz, R. and J. N. Blignaut},
  title =   {{Estimating the Price Elasticity of Demand for Electricity by Sector in {South Africa}}},
  journal = {South African Journal of Economic and Management Sciences},
  year =    {2011},
  volume =  {14},
  number =  {4},
  pages =   {449-465}
}

@Article{Inglesi-Lotz2011a,
  author =  {Inglesi-Lotz, R.},
  title =   {{The Evolution of Price Elasticity of Electricity Demand in {South Africa}: A Kalman Filter Application}},
  journal = {Energy Policy},
  year =    {2011},
  volume =  {39},
  number =  {6},
  pages =   {3690-3696}
}

@article{ioannidis2017power,
  author  = {Ioannidis, John P. A. and Stanley, Tom D. and Doucouliagos, Hristos},
  title   = {{The Power of Bias in Economics Research}},
  journal = {The Economic Journal},
  volume  = {127},
  pages   = {F236--F265},
  year    = {2017}
}

@Article{Iqbal1986,
  author =  {Iqbal, M.},
  title =   {{Substitution of Labour, Capital, and Energy in the Manufacturing Sector in {Pakistan}}},
  journal = {Empirical Economics},
  year =    {1986},
  volume =  {11},
  number =  {2},
  pages =   {81-95}
}

@Article{irsova2023spurious,
  author    = {Irsova, Zuzana and Bom, Pedro RD and Havranek, Tomas and Rachinger, Heiko},
  journal   = {Nature Communications},
  title     = {Spurious precision in meta-analysis of observational research},
  year      = {2025},
  pages     = {8454},
  volume    = {16},
}

@article{ishaque2018revisiting,
  title={Revisiting income and price elasticities of electricity demand in {Pakistan}},
  author={Ishaque, Hanan},
  journal={Economic Research-Ekonomska Istrazivanja},
  volume={31},
  number={1},
  pages={1137--1151},
  year={2018},
  publisher={Taylor and Francis Group i Sveuciliste Jurja Dobrile u Puli, Fakultet~...}
}

@article{ito2014,
  title={Do consumers respond to marginal or average price? Evidence from nonlinear electricity pricing},
  author={Ito, Koichiro},
  journal={American Economic Review},
  volume={104},
  number={2},
  pages={537--563},
  year={2014},
  publisher={American Economic Association 2014 Broadway, Suite 305, Nashville, TN 37203}
}

@article{ito2018moral,
  title={Moral suasion and economic incentives: Field experimental evidence from energy demand},
  author={Ito, Koichiro and Ida, Takanori and Tanaka, Makoto},
  journal={American Economic Journal: Economic Policy},
  volume={10},
  number={1},
  pages={240--267},
  year={2018},
  publisher={American Economic Association}
}

@Article{Jaffee1982,
  author  = {Jaffee, B. L. and Houstan, D. A. and Olshavsky, R. W.},
  journal = {Journal of Consumer Affairs},
  title   = {{Residential Electricity Demand in Rural Areas: the Role of Conservation Actions, Engineering Factors, and Economic Variables}},
  year    = {1982},
  number  = {1},
  pages   = {137-151},
  volume  = {16},
}

@Article{Jamil2011,
  author =  {Jamil, F. and E. Ahmad},
  title =   {{Income and Price Elasticities of Electricity Demand: Aggregate and Sector-Wise Analyses}},
  journal = {Energy Policy},
  year =    {2011},
  volume =  {39},
  number =  {9},
  pages =   {5519--5527}
}

@article{javid2014electricity,
  title={Electricity consumption-GDP nexus in {Pakistan}: A structural time series analysis},
  author={Javid, Muhammad and Qayyum, Abdul},
  journal={Energy},
  volume={64},
  pages={811--817},
  year={2014},
  publisher={Elsevier}
}

@article{jessoe2014knowledge,
  title={Knowledge is (less) power: Experimental evidence from residential energy use},
  author={Jessoe, Katrina and Rapson, David},
  journal={American Economic Review},
  volume={104},
  number={4},
  pages={1417--1438},
  year={2014},
  publisher={American Economic Association 2014 Broadway, Suite 305, Nashville, TN 37203}
}

@article{jin2022elasticity,
  title={The elasticity of residential electricity demand and the rebound effect in 18 {European} Union countries},
  author={Jin, Taeyoung and Kim, Jinsoo},
  journal={Energy Sources, Part B: Economics, Planning, and Policy},
  volume={17},
  number={1},
  pages={2053896},
  year={2022},
  publisher={Taylor \& Francis}
}

@Article{Jones1995,
  author =  {Jones, C. T.},
  title =   {{A Dynamic Analysis of Interfuel Substitution in U.S. Industrial Energy Demand}},
  journal = {Journal of Business and Economic Statistics},
  year =    {1995},
  volume =  {13},
  number =  {4},
  pages =   {459-465}
}

@InProceedings{Jungeilges1986,
  author =    {Jungeilges, J. and C. A. Dahl},
  title =     {{Implications of Functional Form on Estimates of {Japanese} Energy Elasticities}},
  booktitle = {{Energy Decisions for the Future: Challenges and Opportunities}},
  year =      {1986},
  editor =    {Miyata, M. and K. Matsui},
  series =    {Proceeding of the "IAEE 8th Annual International Conference," Tokyo, Japan},
  pages =     {140-158}
}

@Article{Kamerschen2004,
  author =  {Kamerschen, D. and D. Porter},
  title =   {{The Demand for Residential, Industrial and Total Electricity, 1973-1998}},
  journal = {Energy Economics},
  year =    {2004},
  volume =  {26},
  number =  {1},
  pages =   {87-100}
}

@Article{Karbuz1997,
  author =  {Karbuz, S. and F. Birol and N. Guerer},
  title =   {{Electricity Demand in {Turkey}}},
  journal = {Pacific and Asian Journal of Energy},
  year =    {1997},
  volume =  {7},
  number =  {1},
  pages =   {55-62}
}

@Article{Kaserman1985,
  author =  {Kaserman, D. L. and J. W. Mayo},
  title =   {{Advertising and the Residential Demand for Electricity}},
  journal = {The Journal of Business},
  year =    {1985},
  volume =  {58},
  number =  {4},
  pages =   {399-408}
}

@TechReport{KEEI1989,
  author =      {KEEI},
  title =       {{Sectoral Energy Demand in the Republic of {Korea}: Analysis and Outlook}},
  institution = {United Nations Economic and Social Commission for Asia and the Pacific, Korean Energy Economics Institute, Seoul, Korea},
  year =        {1989},
  type =        {ESCAP Working paper}
}

@InProceedings{Keng1991,
  author =    {Keng, C. W. K.},
  title =     {{Forecasting Energy Demand with Variable Elasticity Models}},
  booktitle = {{Energy Developments in the 1990's: Challenges Facing Global/Pacific Markets}},
  year =      {1991},
  editor =    {Fesharaki, F. and J. P. Dorian},
  series =    {Proceedings of the "IAEE 14th Annual International Conference" held on July 8-10, 1991, Honolulu, Hawai},
  pages =     {3-28}
}

@article{khan2009demand,
  title={The demand for electricity in {Pakistan}},
  author={Khan, Muhammad Arshad and Qayyum, Abdul},
  journal={OPEC Energy Review},
  volume={33},
  number={1},
  pages={70--96},
  year={2009},
  publisher={Wiley Online Library}
}

@article{khan2016dynamics,
  title={The dynamics of electricity demand in {Pakistan}: A panel cointegration analysis},
  author={Khan, Muhammad Arshad and Abbas, Faisal},
  journal={Renewable and Sustainable Energy Reviews},
  volume={65},
  pages={1159--1178},
  year={2016},
  publisher={Elsevier}
}

@article{khanna2016effects,
  title={Effects of demand side management on {Chinese} household electricity consumption: Empirical findings from {Chinese} household survey},
  author={Khanna, Nina Zheng and Guo, Jin and Zheng, Xinye},
  journal={Energy Policy},
  volume={95},
  pages={113--125},
  year={2016},
  publisher={Elsevier}
}

@Book{Khazzoom1986,
  title =     {{Econometric Model Integrating Conservation Measures in the Residential Demand for Electricity}},
  publisher = {JAI Press, Inc.: Greenwich},
  year =      {1986},
  author =    {Khazzoom, J. D.}
}

@TechReport{knaut2016hen,
  author      = {Knaut, Andreas and Paulus, Simon},
  title       = {{When are consumers responding to electricity prices? An hourly pattern of demand elasticity}},
  institution = {Institute of Energy Economics at the University of Cologne (EWI)},
  type        = {EWI Working Paper},
  number      = {No 16/07},
  year        = {2016}
}

@Article{Kohler1984,
  author =  {Kohler, D. F. and B. M. Mitchell},
  title =   {{Response to Residential Time-Of Use Electricity Rates; How Transferable Are the Findings?}},
  journal = {Journal of Econometrics},
  year =    {1984},
  volume =  {26},
  number =  {1-2},
  pages =   {141-177}
}

@article{kohler2014differential,
  title={Differential electricity pricing and energy efficiency in {South Africa}},
  author={Kohler, Marcel},
  journal={Energy},
  volume={64},
  pages={524--532},
  year={2014},
  publisher={Elsevier}
}

@Article{Kokkelenberg1993,
  author =  {Kokkelenberg, E. C. and T.D. Mount},
  title =   {{Oil Shocks and the Demand for Electricity}},
  journal = {The Energy Journal},
  year =    {1993},
  volume =  {14},
  number =  {2},
  pages =   {113-139}
}

@Article{Kolstad1993,
  author =  {Kolstad, C. D. and J.-K. Lee},
  title =   {{The Specification of Dynamics in Cost Function and Factor Demand Estimation}},
  journal = {Review of Economics and Statistics},
  year =    {1993},
  volume =  {75},
  number =  {4},
  pages =   {721-736}
}

@article{krishnamurthy2015cross,
  title={A cross-country analysis of residential electricity demand in 11 {OECD}-countries},
  author={Krishnamurthy, Chandra Kiran B and Kristrom, Bengt},
  journal={Resource and Energy Economics},
  volume={39},
  pages={68--88},
  year={2015},
  publisher={Elsevier}
}

@article{kwon2016effects,
  title={Effects of electricity-price policy on electricity demand and manufacturing output},
  author={Kwon, Sanguk and Cho, Seong-Hoon and Roberts, Roland K and Kim, Hyun Jae and Park, Kihyun and Yu, T Edward},
  journal={Energy},
  volume={102},
  pages={324--334},
  year={2016},
  publisher={Elsevier}
}

@article{labandeira2017meta,
  title={A meta-analysis on the price elasticity of energy demand},
  author={Labandeira, Xavier and Labeaga, Jose M and Lopez-Otero, Xiral},
  journal={Energy Policy},
  volume={102},
  pages={549--568},
  year={2017},
  publisher={Elsevier}
}

@article{lanot2021price,
  title={The price elasticity of electricity demand when marginal incentives are very large},
  author={Lanot, Gauthier and Vesterberg, Mattias},
  journal={Energy Economics},
  volume={104},
  pages={105604},
  year={2021},
  publisher={Elsevier}
}

@Article{Lareau1982,
  author =  {Lareau, T. J. and J. Darmstadter},
  title =   {{Energy and Consumer-Expenditure Patterns: Modeling Approaches and Projections}},
  journal = {Annual Review of Energy},
  year =    {1982},
  volume =  {7},
  number =  {1},
  pages =   {261-292}
}

@TechReport{Larsson2004,
  author =      {Larsson, J.},
  title =       {{Four Essays on Factor Demand Modeling}},
  institution = {Department of Economics, Goteborg University, Sweden},
  year =        {2004},
  type =        {Technical report}
}

@PhdThesis{Larsson2006,
  author = {Larsson, J.},
  title =  {{Four Essays on Technology, Productivity and Environment}},
  school = {School of Business, Economics and Law, Gothenburg University, Sweden},
  year =   {2006},
  type =   {Ph.D. thesis}
}

@Article{Laumas1981,
  author =  {Laumas, P. S. and M. Williams},
  title =   {{Energy and Economic Development}},
  journal = {Review of World Economics},
  year =    {1981},
  volume =  {117},
  number =  {4},
  pages =   {706-716}
}

@article{lee2010panel,
  title={A panel data analysis of the demand for total energy and electricity in {OECD} countries},
  author={Lee, Chien-Chiang and Lee, Jun-De},
  journal={The Energy Journal},
  pages={1--23},
  year={2010},
  publisher={JSTOR}
}

@Article{Lee2011,
  author =  {Lee, C.-C. and Y.-B. Chiu},
  title =   {{Electricity Demand Elasticities and Temperature: Evidence from Panel Smooth Transition Regression with Instrumental Variable Approach}},
  journal = {Energy Economics},
  year =    {2011},
  volume =  {33},
  number =  {5},
  pages =   {896-902}
}

@article{liddle2021prices,
  title={How prices, income, and weather shape household electricity demand in high-income and middle-income countries},
  author={Liddle, Brantley and Huntington, Hillard},
  journal={Energy Economics},
  volume={95},
  pages={104995},
  year={2021},
  publisher={Elsevier}
}

@article{lijesen2007real,
  title={The real-time price elasticity of electricity},
  author={Lijesen, Mark G},
  journal={Energy Economics},
  volume={29},
  number={2},
  pages={249--258},
  year={2007},
  publisher={Elsevier}
}

@Article{Lillard1981,
  author =  {Lillard, L. A. and J. P. Acton},
  title =   {{Seasonal Electricity Demand and Pricing Analysis with A Variable Response Model}},
  journal = {The Bell Journal of Economics},
  year =    {1981},
  volume =  {12},
  number =  {1},
  pages =   {71-92}
}

@Article{Lim2014,
  author =  {Lim, K. M. and S. Y. Lim and S. H. Yoo},
  title =   {{Short- and Long-Run Elasticities of Electricity Demand in the {Korean} Service Sector}},
  journal = {Energy Policy},
  year =    {2014},
  volume =  {67},
  pages =   {517-521}
}

@article{lin2014electricity,
  title={Electricity demand and conservation potential in the {Chinese} nonmetallic mineral products industry},
  author={Lin, Boqiang and Ouyang, Xiaoling},
  journal={Energy Policy},
  volume={68},
  pages={243--253},
  year={2014},
  publisher={Elsevier}
}

@article{lin2020chinese,
  title={{Chinese} electricity demand and electricity consumption efficiency: Do the structural changes matter?},
  author={Lin, Boqiang and Zhu, Junpeng},
  journal={Applied Energy},
  volume={262},
  pages={114505},
  year={2020},
  publisher={Elsevier}
}

@TechReport{Liu2005,
  author =      {Liu, G.},
  title =       {{Estimating Energy Demand Elasticities for {OECD} Countries: A Dynamic Panel Data Approach}},
  institution = {Research Department, Statistics Norway, Kongsvinger, Norway},
  year =        {2005},
  type =        {Discussion Paper},
  number =      {373/2004}
}

@TechReport{Lohani1992,
  author =      {Lohani, P. R.},
  title =       {{Electricity Demand in Developing Countries}},
  institution = {Colorado School of Mines, Golden, Colorado},
  year =        {1992},
  type =        {Technical report}
}

@Article{Lyman1994,
  author =  {Lyman, R. A.},
  title =   {{Philippine Electric Demand and Equivalence Scales}},
  journal = {Southern Economic Journal},
  year =    {1994},
  volume =  {60},
  number =  {3},
  pages =   {596-610}
}

@Article{Lynk1989,
  author =  {Lynk, E. L.},
  title =   {{The Demand for Energy by U.K. Manufacturing Industry}},
  journal = {The Manchester School},
  year =    {1989},
  volume =  {57},
  number =  {1},
  pages =   {1-16}
}

@article{ma2016long,
  title={Long-run estimates of interfuel and interfactor elasticities},
  author={Ma, Chunbo and Stern, David I},
  journal={Resource and Energy Economics},
  volume={46},
  pages={114--130},
  year={2016},
  publisher={Elsevier}
}

@TechReport{Macroconsult2001,
  author =      {Macroconsult},
  title =       {{Desarrollo de un Modelo Econometrico de la Demanda de Energia para el Sistema Interconectado Nacional}},
  institution = {Macroconsult S.A., Lima, Peru},
  year =        {2001},
  type =        {Consultoria para Comision de Tarifas de Energia},
  number =      {1-2}
}

@TechReport{Maddala1994,
  author =      {Maddala, G. S. and R. Trost and F. Joutz and H. Li},
  title =       {{Estimation of Short Run and Long Run Elasticities of Energy Demand from Panel Data Using Shrinkage Estimators}},
  institution = {Presented et the "5th Conference on Panel Data" held on May 5, 1994, Universite Paris, Paris, France},
  year =        {1994},
  journal =     {Presented et the 5th Conference on Panel Data, May 5, 1994, Universite Paris}
}

@Article{Maddala1997,
  author =  {Maddala, G. S. and R. Trost, F. Joutz and H. Li},
  title =   {{Estimation of Short Run and Long Run Elasticities of Energy Demand from Panel Data Using Shrinkage Estimators}},
  journal = {Journal of Business and Economic Statistics},
  year =    {1997},
  volume =  {15},
  number =  {1},
  pages =   {90-100}
}

@Article{Maddigan1983,
  author =  {Maddigan, R. J. and W. S. Chern and C. G. Rizy},
  title =   {{Rural Residential Demand for Electricity}},
  journal = {Land Economics},
  year =    {1983},
  volume =  {59},
  number =  {2},
  pages =   {150-161}
}

@Article{Maddock1991,
  author =  {Maddock, R. and E. Castan},
  title =   {{The Welfare Impact of Rising Block Pricing: Electricity in {Colombia}}},
  journal = {The Energy Journal},
  year =    {1991},
  volume =  {12},
  number =  {4},
  pages =   {65-77}
}

@article{madlener2011econometric,
  author  = {Madlener, Reinhard},
  title   = {{Econometric estimation of energy demand elasticities}},
  journal = {E.ON Energy Research Center Series},
  volume  = {3},
  number  = {8},
  pages   = {1--59},
  year    = {2011}
}

@Article{Mahmud1990,
  author =  {Mahmud, F. and S. Chishti},
  title =   {{The Demand for Energy in the Large-Scale Manufacturing Sector of {Pakistan}}},
  journal = {Energy Economics},
  year =    {1990},
  volume =  {12},
  number =  {4},
  pages =   {251-255}
}

@TechReport{Mansur2005,
  author =      {Mansur, E. T. and R. Mendelsohn and W. Morrison},
  title =       {{A Discrete-Continuous Choice Model of Climate Change Impacts on Energy}},
  institution = {Yale School of Management, School of Forestry and Environmental Studies, Yale University, New Haven, Connecticut},
  year =        {2005},
  type =        {Technical report},
  number =      {219/2005}
}

@article{maria2012estimation,
  title={Estimation of elasticities for domestic energy demand in Mozambique},
  author={Arthur, Maria de Fatima S. R. and Bond, Craig A. and Willson, Bryan},
  journal={Energy Economics},
  volume={34},
  number={2},
  pages={398--409},
  year={2012},
  publisher={Elsevier}
}

@article{masike2022time,
  title={The time-varying elasticity of South {African} electricity demand},
  author={Masike, Kabelo and Vermeulen, Cobus},
  journal={Energy},
  volume={238},
  pages={121984},
  year={2022},
  publisher={Elsevier}
}

@InProceedings{Matsui1979,
  author =    {Matsui, K.},
  title =     {{Income and Price Elasticities of Energy Demand in {Japan}}},
  booktitle = {{Energy in Japan}},
  year =      {1979},
  number =    {46/1979},
  series =    {Quarterly Report},
  pages =     {9-24},
  publisher = {Japanese Institute of Economic Research},
  chapter =   {{Income and Price Elasticities of Energy Demand in Japan}}
}

@Article{Matsukawa1993,
  author =  {Matsukawa, I. and Y. Fujii and S. Madono},
  title =   {{Price, Environmental Regulation, and Fuel Demand: Econometric Estimates for {Japanese} Manufacturing Industries}},
  journal = {The Energy Journal},
  year =    {1993},
  volume =  {14},
  number =  {4},
  pages =   {37-56}
}

@TechReport{McFadden1977,
  author =      {McFadden, D. L. and C. Puig and D. Kirshner},
  title =       {{Determinants of the Long-Run Demand for Electricity}},
  institution = {Proceedings of the Business and Economics Section of the American Statistical Association, pp. 109-119},
  year =        {1977}
}

@TechReport{Mchugh1977,
  author =      {Mchugh, W. M.},
  title =       {{Energy Demand Modeling and Forecasting}},
  institution = {Mathematical Sciences Northwest, Inc., Bellevue, Washington},
  year =        {1977},
  type =        {Final report},
  number =      {02/1977}
}

@TechReport{mcrae2016price,
  author      = {McRae, Shaun and Meeks, Robyn},
  title       = {{Price perception and electricity demand with nonlinear tariffs}},
  institution = {National Bureau of Economic Research},
  type        = {NBER Conference Paper},
  number      = {f88094},
  year        = {2016}
}

@article{meher2020estimating,
  title={Estimating and forecasting residential electricity demand in {Odisha}},
  author={Meher, Shibalal},
  journal={Journal of Public Affairs},
  volume={20},
  number={3},
  pages={e2065},
  year={2020},
  publisher={Wiley Online Library}
}

@InProceedings{Mendoza1987,
  author =    {Mendoza, Y. and R. Vargas},
  title =     {{Domestic Energy Demand in Oil Abundant Latin American Countries}},
  booktitle = {{The Changing World Energy Economy}},
  year =      {1987},
  editor =    {Wood, D. O.},
  number =    {19-21},
  series =    {Papers and Proceedings of the EAEE 8th Annual North American Conference},
  pages =     {309-313},
  publisher = {MIT Press: Cambridge, Massachusetts}
}

@article{mikayilov2017modeling,
  title={Modeling of electricity demand for {Azerbaijan}: time-varying coefficient cointegration approach},
  author={Mikayilov, Jeyhun I and Hasanov, Fakhri J and Bollino, Carlo A and Mahmudlu, Ceyhun},
  journal={Energies},
  volume={10},
  number={11},
  pages={1918},
  year={2017},
  publisher={MDPI}
}

@article{miller2016sensitivity,
  title={Sensitivity of price elasticity of demand to aggregation, unobserved heterogeneity, price trends, and price endogeneity: Evidence from {US} Data},
  author={Miller, Mark and Alberini, Anna},
  journal={Energy Policy},
  volume={97},
  pages={235--249},
  year={2016},
  publisher={Elsevier}
}

@PhdThesis{Moghaddam2003,
  author = {Moghaddam, M. R.},
  title =  {{Improving {Iran}'s Domestic Energy Basket}},
  school = {Department of Economics and Business Administration, Tilberg University, Tilberg, Netherlands},
  year =   {2003},
  type =   {Ph.D. thesis}
}

@article{morovat2019estimating,
  title={Estimating the elasticity of electricity demand in {Iran}: A sectoral-Province Approach},
  author={Morovat, Habib and Faridzad, Ali and Lowni, Sahar},
  journal={Iranian Economic Review},
  volume={23},
  number={4},
  pages={861--881},
  year={2019},
  publisher={University of Tehran, Faculty of Economics}
}

@InProceedings{Mount1974,
  author =    {Mount, T. D. and L. D. Chapman and T. J. Tyrrell},
  title =     {{Electricity Demand in the {United States}: An Econometric Analysis}},
  booktitle = {{Energy: Demand, Conservation, and Institutional Problems}},
  year =      {1974},
  editor =    {Macrakis, M. S.},
  series =    {Proceedings from a conference held on February 12--14, 1973, Cambridge, Massachusetts},
  pages =     {318-329},
  publisher = {MIT Press: Cambridge, Massachusetts}
}

@InProceedings{Mount1979,
  author =    {Mount, T. D. and L. D. Chapman},
  title =     {{Electricity Demand, Sulfur Emissions and Health: An Econometric Analysis of Power Generation in the {United States}}},
  booktitle = {{International Studies of the Demand for Energy}},
  year =      {1979},
  editor =    {Nordhaus, W. D.},
  publisher = {North-Holland Publishing Company: Amsterdam}
}

@Article{Mountain1989,
  author =  {Mountain, D. C. and C. Hsiao},
  title =   {{A Combined Structural and Flexible Functional Approach for Modeling Energy Substitution}},
  journal = {Journal of the American Statistical Association},
  year =    {1989},
  volume =  {84},
  number =  {405},
  pages =   {76-87}
}

@Article{Mountain1989a,
  author =  {Mountain, D. C.},
  title =   {{A Quadratic Quasi Cobb-Douglas Extension of the Multi-Input CES Formulation}},
  journal = {European Economic Review},
  year =    {1989},
  volume =  {33},
  number =  {1},
  pages =   {143-158}
}

@Article{Mountain1989b,
  author =  {Mountain, D. C. and B. P. Stipdonk and C. J. Warren},
  title =   {{Technological Innovation and A Changing Energy Mix: A Parametric and Flexible Approach to Modeling {Ontario} Manufacturing}},
  journal = {The Energy Journal},
  year =    {1989},
  volume =  {10},
  number =  {4},
  pages =   {139-158}
}

@Article{Munley1990,
  author =  {Munley, V. G. and L. W. Taylor and J. P. Formby},
  title =   {{Electricity Demand in Multi-Family, Renter-Occupied Residences}},
  journal = {Southern Economic Journal},
  year =    {1990},
  volume =  {57},
  number =  {1},
  pages =   {178-194}
}

@Article{Murray1978,
  author =  {Murray, M. P. and R. Spann and L. Pulley and E. Beauvais},
  title =   {{The Demand for Electricity in Virginia}},
  journal = {Review of Economics and Statistics},
  year =    {1978},
  volume =  {60},
  number =  {4},
  pages =   {585-600}
}

@Article{Nagata2001,
  author =  {Nagata, Y.},
  title =   {{A Forecast of Energy Demand in {Japan} Considering the Asymmetric Price Elasticities}},
  journal = {Energy Studies Review},
  year =    {2001},
  volume =  {10},
  number =  {1},
  pages =   {17-26}
}

@article{nakajima2010change,
  title={Change in consumer sensitivity to electricity prices in response to retail deregulation: A panel empirical analysis of the residential demand for electricity in the {United States}},
  author={Nakajima, Tadahiro and Hamori, Shigeyuki},
  journal={Energy Policy},
  volume={38},
  number={5},
  pages={2470--2476},
  year={2010},
  publisher={Elsevier}
}

@article{nakajima2010residential,
  title={The residential demand for electricity in {Japan}: an examination using empirical panel analysis techniques},
  author={Nakajima, Tadahiro},
  journal={Journal of Asian Economics},
  volume={21},
  number={4},
  pages={412--420},
  year={2010},
  publisher={Elsevier}
}

@Article{Narayan2005,
  author =  {Narayan, P. K. and R. Smyth},
  title =   {{Residential Demand for Electricity in {Australia}: An Application of the Bounds Testing Approach to Cointegration}},
  journal = {Energy Policy},
  year =    {2005},
  volume =  {33},
  number =  {4},
  pages =   {457-464}
}

@Article{Narayan2007,
  author =  {Narayan, K. P. and R. Smyth and A. Prasad},
  title =   {{Electricity Consumption in G7 Countries: A Panel Cointegration Analysis of Residential Demand Elasticities}},
  journal = {Energy Policy},
  year =    {2007},
  volume =  {35},
  number =  {9},
  pages =   {4485-4494}
}

@article{nasir2008residential,
  title={Residential demand for electricity in {Pakistan}},
  author={Nasir, Muhammad and Tariq, Muhammad Salman and Arif, Ankasha},
  journal={The Pakistan Development Review},
  pages={457--467},
  year={2008},
  publisher={JSTOR}
}

@Article{Okajima2013,
  author =  {Okajima, S. and H. Okajima},
  title =   {{Estimation of {Japanese} Price Elasticities of Residential Electricity Demand, 1990-2007}},
  journal = {Energy Economics},
  year =    {2013},
  volume =  {40},
  pages =   {433-440}
}

@TechReport{Oliveira1993,
  author =      {Oliveira, R. A.},
  title =       {{A Pooled Cross-Section, Time-Series Econometric Analysis of Residential Electricity Demand in Oregon: Preliminary Findings}},
  institution = {Oregon Public Utility Commission, Salem, Oregon},
  year =        {1993}
}
\end{multicols}
\end{singlespace}

\newpage
\begin{appendices}
\singlespacing  

\FloatBarrier
\section{Literature search and summary statistics}\label{app:search}

\begin{figure}[p]\centering
\caption{PRISMA flow diagram of the literature search}\label{fig:prisma}
\smallskip
\begin{tikzpicture}[
  rect/.style={rectangle, draw=black!60, minimum height=3.0em, font={\scriptsize\hyphenpenalty=10000\exhyphenpenalty=10000}, text width=8.6em, align=center, inner sep=2.5pt},
  exc/.style ={rectangle, draw=black!40, minimum height=2.6em, font={\scriptsize\hyphenpenalty=10000\exhyphenpenalty=10000}, text width=6.8em, align=center, inner sep=2.5pt},
  band/.style={draw=cyan!55, semithick, rectangle, fill=cyan!15, minimum width=0.5cm, inner xsep=2pt, inner ysep=0pt},
  >=Latex, node distance=0.9cm and 0.6cm]
\node[rect, inner ysep=6pt] (db) {8{,}240 records identified by database and citation searching (Google Scholar and RePEc; concluded 28 March 2026)};
\node[rect, below=of db] (scr) {8{,}240 records screened on title and abstract};
\node[exc, right=0.6cm of scr] (scrx) {7{,}470 excluded (off-topic or duplicate)};
\node[rect, below=2.6cm of scr] (el) {770 records assessed for eligibility};
\node[exc, right=0.6cm of el] (elx) {357 excluded (not an own-price elasticity, or a superseded version)};
\node[rect, below=2.6cm of el] (dbinc) {413 studies included (database $+$ citation)};
\node[rect, anchor=west, inner ysep=6pt] (omid) at ([xshift=0.8cm]scrx.east |- db) {918 records identified via other methods (citation snowballing; prior meta-analyses' datasets; studies published 2022--2026)};
\node[rect, anchor=west] (omscr) at (omid.west |- scr) {918 records screened on title and abstract};
\node[exc, right=0.6cm of omscr] (omscrx) {786 excluded (off-topic or duplicate)};
\node[rect, anchor=west] (omas) at (omid.west |- el) {132 records assessed for eligibility};
\node[exc, right=0.6cm of omas] (omelx) {63 excluded (other outcome, duplicate, or out of scope)};
\node[rect, anchor=west] (omin) at (omid.west |- dbinc) {69 studies included (other methods)};
\node[rect, text width=20em, fill=black!4] (total) at ($(dbinc.south)!0.5!(omin.south)+(0,-1.4cm)$) {482 studies carry own-price estimates; the 462 reporting a usable standard error contribute the 4{,}720 estimates analyzed};
\coordinate (lx) at ([xshift=-0.5cm]db.west);
\coordinate (Bb) at ($(db.south)!0.5!(scr.north)$);
\coordinate (Bc) at ($(scr.south)!0.5!(el.north)$);
\coordinate (Bd) at ($(el.south)!0.5!(dbinc.north)$);
\node[band, fit={(lx|-db.north) (lx|-Bb)}]    (bID) {};
\node[band, fit={(lx|-Bb) (lx|-Bc)}]          (bSC) {};
\node[band, fit={(lx|-Bc) (lx|-Bd)}]          (bEL) {};
\node[band, fit={(lx|-Bd) (lx|-total.south)}] (bIN) {};
\node[rotate=90, font=\scriptsize\bfseries, text=black!75] at (bID.center) {Identification};
\node[rotate=90, font=\scriptsize\bfseries, text=black!75] at (bSC.center) {Screening};
\node[rotate=90, font=\scriptsize\bfseries, text=black!75] at (bEL.center) {Eligibility};
\node[rotate=90, font=\scriptsize\bfseries, text=black!75] at (bIN.center) {Included};
\draw[->] (db) -- (scr);
\draw[->] (scr) -- (scrx);   \draw[->] (scr) -- (el);
\draw[->] (el) -- (elx);     \draw[->] (el) -- (dbinc);
\draw[->] (omid) -- (omscr);
\draw[->] (omscr) -- (omscrx); \draw[->] (omscr) -- (omas);
\draw[->] (omas) -- (omelx);   \draw[->] (omas) -- (omin);
\draw[->] (dbinc) |- (total.west);
\draw[->] (omin) |- (total.east);
\end{tikzpicture}
\\[0.4cm]
\begin{minipage}{0.96\textwidth}\footnotesize\textit{Notes:} Two-stream PRISMA flow for the electricity-demand price-elasticity corpus; reporting follows the meta-analysis guidelines of \citet{havranek2020guidelines} and, for AI use, \citet{Cook2026b} and \citet{Cook2026}. \emph{Database and citation arm:} a systematic Google Scholar and RePEc search (own-price elasticity of electricity demand paired with residential, industrial, tariff, and estimation-method terms) with forward and backward citation searching, concluded 28 March 2026, retrieves 8{,}240 records; 7{,}470 are removed at de-duplication and title/abstract screening and 770 are assessed in full text, of which 357 are excluded (a cross-price or income elasticity rather than own-price, or a superseded or out-of-scope version), leaving 413 studies. \emph{Other-methods arm:} backward and forward citation snowballing on the most-cited estimates, the datasets of prior electricity-demand meta-analyses \citep{dahl2011,espey2004,labandeira2017meta,zhu2018meta,zabaloy2022household,fatima2023price}, and studies published between 2022 and 2026 identify 918 further records; 132 are assessed and 69 are included. Together the two arms give $413+69=482$ studies with own-price estimates; applying the inclusion rule (a usable standard error, $t$-statistic, or $p$-value) leaves 462 studies contributing 4{,}720 estimates. PRISMA~=~Preferred Reporting Items for Systematic Reviews and Meta-Analyses \citep{page2021prisma}. Data, code, and additional materials are available in an online appendix at \url{https://meta-analysis.cz/electricity}.\end{minipage}
\end{figure}
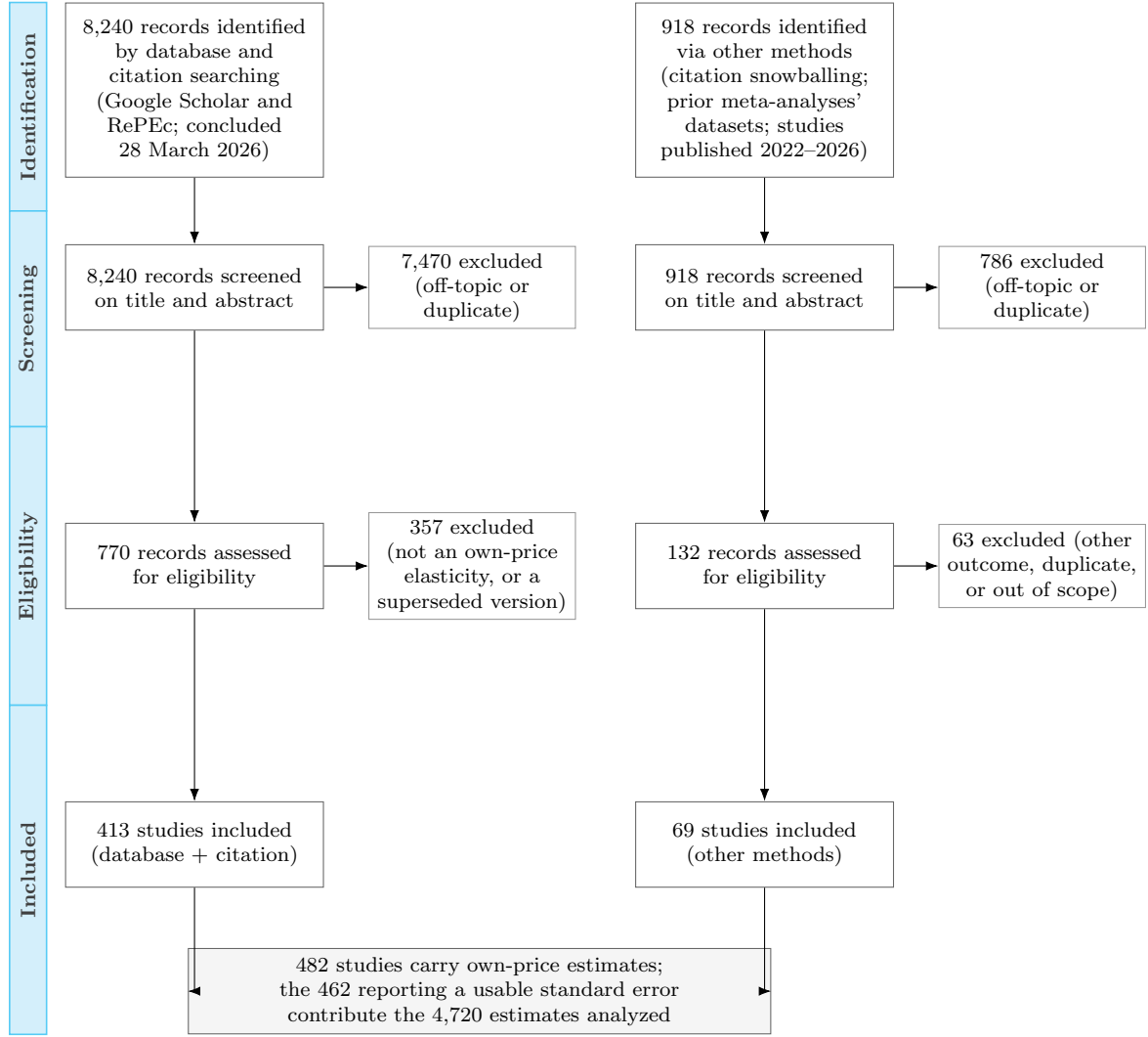

The corpus was assembled by systematic database search and citation tracking; \autoref{fig:prisma} reports the flow. The database-and-citation arm builds on a systematic Google Scholar and RePEc search concluded on 28 March 2026, which produced a 413-study base collection; the other-methods arm adds backward and forward citation snowballing, the datasets of prior electricity-demand meta-analyses, and studies published through early 2026, extending the collection to 482 studies. Reporting follows the meta-analysis guidelines of \citet{havranek2020guidelines}; for the use of AI, we also follow the guidance of \citet{Cook2026b} and \citet{Cook2026}. To qualify for the analysis, a study had to report an own-price elasticity of electricity demand, not a cross-price or income elasticity, together with a usable measure of uncertainty: a standard error, a $t$-statistic, or a $p$-value from which a standard error could be recovered. The 462 studies meeting this rule contribute the 4,720 own-price elasticity estimates that make up the corpus; the remaining 20 studies report no usable measure of uncertainty and are excluded. We convert the 745 compensated estimates that report both an income elasticity and a usable standard error; together with the 2,579 directly reported Marshallian estimates they make up the 3,324-estimate baseline. The 1,396 estimates held out of that baseline are of three kinds: 859 (from 77 studies) that cannot be placed on the Marshallian footing for want of a usable income elasticity to convert, 420 whose elasticity is indirectly derived from an inverted formula, and 117 whose standard error is imputed from a reported significance level (all from a single study).

\begin{figure}[t]
\centering
\caption{Distribution of the price-elasticity estimates}\label{fig:distest}
\includegraphics[width=\textwidth]{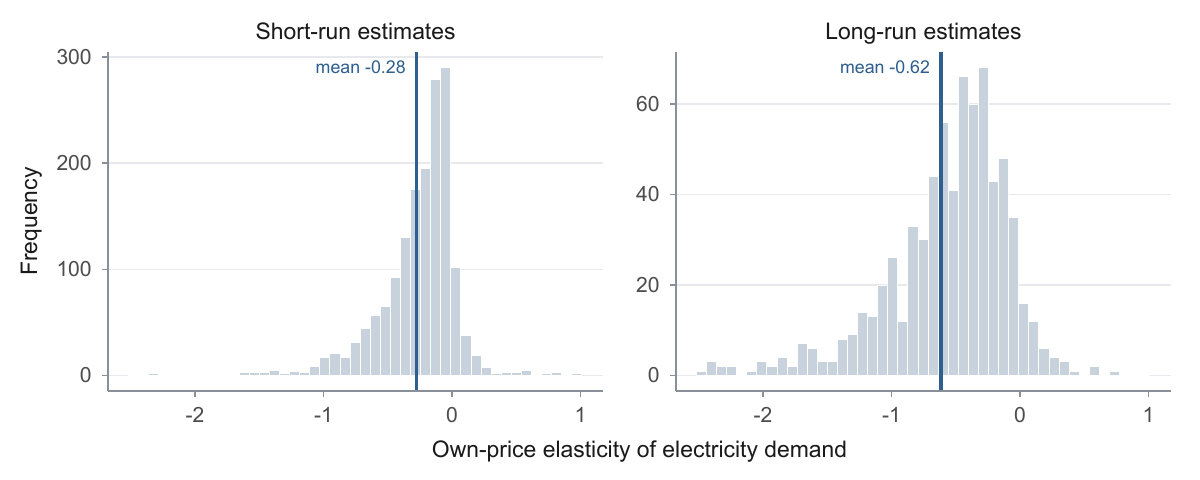}
\par\smallskip
\begin{minipage}{\linewidth}\footnotesize\textit{Notes:} Histograms of the own-price elasticity of electricity demand (Marshallian-equivalent, reported values), for the 1{,}647 short-run and 723 long-run headline estimates (usable standard error); the two panels share a common horizontal scale, truncated to $[-2.5,\,1]$ so that 8 short-run and 13 long-run outlying estimates fall outside the frame (trimmed from the plot only, and kept in every statistic). The vertical line marks the simple (unweighted) mean of each subsample, printed alongside. The long-run estimates sit farther from zero (more elastic) than the short-run estimates, reproducing the horizon gradient of \autoref{sec:horizon}.\end{minipage}
\end{figure}

\autoref{fig:distest} plots the distribution of the estimates, and \autoref{tab:sumsr} and \autoref{tab:sumlr} summarize them across the main subsamples for the short and long run separately. The estimates are overwhelmingly negative, as demand theory requires; wrong-signed (positive) estimates are rare, the mass concentrating in a narrow band between roughly $-0.5$ and $0$ in the short run and shifting away from zero at the long-run horizon.

\begin{figure}[t]
\centering
\caption{The adjusted ladder (the key result of \autoref{sec:ladder})}
\label{fig:ladderfig}
\begin{threeparttable}
\includegraphics[width=.82\textwidth]{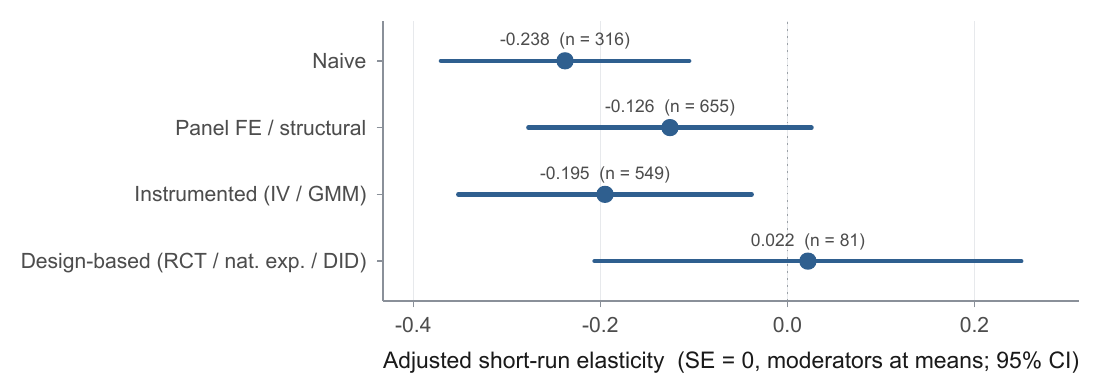}
\begin{tablenotes}[flush]
\footnotesize
\item \textit{Notes:} Graphical version of \autoref{tab:ladder}: adjusted short-run elasticity by tier with 95\% study-clustered intervals.
\end{tablenotes}
\end{threeparttable}
\end{figure}

\begin{table}[t]
\centering
\caption{Summary of the short-run estimates by subsample}\label{tab:sumsr}
\footnotesize
\begin{tabular*}{\textwidth}{@{\extracolsep{\fill}}l rr rc rc@{}}
\toprule
 & & & \multicolumn{2}{c}{Unweighted} & \multicolumn{2}{c}{Weighted} \\
\cmidrule(lr){4-5}\cmidrule(lr){6-7}
Subsample & $N$ & Studies & Mean & 95\% CI & Mean & 95\% CI \\
\midrule
All estimates & 1,647 & 226 & $-0.27$ & $[-0.28,\,-0.25]$ & $-0.11$ & $[-0.11,\,-0.11]$ \\
\addlinespace
\multicolumn{7}{@{}l}{\emph{Identification tier}} \\
\hspace{1em}Design-based (tiers 1--3) & 81 & 8 & $-0.09$ & $[-0.15,\,-0.03]$ & $-0.08$ & $[-0.08,\,-0.07]$ \\
\hspace{1em}Instrumented (IV) & 549 & 52 & $-0.30$ & $[-0.33,\,-0.28]$ & $-0.10$ & $[-0.11,\,-0.10]$ \\
\hspace{1em}Panel FE / structural & 655 & 98 & $-0.21$ & $[-0.23,\,-0.18]$ & $-0.15$ & $[-0.15,\,-0.15]$ \\
\hspace{1em}Naive & 316 & 70 & $-0.37$ & $[-0.41,\,-0.32]$ & $-0.11$ & $[-0.11,\,-0.11]$ \\
\addlinespace
\multicolumn{7}{@{}l}{\emph{Demand sector}} \\
\hspace{1em}Residential & 1,048 & 165 & $-0.29$ & $[-0.31,\,-0.27]$ & $-0.18$ & $[-0.18,\,-0.18]$ \\
\hspace{1em}Industrial & 220 & 56 & $-0.34$ & $[-0.40,\,-0.27]$ & $-0.07$ & $[-0.08,\,-0.07]$ \\
\hspace{1em}Commercial & 162 & 43 & $-0.24$ & $[-0.29,\,-0.19]$ & $-0.09$ & $[-0.09,\,-0.08]$ \\
\addlinespace
\multicolumn{7}{@{}l}{\emph{Data structure}} \\
\hspace{1em}Time-series & 555 & 104 & $-0.21$ & $[-0.24,\,-0.18]$ & $-0.04$ & $[-0.05,\,-0.04]$ \\
\hspace{1em}Cross-section & 128 & 26 & $-0.48$ & $[-0.55,\,-0.41]$ & $-0.26$ & $[-0.26,\,-0.25]$ \\
\hspace{1em}Panel & 964 & 106 & $-0.27$ & $[-0.29,\,-0.25]$ & $-0.13$ & $[-0.13,\,-0.13]$ \\
\addlinespace
\multicolumn{7}{@{}l}{\emph{Price measurement}} \\
\hspace{1em}Average price & 1,030 & 166 & $-0.26$ & $[-0.28,\,-0.25]$ & $-0.11$ & $[-0.11,\,-0.11]$ \\
\hspace{1em}Marginal price & 401 & 48 & $-0.30$ & $[-0.34,\,-0.26]$ & $-0.05$ & $[-0.06,\,-0.05]$ \\
\addlinespace
\multicolumn{7}{@{}l}{\emph{Tariff regime}} \\
\hspace{1em}Time-of-use & 210 & 19 & $-0.17$ & $[-0.22,\,-0.12]$ & $-0.01$ & $[-0.01,\,-0.01]$ \\
\hspace{1em}Increasing-block & 405 & 40 & $-0.31$ & $[-0.34,\,-0.29]$ & $-0.16$ & $[-0.16,\,-0.16]$ \\
\hspace{1em}Decreasing-block & 118 & 23 & $-0.27$ & $[-0.33,\,-0.22]$ & $-0.10$ & $[-0.10,\,-0.09]$ \\
\addlinespace
\multicolumn{7}{@{}l}{\emph{Geographic aggregation}} \\
\hspace{1em}Country-level & 396 & 87 & $-0.22$ & $[-0.25,\,-0.18]$ & $-0.08$ & $[-0.08,\,-0.07]$ \\
\hspace{1em}Sub-country level & 1,244 & 142 & $-0.28$ & $[-0.30,\,-0.27]$ & $-0.14$ & $[-0.14,\,-0.14]$ \\
\addlinespace
\multicolumn{7}{@{}l}{\emph{Region}} \\
\hspace{1em}United States & 634 & 70 & $-0.25$ & $[-0.27,\,-0.22]$ & $-0.09$ & $[-0.10,\,-0.09]$ \\
\hspace{1em}Europe & 273 & 37 & $-0.26$ & $[-0.30,\,-0.23]$ & $-0.08$ & $[-0.08,\,-0.08]$ \\
\hspace{1em}Other & 781 & 127 & $-0.30$ & $[-0.32,\,-0.27]$ & $-0.17$ & $[-0.17,\,-0.17]$ \\
\bottomrule
\end{tabular*}
\par\smallskip
\begin{minipage}{\linewidth}\footnotesize\textit{Notes:} Headline short-run sample (Marshallian-equivalent, usable standard error; 1{,}647 estimates from 226 studies). \emph{Unweighted} is the simple mean of the 1\%-winsorized elasticity with its analytic 95\% interval ($\pm 1.96\,\mathrm{s.d.}/\sqrt{N}$); \emph{Weighted} is the inverse-variance (precision-)weighted mean with its fixed-effect 95\% interval. Subsamples are defined by indicator variables and are not mutually exclusive, so they need not partition the full sample (an estimate can be, for example, both residential and panel). Tiers 1--3 are pooled as design-based; ``Sub-country level'' pools regional, city, and micro-level estimates.\end{minipage}
\end{table}

\begin{table}[t]
\centering
\caption{Summary of the long-run estimates by subsample}\label{tab:sumlr}
\footnotesize
\begin{tabular*}{\textwidth}{@{\extracolsep{\fill}}l rr rc rc@{}}
\toprule
 & & & \multicolumn{2}{c}{Unweighted} & \multicolumn{2}{c}{Weighted} \\
\cmidrule(lr){4-5}\cmidrule(lr){6-7}
Subsample & $N$ & Studies & Mean & 95\% CI & Mean & 95\% CI \\
\midrule
All estimates & 723 & 151 & $-0.58$ & $[-0.62,\,-0.55]$ & $-0.38$ & $[-0.38,\,-0.38]$ \\
\addlinespace
\multicolumn{7}{@{}l}{\emph{Identification tier}} \\
\hspace{1em}Design-based (tiers 1--3) & 4 & 1 & $-0.32$ & $[-0.36,\,-0.28]$ & $-0.29$ & $[-0.35,\,-0.23]$ \\
\hspace{1em}Instrumented (IV) & 190 & 36 & $-0.68$ & $[-0.76,\,-0.61]$ & $-0.24$ & $[-0.25,\,-0.23]$ \\
\hspace{1em}Panel FE / structural & 397 & 81 & $-0.55$ & $[-0.60,\,-0.50]$ & $-0.25$ & $[-0.26,\,-0.25]$ \\
\hspace{1em}Naive & 121 & 36 & $-0.54$ & $[-0.62,\,-0.47]$ & $-0.50$ & $[-0.50,\,-0.50]$ \\
\addlinespace
\multicolumn{7}{@{}l}{\emph{Demand sector}} \\
\hspace{1em}Residential & 464 & 106 & $-0.59$ & $[-0.64,\,-0.55]$ & $-0.43$ & $[-0.43,\,-0.43]$ \\
\hspace{1em}Industrial & 114 & 43 & $-0.50$ & $[-0.60,\,-0.40]$ & $-0.17$ & $[-0.19,\,-0.16]$ \\
\hspace{1em}Commercial & 73 & 29 & $-0.57$ & $[-0.70,\,-0.44]$ & $-0.28$ & $[-0.31,\,-0.26]$ \\
\addlinespace
\multicolumn{7}{@{}l}{\emph{Data structure}} \\
\hspace{1em}Time-series & 262 & 80 & $-0.56$ & $[-0.64,\,-0.49]$ & $-0.09$ & $[-0.10,\,-0.09]$ \\
\hspace{1em}Cross-section & 122 & 16 & $-0.64$ & $[-0.70,\,-0.58]$ & $-0.49$ & $[-0.49,\,-0.48]$ \\
\hspace{1em}Panel & 339 & 59 & $-0.58$ & $[-0.63,\,-0.53]$ & $-0.31$ & $[-0.31,\,-0.30]$ \\
\addlinespace
\multicolumn{7}{@{}l}{\emph{Price measurement}} \\
\hspace{1em}Average price & 587 & 126 & $-0.57$ & $[-0.61,\,-0.53]$ & $-0.42$ & $[-0.42,\,-0.42]$ \\
\hspace{1em}Marginal price & 52 & 17 & $-0.70$ & $[-0.83,\,-0.57]$ & $-0.49$ & $[-0.50,\,-0.47]$ \\
\addlinespace
\multicolumn{7}{@{}l}{\emph{Tariff regime}} \\
\hspace{1em}Time-of-use & 30 & 6 & $-1.00$ & $[-1.27,\,-0.72]$ & $-0.29$ & $[-0.32,\,-0.26]$ \\
\hspace{1em}Increasing-block & 152 & 23 & $-0.53$ & $[-0.60,\,-0.46]$ & $-0.47$ & $[-0.48,\,-0.47]$ \\
\hspace{1em}Decreasing-block & 45 & 10 & $-0.79$ & $[-0.92,\,-0.65]$ & $-0.85$ & $[-0.87,\,-0.84]$ \\
\addlinespace
\multicolumn{7}{@{}l}{\emph{Geographic aggregation}} \\
\hspace{1em}Country-level & 265 & 71 & $-0.52$ & $[-0.59,\,-0.46]$ & $-0.40$ & $[-0.40,\,-0.39]$ \\
\hspace{1em}Sub-country level & 449 & 80 & $-0.62$ & $[-0.67,\,-0.58]$ & $-0.37$ & $[-0.37,\,-0.37]$ \\
\addlinespace
\multicolumn{7}{@{}l}{\emph{Region}} \\
\hspace{1em}United States & 177 & 30 & $-0.57$ & $[-0.63,\,-0.51]$ & $-0.54$ & $[-0.55,\,-0.53]$ \\
\hspace{1em}Europe & 200 & 38 & $-0.58$ & $[-0.66,\,-0.50]$ & $-0.27$ & $[-0.28,\,-0.27]$ \\
\hspace{1em}Other & 353 & 94 & $-0.60$ & $[-0.65,\,-0.54]$ & $-0.41$ & $[-0.41,\,-0.41]$ \\
\bottomrule
\end{tabular*}
\par\smallskip
\begin{minipage}{\linewidth}\footnotesize\textit{Notes:} Headline long-run sample (Marshallian-equivalent, usable standard error; 723 estimates from 151 studies). Columns and conventions are as in \autoref{tab:sumsr}; a subsample present at the short but not the long horizon is omitted.\end{minipage}
\end{table}

\autoref{tab:premise} collects the statements of the rising-flexibility premise that this paper tests, in the words of those who hold it: five sources across four decades, from the earliest time-of-use pricing experiments to the field's most recent quantitative synthesis, none of which estimates a time trend in price responsiveness.

\begin{table}[t]
\centering
\begin{threeparttable}
\caption{The rising-flexibility premise, 1985--2025}
\label{tab:premise}
\footnotesize
\begin{tabularx}{\textwidth}{lX}
\toprule
Source & The premise, in the source's words \\
\midrule
\citet[p.~39]{aigner1985residential} & ``[A]ppliance choices will be made with an eye to TOU response; new appliances will become widely available\ldots{} the pricing strategy surely must be even more desirable in the long-run.'' \\
\citet{faruqui2010household} & Roughly 40\% of the \$40 billion advanced-metering investment ``could be covered by reductions in power generation costs that could be brought about through demand response''; dynamic pricing ``will have to be instituted.'' \\
\citet{torriti2014time} & ``[M]uch will change with the introduction of In Home Displays (or `smart metres') in every home.'' \\
\citet{huntington2019industrializing} & ``[I]mportant transitions may eventually reshape future energy demand responses.'' \\
\citet{kahnlang2025slices} & Time-based rates ``may be a useful and efficient tool for encouraging demand-side flexibility''; the response ``may be increasingly important as the power sector becomes more decarbonized.'' \\
\bottomrule
\end{tabularx}
\par\smallskip
\begin{minipage}{\textwidth}
\footnotesize\textit{Notes:} Verbatim quotations from the cited works. The statements span four decades and include the field's most recent quantitative synthesis, which treats the future path of responsiveness as an open prediction problem. The \citet{aigner1985residential} entry records the profession's prevailing expectation at the time rather than his own belief; he himself doubted that the effect would grow. None of these sources estimates a time trend in price responsiveness.
\end{minipage}
\end{threeparttable}
\end{table}

\FloatBarrier
\section{Studies included in the meta-analysis}\label{app:studies}

\autoref{tab:studies} lists every study contributing estimates to the corpus. Studies are ordered alphabetically by first author; each entry links by hyperlink to its full reference in the bibliography.

\begin{singlespace}\scriptsize
\begin{longtable}{@{}l@{\hskip 1.3em}l@{\hskip 1.3em}l@{}}
\caption{Studies included in the meta-analysis}\label{tab:studies}\\
\toprule\endfirsthead
\multicolumn{3}{@{}l}{\emph{\autoref{tab:studies} continued}}\\
\toprule\endhead
\bottomrule\multicolumn{3}{r}{\scriptsize Continued on next page}\\\endfoot
\bottomrule
\multicolumn{3}{@{}>{\scriptsize}p{0.95\linewidth}@{}}{\emph{Notes:} The 462 studies that make up the meta-analysis corpus, ordered alphabetically by first author and reading down each column. Each entry links to the reference list by hyperlink.}\\
\endlastfoot
\cite{ackah2014demand} & \cite{egorova2004sectoral} & \cite{mamkhezri2025assessing} \\
\cite{Acton1976} & \cite{ekpo2011dynamics} & \cite{Mansur2005} \\
\cite{Acton1980} & \cite{elshazly2013electricity} & \cite{maria2012estimation} \\
\cite{ADNDE1981} & \cite{Eltony1993} & \cite{martins2023price} \\
\cite{adom2017long} & \cite{Eltony1995} & \cite{masike2022time} \\
\cite{agostini2015elasticities} & \cite{Eltony1996} & \cite{Matsui1979} \\
\cite{Akmal2001} & \cite{Eltony1999} & \cite{Matsukawa1993} \\
\cite{Al-Faris2002} & \cite{Eltony2004} & \cite{mattos2005demanda} \\
\cite{Al-Sahlawi1999} & \cite{Eltony2006} & \cite{McFadden1977} \\
\cite{al2018estimating} & \cite{Eltony2007} & \cite{Mchugh1977} \\
\cite{al2018exploring} & \cite{enrich2024measuring} & \cite{mcrae2016price} \\
\cite{alberini2011residential} & \cite{Erickson1973} & \cite{meher2020estimating} \\
\cite{alberini2011response} & \cite{Eskeland1994} & \cite{Mendoza1987} \\
\cite{alberini2019response} & \cite{eskeland2010electricity} & \cite{mikayilov2017modeling} \\
\cite{alberini2022wild} & \cite{fan2019impacts} & \cite{mikayilov2020electricity} \\
\cite{alter2011empirical} & \cite{Fatai2003} & \cite{miller2016sensitivity} \\
\cite{amarawickrama2008electricity} & \cite{fell2014new} & \cite{modiano1984elasticidade} \\
\cite{amusa2009aggregate} & \cite{Filippini1995a} & \cite{Moghaddam2003} \\
\cite{Anderson1971} & \cite{Filippini1995b} & \cite{morovat2019estimating} \\
\cite{Anderson1973a} & \cite{Filippini1999} & \cite{Mount1974} \\
\cite{Anderson1973b} & \cite{Filippini2004} & \cite{Mount1979} \\
\cite{Anderson1974} & \cite{Filippini2011} & \cite{Mountain1989} \\
\cite{andrade1997elasticidade} & \cite{filippini2018habits} & \cite{Mountain1989a} \\
\cite{Andrikopoulos1989} & \cite{Fisher1962} & \cite{Mountain1989b} \\
\cite{Ang1992} & \cite{Fouquet1995} & \cite{Munley1990} \\
\cite{Apte1983} & \cite{frondel2019} & \cite{Murray1978} \\
\cite{Archibald1982} & \cite{fullerton2012residential} & \cite{Nagata2001} \\
\cite{Arisoy2014} & \cite{gabreyohanne2010nonlinear} & \cite{nakajima2010change} \\
\cite{aroonruengsawat2012impact} & \cite{gam2012electricity} & \cite{nakajima2010residential} \\
\cite{Arsenault1995} & \cite{Garbacz1983a} & \cite{Narayan2005} \\
\cite{Asadoorian2006} & \cite{Garbacz1983b} & \cite{Narayan2007} \\
\cite{asadoorian2008modeling} & \cite{Garbacz1983c} & \cite{nasir2008residential} \\
\cite{aslam2023untangling} & \cite{Garbacz1984a} & \cite{Okajima2013} \\
\cite{Atakhanova2005} & \cite{Garbacz1984b} & \cite{Oliveira1993} \\
\cite{atalla2016modelling} & \cite{Garbacz1984c} & \cite{Olivia2008} \\
\cite{athukorala2010estimating} & \cite{Garbacz1986} & \cite{ortizvelazqu2017analisis} \\
\cite{athukorala2019household} & \cite{Garcia-Cerrutti2000} & \cite{ota2018demographic} \\
\cite{Atkinson1979a} & \cite{gautam2018estimating} & \cite{OteroPrada1984} \\
\cite{Atkinson1979b} & \cite{ghaderi2006electricity} & \cite{otsuka2016determinants} \\
\cite{auray2019price} & \cite{Gill1978} & \cite{otsuka2017determinants} \\
\cite{azevedo2011residential} & \cite{Glakpe1985} & \cite{otsuka2023lessons} \\
\cite{babatunde2009demand} & \cite{Gollnick1975} & \cite{Parfomak1996} \\
\cite{Badri1992} & \cite{goo2019urban} & \cite{Parhizgari1978} \\
\cite{Balabanoff1994} & \cite{Green1986} & \cite{Park1984} \\
\cite{Banda2007} & \cite{Gundimeda2008} & \cite{Parti1980} \\
\cite{Barnes1981} & \cite{Guo1994} & \cite{paul2009partial} \\
\cite{barrientos2018estimation} & \cite{halicioglu2007residential} & \cite{pellini2021estimating} \\
\cite{Basu1976} & \cite{Hall1986} & \cite{Pesaran1999} \\
\cite{Baughman1979} & \cite{Halvorsen1975} & \cite{pielow2012modeling} \\
\cite{Beenstock1999} & \cite{Halvorsen1976} & \cite{Pindyck1979} \\
\cite{Beierlein1981} & \cite{Halvorsen1977} & \cite{Pindyck1980} \\
\cite{bekhet2011assessing} & \cite{Halvorsen1979} & \cite{Pitt1985} \\
\cite{Belanger1990} & \cite{Halvorsen2001} & \cite{polemis2007modeling} \\
\cite{benavente2005estimando} & \cite{ham1997time} & \cite{polemis2013electricity} \\
\cite{Bernard1987} & \cite{Hartman1981} & \cite{pourazarm2013estimating} \\
\cite{Bernard1996} & \cite{hasanov2016modeling} & \cite{qi2008application} \\
\cite{bernard2011pseudo} & \cite{Hausman1979} & \cite{Rahman1982} \\
\cite{Berndt1980} & \cite{Hawkins1975} & \cite{rai2014price} \\
\cite{Berndt1984} & \cite{Hawkins1977} & \cite{Ramcharran1988} \\
\cite{Bernstein2006} & \cite{Hawkins1978} & \cite{ramcharran1990electricity} \\
\cite{bernstein2011responsiveness} & \cite{He2004} & \cite{rapson2014durable} \\
\cite{bernstein2015short} & \cite{Henderson1983} & \cite{romerojordan2014household} \\
\cite{Betancourt1981} & \cite{Henriksson2014} & \cite{rosasflores2017elements} \\
\cite{bianco2009electricity} & \cite{Henson1984} & \cite{Rossi1989} \\
\cite{bianco2010analysis} & \cite{Herriges1994} & \cite{Roth1981} \\
\cite{Bigano2006} & \cite{Hesse1986} & \cite{rouhani2022estimating} \\
\cite{bigerna2014electricity} & \cite{Hieronymus1976} & \cite{Roy1986} \\
\cite{bilgili2010short} & \cite{Hill1983} & \cite{Ryan1996} \\
\cite{biswas2026electricity} & \cite{hirth2022very} & \cite{ryu2021household} \\
\cite{Bjerkholt1983} & \cite{Hogan1989} & \cite{sa2009electricity} \\
\cite{Bjorner2001} & \cite{Holtedahl2004} & \cite{sabir2013demand} \\
\cite{blasch2017explaining} & \cite{hondroyiannis2004estimating} & \cite{Saddler1980} \\
\cite{blazquez2013residential} & \cite{Horowitz2007} & \cite{Saddler1983} \\
\cite{Blundell1999} & \cite{hosoe2009regional} & \cite{saha2018analysis} \\
\cite{boogen2017dynamic} & \cite{Houston1982} & \cite{salari2016residential} \\
\cite{borenstein2009electricity} & \cite{Houthakker1951} & \cite{saunoris2013dynamics} \\
\cite{Bose1999} & \cite{Houthakker1974} & \cite{schulte2017price} \\
\cite{Botero1990} & \cite{Houthakker1980} & \cite{Schwarz1984} \\
\cite{Branch1993} & \cite{Hsiao1989} & \cite{setshedi2025electricity} \\
\cite{Brenton1997} & \cite{Hsiao1994} & \cite{shaffer2020misunderstanding} \\
\cite{burke2018electricity} & \cite{hsing1994estimation} & \cite{Shi2012} \\
\cite{burke2018price} & \cite{Hsueh1986} & \cite{Shin1981} \\
\cite{bushnell2005consumption} & \cite{hu2019energy} & \cite{Shin1985} \\
\cite{byrne2021experimental} & \cite{huang2024comprehensive} & \cite{Silk1997} \\
\cite{cabral2020elasticity} & \cite{Hughes-Cromwick1985} & \cite{silva2018electricity} \\
\cite{campbell2018price} & \cite{hung2015dynamic} & \cite{siqueira2006demanda} \\
\cite{cao2019chinese} & \cite{Huntington1982} & \cite{Smith1980} \\
\cite{cao2023experiment} & \cite{Hyndman1980} & \cite{solaymani2015aggregate} \\
\cite{Cargill1971} & \cite{idso2024price} & \cite{Sterner1985} \\
\cite{Carlevaro1983} & \cite{IEEJ1986} & \cite{Sterner1989} \\
\cite{casarin2011price} & \cite{Ilmakunnas1989} & \cite{su2018electricity} \\
\cite{Cavoulacos1983} & \cite{Inglesi-Lotz2011} & \cite{sun2013reforming} \\
\cite{cebula2009empirical} & \cite{Inglesi-Lotz2011a} & \cite{Sutherland1983a} \\
\cite{cebula2012us} & \cite{inglesi2010aggregate} & \cite{Sutherland1983b} \\
\cite{cetinkaya2015electricity} & \cite{Iqbal1986} & \cite{SZConsultores1999} \\
\cite{chama2012econometric} & \cite{irffi2006dynamic} & \cite{talbi2022does} \\
\cite{Chang1981a} & \cite{ishaque2018revisiting} & \cite{tambe2014estimating} \\
\cite{Chang1981b} & \cite{ito2014} & \cite{tanishita2019price} \\
\cite{Chang1991} & \cite{ito2018moral} & \cite{tatla2017short} \\
\cite{chang2003electricity} & \cite{Jaffee1982} & \cite{Taylor1977} \\
\cite{chang2014time} & \cite{Jamil2011} & \cite{Taylor1979} \\
\cite{Chaudhary1999} & \cite{javid2014electricity} & \cite{taylor2005hourly} \\
\cite{chaudhry2010panel} & \cite{jeong2011household} & \cite{Terza1986} \\
\cite{chen2025modeling} & \cite{jeong2025analysis} & \cite{THEC1983} \\
\cite{Chern1975} & \cite{jessoe2014knowledge} & \cite{Tiwari2000} \\
\cite{Chern1978} & \cite{jia2021elasticities} & \cite{tiwari2019time} \\
\cite{Chern1988} & \cite{jin2022elasticity} & \cite{tran2023estimation} \\
\cite{chindarkar2019price} & \cite{johnsen2001demand} & \cite{Tserkezos1992} \\
\cite{Chishti1993} & \cite{Jones1995} & \cite{turkekul2011co} \\
\cite{Choi2002} & \cite{Jungeilges1986} & \cite{twerefou2020efficiency} \\
\cite{Christodoulakis1997} & \cite{Kamerschen2004} & \cite{uhr2019estimation} \\
\cite{Christopoulos2000} & \cite{Karbuz1997} & \cite{Urga2003} \\
\cite{Chung1981} & \cite{Kaserman1985} & \cite{Uri1977} \\
\cite{cialani2018household} & \cite{KEEI1989} & \cite{Uri1978} \\
\cite{cicchetti1975alternative} & \cite{Keng1991} & \cite{Uri1979b} \\
\cite{CISEPA1998} & \cite{khan2009demand} & \cite{Uri1979c} \\
\cite{Considine2000} & \cite{khan2016dynamics} & \cite{Uri1979d} \\
\cite{costa2010california} & \cite{khanna2016effects} & \cite{Uri1979e} \\
\cite{costa2011electricity} & \cite{Khazzoom1986} & \cite{Uri1982} \\
\cite{Coughlin1995} & \cite{knaut2016hen} & \cite{Uri1983} \\
\cite{cuddington2015estimating} & \cite{Kohler1984} & \cite{Vashist1984} \\
\cite{Dahan1996} & \cite{kohler2014differential} & \cite{Veall1983} \\
\cite{damien2019impacts} & \cite{Kokkelenberg1993} & \cite{Veall1987} \\
\cite{davis2008durable} & \cite{Kolstad1993} & \cite{Velez1987} \\
\cite{DeCian2007} & \cite{krishnamurthy2015cross} & \cite{Verleger1973} \\
\cite{delfino1979demanda} & \cite{kwon2016effects} & \cite{Vlachou1986} \\
\cite{Delfino1995} & \cite{lanot2021price} & \cite{volland2018price} \\
\cite{Denton1999} & \cite{Lareau1982} & \cite{wakashiro2019estimating} \\
\cite{Denton2003} & \cite{Larsson2004} & \cite{walfridson1987dynamic} \\
\cite{Dergiades2008} & \cite{Larsson2006} & \cite{Walker1979} \\
\cite{deryugina2020long} & \cite{Laumas1981} & \cite{Wang1985} \\
\cite{DeVita2006} & \cite{lavin2011impact} & \cite{wang2017industrial} \\
\cite{Diabli1998} & \cite{lee2010panel} & \cite{wang2020price} \\
\cite{dicembrino2013structural} & \cite{Lee2011} & \cite{wang2021performance} \\
\cite{dilaver2011industrial} & \cite{li2020price} & \cite{Westley1984a} \\
\cite{dilaver2011turkish} & \cite{liddle2021prices} & \cite{Westley1984b} \\
\cite{Dobozi1988} & \cite{lijesen2007real} & \cite{Westley1989a} \\
\cite{Dodgson1990} & \cite{Lillard1981} & \cite{Westley1989b} \\
\cite{Donatos1989} & \cite{Lim2014} & \cite{Westley1992} \\
\cite{Donatos1991} & \cite{lin2014electricity} & \cite{Wijemanne1987} \\
\cite{dong2018price} & \cite{lin2020chinese} & \cite{Wilson1974} \\
\cite{dong2020estimating} & \cite{Liu2005} & \cite{woo2018price} \\
\cite{Donnelly1984} & \cite{Lohani1992} & \cite{Yang1978} \\
\cite{Donnelly1984a} & \cite{loi2018analysing} & \cite{yang2023environmental} \\
\cite{Donnelly1984b} & \cite{Lyman1994} & \cite{yin2016long} \\
\cite{Donnelly1985} & \cite{Lynk1989} & \cite{Yoo2007} \\
\cite{Donnelly1987} & \cite{ma2016long} & \cite{Young1983} \\
\cite{Douthitt1989} & \cite{Macroconsult2001} & \cite{yu2020demand} \\
\cite{Dubin1985} & \cite{Maddala1994} & \cite{Zachariadis2007} \\
\cite{dulleck2004customer} & \cite{Maddala1997} & \cite{zarnikau1990customer} \\
\cite{Duncan1976} & \cite{Maddigan1983} & \cite{zhai2023price} \\
\cite{Dunstan1988} & \cite{Maddock1991} & \cite{zhou2013estimation} \\
\cite{durant1990residential} & \cite{madlener2011econometric} & \cite{Ziramba2008} \\
\cite{durmaz2020estimation} & \cite{Mahmud1990} & \cite{ziramba2012long} \\
\end{longtable}
\end{singlespace}

\FloatBarrier
\section{Coding the identification ladder}\label{app:coding}

Each estimate's price-identification strategy is coded from the primary study's description of its price variation, not from its estimator label alone. Tier 1 (randomized): experimental assignment of tariffs or randomized encouragement (e.g., \citealp{wolak2011residential,jessoe2014knowledge,ito2018moral}). Tier 2 (natural experiment): plausibly exogenous price variation from a mandated change or a spatial boundary design (e.g., \citealp{ito2014}, exploiting spatial discontinuities across utility service-area boundaries). Tier 3: difference-in-differences designs against untreated comparison groups (e.g., \citealp{deryugina2020long}, which tracks municipally-aggregated communities against untreated neighbors). Tier 4: instrumented specifications (instrumental variables, IV; two- and three-stage least squares, 2SLS and 3SLS; and the generalized method of moments, GMM), including supply-shifter and Lewbel-type instruments (e.g., \citealp{alberini2019response}). Of the 1,068 Tier 4 estimates from 91 studies in the full corpus, only 3 studies report any recoverable first-stage instrument-strength diagnostic (a first-stage $F$, Kleibergen-Paap, or comparable statistic), and none does so in a systematically coded field; we therefore cannot audit instrument strength across the tier and note the coverage rather than adjudicate it. Tier 5: panel fixed effects, cointegration systems, and structural demand systems without price instrumentation (e.g., \citealp{Bernstein2006}). Tier 6: naive OLS (ordinary least squares) or time-series regressions of quantity on price; \citet{Barnes1981}, for instance, report such an estimate alongside an instrumented one and trace the gap between them to the simultaneity that block pricing induces between price and quantity. Estimates whose strategy cannot be determined are excluded from ladder analyses (878 estimates spanning 61 studies, of which 91 carry a usable Marshallian-equivalent estimate and so enter the headline pool that feeds every non-ladder analysis: the horizon path, the two-clock test, and the publication-bias battery). The full per-study mapping ships with the replication package. In the short-run headline sample the tiers contain 81 (design-based), 549 (IV), 655 (FE/structural), and 316 (naive) estimates. Coding in this literature, ours included, has historically prioritized each study's preferred estimates; the corpus carries a \texttt{preferred} flag (1{,}204 estimates), and all headline results are robust to restricting the sample to these flagged estimates (\autoref{sec:selection}).

The tier assignments were made by three coders from each study's description of its price variation, and the full dataset then underwent a study-by-study audit against the primary sources, including a re-check of the estimates drawn from the earlier compilation of \citet{dahl2011}. Three features make the classification dependable. First, the complete per-study tier mapping ships with the replication package, so every assignment can be checked against the primary source. Second, the classification that carries the argument, design-based (Tiers 1--3) versus not, turns on observable, low-discretion features of a study's design (random assignment, a mandated or plausibly exogenous price change, or an explicit untreated comparison group) rather than on a subjective judgment of estimator quality. Third, the headline analyses collapse the six tiers into four groups (design-based, instrumented, fixed-effects/structural, and naive), so boundary calls within a group do not bear on the design-based-versus-naive contrast that anchors the paper.

\FloatBarrier
\section{Within-study contrasts and robustness}\label{app:robust}

The imputed data-year offset is immaterial to the calendar-time results. Dropping all 23 estimates with no reported data years (assigned publication year minus three in the baseline) and re-estimating leaves the composition-adjusted short-run data-clock trend at $+0.0168$ per decade, against the reported $+0.017$ (unchanged across every horizon and specification at the reported precision); the short- and long-run two-clock data-clock coefficients are likewise unchanged at $-0.0334$ and $-0.0919$.

\begin{figure}[H]
\centering
\caption{Within-study tier contrasts}
\label{fig:withinstudy}
\begin{threeparttable}
\includegraphics[width=.72\textwidth]{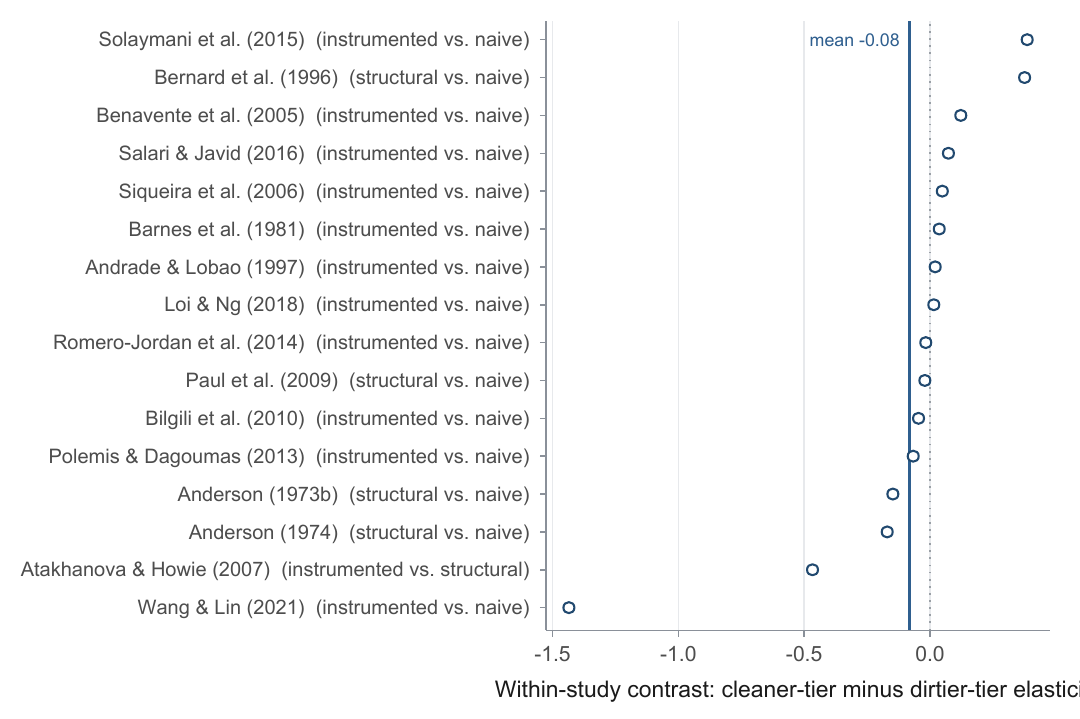}
\begin{tablenotes}[flush]
\footnotesize
\item \textit{Notes:} For each of the sixteen studies reporting estimates on more than one identification tier (tier pair in parentheses), the mean estimate on the cleaner tier minus the mean estimate on the lower tier, same study and data. The mean contrast is $-0.081$ (cleaner estimates more elastic, driven by IV-vs-naive pairs), with a paired $t$ of $-0.79$ and positive contrasts in 50\% of studies: no significant systematic within-study shift, consistent with the IV overshoot in \autoref{tab:ladder}.
\end{tablenotes}
\end{threeparttable}
\end{figure}

\begin{figure}[H]
\centering
\caption{The adjustment path as a continuous curve (illustrative)}
\label{fig:metairf}
\begin{threeparttable}
\includegraphics[width=.9\textwidth]{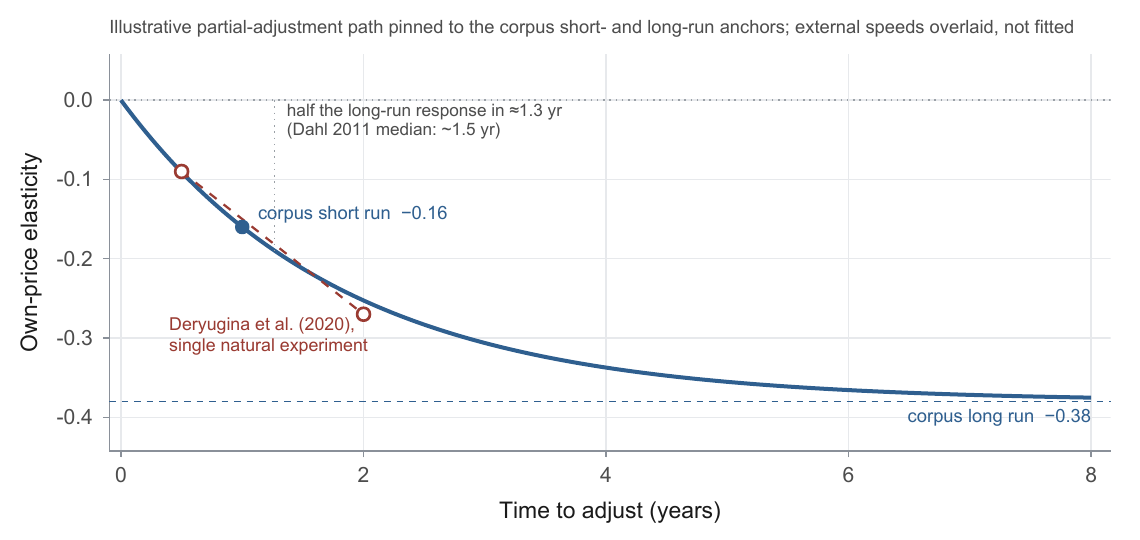}
\begin{tablenotes}[flush]
\footnotesize
\item \textit{Notes:} Illustrative partial-adjustment path $\varepsilon(t)=\varepsilon_{\text{LR}}(1-e^{-\lambda t})$ with $\varepsilon_{\text{LR}}=-0.38$ (the corrected long-run elasticity) and $\lambda$ set so $\varepsilon(1\text{ yr})=-0.16$ (the corrected short-run elasticity); the implied half-life of adjustment is about $1.3$ years. Overlaid, not fitted: the median dynamic-model adjustment speed of \citet{dahl2011} (half the long-run response in about $1.5$ years) and the natural-experiment path of \citet{deryugina2020long} ($-0.09$ at six months, $-0.27$ at two years). The curve illustrates the discrete gradient of \autoref{fig:horizon}; it is not a structural estimate.
\end{tablenotes}
\end{threeparttable}
\end{figure}

\medskip\noindent\textbf{The adjustment path as a continuous curve.} The horizon result of \autoref{sec:horizon} is a three-point gradient, short, intermediate, and long run. A single smooth adjustment path reproduces it (\autoref{fig:metairf}). We take a textbook partial-adjustment form $\varepsilon(t)=\varepsilon_{\text{LR}}\,(1-e^{-\lambda t})$, fix the long-run asymptote at the corrected value ($-0.38$), and set the speed $\lambda$ so the curve reaches the corrected short-run value ($-0.16$) at one year. Nothing in this path is fitted to dynamics in our data, which record only discrete horizon flags, yet it lands on two independent external benchmarks: the median speed of adjustment across dynamic models, half the long-run response in about a year and a half \citep{dahl2011}, and the within-experiment trajectory of \citet{deryugina2020long}, whose six-month and two-year points fall almost on the curve. We read the agreement as corroboration that the horizon gradient is a genuine speed of adjustment rather than composition across horizon bins. We present the curve as an illustration only: it imposes a functional form the rest of the paper avoids, and a true meta-analytic impulse response would require each dynamic study's own adjustment coefficient, which the final dataset does not record.

\noindent\textbf{Sample robustness: the pure-Marshallian core.} The baseline pools directly reported Marshallian estimates with Slutsky-converted compensated (Hicksian) ones (\autoref{sec:data}). Restricting instead to the 2,579 directly reported Marshallian estimates alone (from 272 studies) leaves the picture unchanged: short-run PET $-0.172$ (se $0.028$), long-run PET $-0.356$ (se $0.041$), funnel asymmetry $p<0.001$ in both. The calendar-time trend is unchanged as well: on this native-Marshallian sample the fully adjusted short-run data-clock coefficient is $+0.043$ per decade ($p=0.234$), the same sign and the same statistical indistinguishability from zero as the pooled headline, and the time-of-use coefficient stays positive and significant ($+0.19$, $p<0.001$). The harmonization is numerically inconsequential in either direction: because electricity's budget share is small the Slutsky shift is about $0.012$ per estimate, varying the budget share $s$ over $[0.02, 0.08]$ moves the short-run corrected value by less than $0.01$, and the \emph{converted-from-Hicksian} moderator in the model average (\autoref{app:bma}) is robustly selected only in the long-run model (posterior mean $-0.18$) and does not overturn the pooled short- and long-run corrections. One-preferred-estimate-per-study and sample-size weighting (no standard errors in the weights) likewise leave the short-run correction in the same range. The long-run ladder preserves the short-run IV overshoot (design-based cell too thin to report): IV $-0.41$ against FE $-0.33$, with naive at $-0.23$ adjusted.

\medskip\noindent\textbf{The estimates held out of the baseline.} The 1,396 estimates held out of the Marshallian-equivalent baseline (\autoref{app:search} gives their composition) are statistically indistinguishable from the estimation sample: \autoref{tab:heldout} shows their raw and bias-corrected means and their funnel asymmetry track the baseline, and in a pooled funnel regression neither the level of the held-out estimates ($-0.03$, $p=0.62$) nor their funnel slope ($-0.01$, $p=0.97$) differs from it. Re-running the correction on the full corpus of 4,720 leaves the bias-corrected elasticity at $-0.25$, the value the baseline delivers. The restriction to the estimation sample is a matter of placing estimates on a common scale, not a selection that shapes the result.

\FloatBarrier
\section{Publication bias: the full battery}\label{app:pb}

\begin{figure}[!ht]
\centering
\caption{Funnel plots of the short-run and long-run elasticities}\label{fig:funnels}
\begin{threeparttable}
\includegraphics[width=\textwidth]{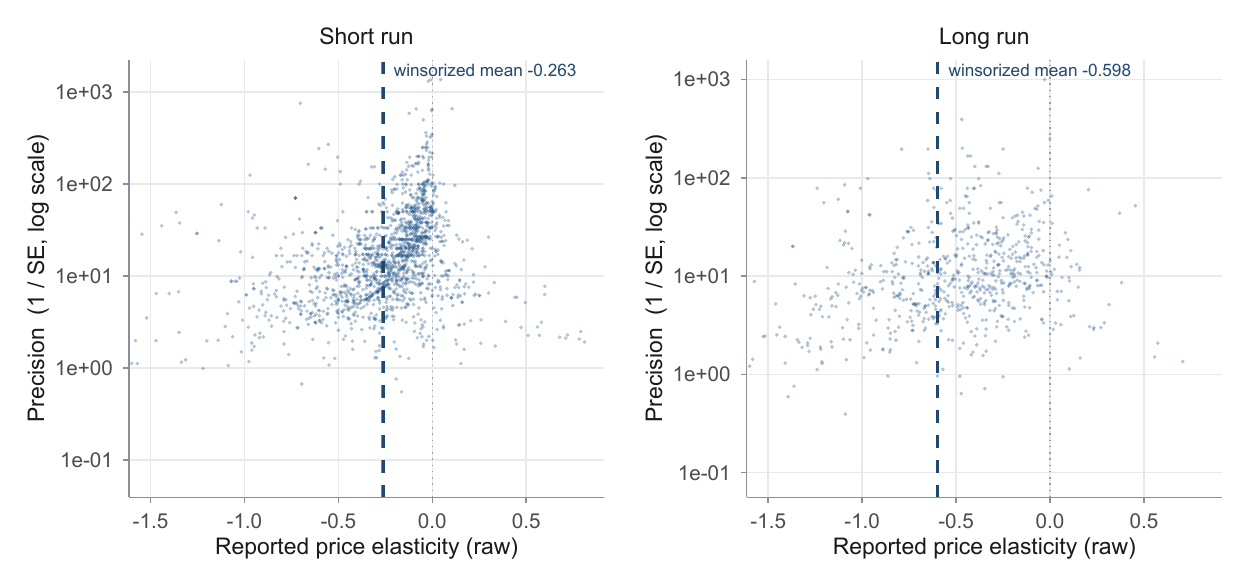}
\begin{tablenotes}[flush]
\footnotesize
\item \textit{Notes:} Raw reported own-price elasticities against their precision ($1/\mathrm{SE}$), for the short-run and long-run headline samples; the dashed line marks the 1\%-winsorized simple mean of each sample ($-0.26$ short run, $-0.59$ long run). The leftward asymmetry, a thinning of imprecise estimates on the positive side, is the funnel signature of selection against small or wrong-signed elasticities. The vertical axis (precision, $1/\mathrm{SE}$) is drawn on a logarithmic scale, which spreads the dense cluster of imprecise estimates near the bottom and makes the asymmetry easier to see; the horizontal axis is clipped to $[-1.5, 0.8]$ for legibility, and a small number of outliers fall outside the frame but enter every calculation.
\end{tablenotes}
\end{threeparttable}
\end{figure}

The main text corrects the short-run elasticity with a compact battery (\autoref{tab:battery}); here we re-run the \emph{full} battery: the linear FAT-PET family (Panel~A), the nonlinear correctors, including the Bom--Rachinger endogenous kink (Panel~B), and the endogeneity- and selection-robust estimators (MAIVE, $p$-uniform*, and a selection model; Panel~C), on each sample the paper leans on, so the correction can be read off estimator by estimator. \autoref{fig:funnels} plots the raw short- and long-run estimates against their precision. \autoref{tab:pbhorizon} runs the battery on the short-run and long-run headline samples; \autoref{tab:pbtier1} and \autoref{tab:pbtier2} split the short-run sample by identification tier (naive and panel fixed-effects; instrumented and design-based); and \autoref{tab:pbprice} splits it by price regime (time-of-use tariffs and marginal price). The message is broadly uniform: in nearly every cell the correctors pull the reported mean toward zero, the funnel-asymmetry (FAT) slope is negative wherever the sample is observational, and the least-selected cell, the design-based sample (baseline FAT slope $+0.029$), carries one of the least price-responsive corrected values; the time-of-use sample is strongly selected (its FAT slope is among the largest in the battery) yet its corrected value is still near zero, which strengthens rather than weakens the low-elasticity conclusion. One further corrector, RTMA (right-truncated meta-analysis; \citealp{Mathur2024}), is not identified in this literature and we do not report it: nearly every estimate is \emph{affirmative} (statistically significant with the expected negative sign), so the small residue of non-affirmative estimates on which the right-truncated likelihood relies is too thin to pin down the model, which then over-corrects implausibly past the raw mean (to $+0.26$ in the short run).

\begin{table}[!ht]
\centering
\begin{threeparttable}
\caption{Estimates held out of the estimation sample}
\label{tab:heldout}
\small
\begin{tabular}{@{}l r r r r c r@{}}
\toprule
 & $N$ & Studies & Mean & Median & PET (SE) & FAT \\
\midrule
Baseline (estimated) & 3{,}324 & 366 & $-0.38$ & $-0.28$ & $-0.24$ ($0.02$) & $-0.92$ \\
\addlinespace
\multicolumn{7}{@{}l}{\emph{Held out of estimation:}} \\
\hspace{0.4cm}Not Marshallian-convertible & 859 & 77 & $-0.41$ & $-0.24$ & $-0.27$ ($0.05$) & $-0.96$ \\
\hspace{0.4cm}Inverted-formula & 420 & 32 & $-0.47$ & $-0.31$ & $-0.37$ ($0.10$) & $-0.60$ \\
\hspace{0.4cm}Imputed standard error & 117 & 1 & $-0.11$ & $-0.09$ & -- & -- \\
\hspace{0.4cm}All held out & 1{,}396 & 110 & $-0.40$ & $-0.23$ & $-0.27$ ($0.05$) & $-0.93$ \\
\addlinespace
Full corpus & 4{,}720 & 462 & $-0.39$ & $-0.26$ & $-0.25$ ($0.02$) & $-0.93$ \\
\bottomrule
\end{tabular}
\begin{tablenotes}[flushleft]\footnotesize
\item \emph{Notes:} Own-price elasticities (1\%-winsorized), all horizons pooled. ``Not Marshallian-convertible'' estimates are compensated (Hicksian) elasticities that lack a usable income elasticity for the Slutsky conversion. PET is the FAT-PET bias-corrected mean with study-clustered standard error; FAT is the funnel-asymmetry slope. The imputed-standard-error group is a single study, so its clustered FAT-PET is not identified. In a pooled funnel regression with a held-out indicator, the held-out estimates differ from the baseline neither in level ($-0.03$, $p=0.62$) nor in funnel slope ($-0.01$, $p=0.97$).
\end{tablenotes}
\end{threeparttable}
\end{table}

\clearpage
\begin{singlespace}\footnotesize\renewcommand{\arraystretch}{0.9}\setlength{\tabcolsep}{4pt}
\begin{longtable}{@{}>{\raggedright\arraybackslash}p{\dimexpr\linewidth-2.2cm*5-\tabcolsep*10\relax}*{5}{>{\centering\arraybackslash}p{2.2cm}}@{}}
\caption{Publication-bias battery: short-run and long-run elasticities}\label{tab:pbhorizon}\\
\endfirsthead
\multicolumn{6}{@{}l}{\emph{\autoref{tab:pbhorizon} continued}}\\
\endhead
\midrule\multicolumn{6}{r}{\scriptsize\emph{Continued on next page}}\\
\endfoot
\bottomrule
\multicolumn{6}{@{}>{\scriptsize}p{\linewidth}@{}}{\emph{Notes:} Publication-bias--corrected mean price elasticity from each estimator, on the 1\%-winsorized headline sample. \emph{Panel A} is the FAT-PET meta-regression $\varepsilon_{is}=\varepsilon_{0}+\gamma\,SE(\varepsilon_{is})+u_{is}$: ``Publication bias'' is the slope $\gamma$ and ``Effect beyond bias'' the intercept $\varepsilon_{0}$ (each coefficient's standard error below it); the columns are OLS unweighted, FE inverse-variance weighted, RE random effects (DerSimonian--Laird), Study one median estimate per study, and Precision inverse-SE weighted. Standard errors are two-way clustered by study and by underlying database wherever a regression is estimated (OLS, FE, Precision, PEESE, and EK); RE reports the DerSimonian--Laird standard error and the study-level techniques are already one observation per study. \emph{Panel B}: PEESE \citep{stanley2014meta}, WAAP \citep{ioannidis2017power}, the top-10\% estimator, STEM \citep{furukawa2019publication}, and EK, the endogenous-kink meta-regression of \citet{bom2019kinked} whose kink point $a_1=|\varepsilon_0|/1.96$ is iterated to convergence. \emph{Panel C}: MAIVE \citep{irsova2023spurious} instruments the estimate variance with the inverse sample size (top 1\% of SEs dropped) and is reported only where the first-stage instrument is adequate ($F\ge10$; otherwise only the $F$ is shown); $p$-uniform* \citep{van2021correcting} works from study medians. ``Selection'' is a step-function selection model (a Vevea--Hedges / Andrews--Kasy-type maximum-likelihood correction \citep{vevea1995general,andrews2019identification} with a cut at the one-sided 2.5\% threshold). ``Observations'' is the number of estimates in the sample, all of which enter every test; the study-level techniques (Study, STEM, $p$-uniform*) then aggregate them to one value per study, and WAAP and the top-10\% estimator average a precision-selected subset of the same estimates. $^{***}$, $^{**}$, and $^{*}$ denote statistical significance at the 1\%, 5\%, and 10\% level.}\\
\endlastfoot
\toprule
\multicolumn{6}{@{}l}{\textbf{Part 1: Short-run elasticities}}\\
\midrule
\multicolumn{6}{@{}l}{\emph{Panel A: Linear techniques}}\\
\addlinespace[2pt]
 & OLS & FE & RE & Study & Precision\\
\midrule
Publication bias & -0.851$^{***}$ & -2.055$^{***}$ & -0.953$^{***}$ & -1.281$^{***}$ & -1.171$^{***}$\\
\emph{\hspace{2mm}(Standard error)} & (0.154) & (0.708) & (0.072) & (0.128) & (0.202)\\
\addlinespace[2pt]
Effect beyond bias & -0.163$^{***}$ & -0.093$^{***}$ & -0.158$^{***}$ & -0.143$^{***}$ & -0.124$^{***}$\\
\emph{\hspace{2mm}(Constant)} & (0.025) & (0.034) & (0.007) & (0.019) & (0.024)\\
\addlinespace[2pt]
Observations & 1,647 & 1,647 & 1,647 & 1,647 & 1,647\\
\midrule
\multicolumn{6}{@{}l}{\emph{Panel B: Nonlinear techniques}}\\
\addlinespace[2pt]
 & Top 10\% & WAAP & PEESE & STEM & EK\\
\midrule
Effect beyond bias & -0.104$^{***}$ & -0.111$^{***}$ & -0.231$^{***}$ & -0.239$^{***}$ & -0.113$^{***}$\\
 & (0.000) & (0.000) & (0.020) & (0.019) & (0.029)\\
\addlinespace[2pt]
Observations & 1,647 & 1,647 & 1,647 & 1,647 & 1,647\\
\midrule
\multicolumn{6}{@{}l}{\emph{Panel C: Endogeneity- and selection-robust techniques}}\\
\addlinespace[2pt]
 & & & p-uniform* & MAIVE & Selection \\
\midrule
Effect beyond bias &  &  & -0.145$^{***}$ & -0.181$^{***}$ & -0.155$^{***}$\\
 &  &  & (0.027) & (0.022) & (0.011)\\
\addlinespace[2pt]
First-stage \emph{F} (MAIVE) &  &  &  & 23 & \\
Observations &  &  & 1,647 & 1,647 & 1,647\\
\addlinespace[2pt]
\addlinespace[2pt]
\toprule
\multicolumn{6}{@{}l}{\textbf{Part 2: Long-run elasticities}}\\
\midrule
\multicolumn{6}{@{}l}{\emph{Panel A: Linear techniques}}\\
\addlinespace[2pt]
 & OLS & FE & RE & Study & Precision\\
\midrule
Publication bias & -0.993$^{***}$ & -1.337 & -1.055$^{***}$ & -1.095$^{***}$ & -0.979$^{***}$\\
\emph{\hspace{2mm}(Standard error)} & (0.160) & (1.096) & (0.104) & (0.160) & (0.220)\\
\addlinespace[2pt]
Effect beyond bias & -0.377$^{***}$ & -0.359$^{***}$ & -0.365$^{***}$ & -0.351$^{***}$ & -0.380$^{***}$\\
\emph{\hspace{2mm}(Constant)} & (0.036) & (0.091) & (0.018) & (0.046) & (0.045)\\
\addlinespace[2pt]
Observations & 723 & 723 & 723 & 723 & 723\\
\midrule
\multicolumn{6}{@{}l}{\emph{Panel B: Nonlinear techniques}}\\
\addlinespace[2pt]
 & Top 10\% & WAAP & PEESE & STEM & EK\\
\midrule
Effect beyond bias & -0.372$^{***}$ & -0.376$^{***}$ & -0.498$^{***}$ & -0.424$^{***}$ & -0.378$^{***}$\\
 & (0.001) & (0.001) & (0.032) & (0.114) & (0.076)\\
\addlinespace[2pt]
Observations & 723 & 723 & 723 & 723 & 723\\
\midrule
\multicolumn{6}{@{}l}{\emph{Panel C: Endogeneity- and selection-robust techniques}}\\
\addlinespace[2pt]
 & & & p-uniform* & MAIVE & Selection \\
\midrule
Effect beyond bias &  &  & -0.385$^{***}$ &  & -0.420$^{***}$\\
 &  &  & (0.052) &  & (0.023)\\
\addlinespace[2pt]
First-stage \emph{F} (MAIVE) &  &  &  & $<1$ & \\
Observations &  &  & 723 & 723 & 723\\
\end{longtable}
\end{singlespace}
\clearpage
\begin{singlespace}\footnotesize\renewcommand{\arraystretch}{0.9}\setlength{\tabcolsep}{4pt}
\begin{longtable}{@{}>{\raggedright\arraybackslash}p{\dimexpr\linewidth-2.2cm*5-\tabcolsep*10\relax}*{5}{>{\centering\arraybackslash}p{2.2cm}}@{}}
\caption{Publication-bias battery: naive and panel fixed-effects tiers}\label{tab:pbtier1}\\
\endfirsthead
\multicolumn{6}{@{}l}{\emph{\autoref{tab:pbtier1} continued}}\\
\endhead
\midrule\multicolumn{6}{r}{\scriptsize\emph{Continued on next page}}\\
\endfoot
\bottomrule
\multicolumn{6}{@{}>{\scriptsize}p{\linewidth}@{}}{\emph{Notes:} Publication-bias--corrected mean price elasticity from each estimator, on the 1\%-winsorized headline sample. \emph{Panel A} is the FAT-PET meta-regression $\varepsilon_{is}=\varepsilon_{0}+\gamma\,SE(\varepsilon_{is})+u_{is}$: ``Publication bias'' is the slope $\gamma$ and ``Effect beyond bias'' the intercept $\varepsilon_{0}$ (each coefficient's standard error below it); the columns are OLS unweighted, FE inverse-variance weighted, RE random effects (DerSimonian--Laird), Study one median estimate per study, and Precision inverse-SE weighted. Standard errors are two-way clustered by study and by underlying database wherever a regression is estimated (OLS, FE, Precision, PEESE, and EK); RE reports the DerSimonian--Laird standard error and the study-level techniques are already one observation per study. \emph{Panel B}: PEESE \citep{stanley2014meta}, WAAP \citep{ioannidis2017power}, the top-10\% estimator, STEM \citep{furukawa2019publication}, and EK, the endogenous-kink meta-regression of \citet{bom2019kinked} whose kink point $a_1=|\varepsilon_0|/1.96$ is iterated to convergence. \emph{Panel C}: MAIVE \citep{irsova2023spurious} instruments the estimate variance with the inverse sample size (top 1\% of SEs dropped) and is reported only where the first-stage instrument is adequate ($F\ge10$; otherwise only the $F$ is shown); $p$-uniform* \citep{van2021correcting} works from study medians. ``Selection'' is a step-function selection model (a Vevea--Hedges / Andrews--Kasy-type maximum-likelihood correction \citep{vevea1995general,andrews2019identification} with a cut at the one-sided 2.5\% threshold). ``Observations'' is the number of estimates in the sample, all of which enter every test; the study-level techniques (Study, STEM, $p$-uniform*) then aggregate them to one value per study, and WAAP and the top-10\% estimator average a precision-selected subset of the same estimates. $^{***}$, $^{**}$, and $^{*}$ denote statistical significance at the 1\%, 5\%, and 10\% level.}\\
\endlastfoot
\toprule
\multicolumn{6}{@{}l}{\textbf{Part 1: Naive}}\\
\midrule
\multicolumn{6}{@{}l}{\emph{Panel A: Linear techniques}}\\
\addlinespace[2pt]
 & OLS & FE & RE & Study & Precision\\
\midrule
Publication bias & -0.750$^{***}$ & -3.187$^{*}$ & -0.736$^{***}$ & -1.227$^{***}$ & -1.330$^{***}$\\
\emph{\hspace{2mm}(Standard error)} & (0.204) & (1.752) & (0.170) & (0.206) & (0.418)\\
\addlinespace[2pt]
Effect beyond bias & -0.245$^{***}$ & -0.081 & -0.248$^{***}$ & -0.188$^{***}$ & -0.152$^{**}$\\
\emph{\hspace{2mm}(Constant)} & (0.049) & (0.093) & (0.023) & (0.044) & (0.073)\\
\addlinespace[2pt]
Observations & 316 & 316 & 316 & 316 & 316\\
\midrule
\multicolumn{6}{@{}l}{\emph{Panel B: Nonlinear techniques}}\\
\addlinespace[2pt]
 & Top 10\% & WAAP & PEESE & STEM & EK\\
\midrule
Effect beyond bias & -0.083$^{***}$ & -0.106$^{***}$ & -0.318$^{***}$ & -0.300$^{***}$ & -0.106\\
 & (0.001) & (0.001) & (0.041) & (0.052) & (0.085)\\
\addlinespace[2pt]
Observations & 316 & 316 & 316 & 316 & 316\\
\midrule
\multicolumn{6}{@{}l}{\emph{Panel C: Endogeneity- and selection-robust techniques}}\\
\addlinespace[2pt]
 & & & p-uniform* & MAIVE & Selection \\
\midrule
Effect beyond bias &  &  & -0.199$^{***}$ &  & -0.226$^{***}$\\
 &  &  & (0.066) &  & (0.033)\\
\addlinespace[2pt]
First-stage \emph{F} (MAIVE) &  &  &  & $<1$ & \\
Observations &  &  & 316 & 316 & 316\\
\addlinespace[2pt]
\addlinespace[2pt]
\toprule
\multicolumn{6}{@{}l}{\textbf{Part 2: Panel FE / structural}}\\
\midrule
\multicolumn{6}{@{}l}{\emph{Panel A: Linear techniques}}\\
\addlinespace[2pt]
 & OLS & FE & RE & Study & Precision\\
\midrule
Publication bias & -0.965$^{***}$ & 0.032 & -0.885$^{***}$ & -1.589$^{***}$ & -0.683$^{**}$\\
\emph{\hspace{2mm}(Standard error)} & (0.174) & (1.165) & (0.124) & (0.146) & (0.316)\\
\addlinespace[2pt]
Effect beyond bias & -0.088$^{***}$ & -0.150$^{**}$ & -0.098$^{***}$ & -0.085$^{***}$ & -0.123$^{***}$\\
\emph{\hspace{2mm}(Constant)} & (0.026) & (0.069) & (0.014) & (0.023) & (0.041)\\
\addlinespace[2pt]
Observations & 655 & 655 & 655 & 655 & 655\\
\midrule
\multicolumn{6}{@{}l}{\emph{Panel B: Nonlinear techniques}}\\
\addlinespace[2pt]
 & Top 10\% & WAAP & PEESE & STEM & EK\\
\midrule
Effect beyond bias & -0.157$^{***}$ & -0.151$^{***}$ & -0.164$^{***}$ & -0.028 & -0.149$^{***}$\\
 & (0.001) & (0.001) & (0.020) & (0.134) & (0.057)\\
\addlinespace[2pt]
Observations & 655 & 655 & 655 & 655 & 655\\
\midrule
\multicolumn{6}{@{}l}{\emph{Panel C: Endogeneity- and selection-robust techniques}}\\
\addlinespace[2pt]
 & & & p-uniform* & MAIVE & Selection \\
\midrule
Effect beyond bias &  &  & -0.128$^{***}$ &  & -0.115$^{***}$\\
 &  &  & (0.031) &  & (0.013)\\
\addlinespace[2pt]
First-stage \emph{F} (MAIVE) &  &  &  & 1 & \\
Observations &  &  & 655 & 655 & 655\\
\end{longtable}
\end{singlespace}
\clearpage
\begin{singlespace}\footnotesize\renewcommand{\arraystretch}{0.9}\setlength{\tabcolsep}{4pt}
\begin{longtable}{@{}>{\raggedright\arraybackslash}p{\dimexpr\linewidth-2.2cm*5-\tabcolsep*10\relax}*{5}{>{\centering\arraybackslash}p{2.2cm}}@{}}
\caption{Publication-bias battery: instrumented and design-based tiers}\label{tab:pbtier2}\\
\endfirsthead
\multicolumn{6}{@{}l}{\emph{\autoref{tab:pbtier2} continued}}\\
\endhead
\midrule\multicolumn{6}{r}{\scriptsize\emph{Continued on next page}}\\
\endfoot
\bottomrule
\multicolumn{6}{@{}>{\scriptsize}p{\linewidth}@{}}{\emph{Notes:} Publication-bias--corrected mean price elasticity from each estimator, on the 1\%-winsorized headline sample. \emph{Panel A} is the FAT-PET meta-regression $\varepsilon_{is}=\varepsilon_{0}+\gamma\,SE(\varepsilon_{is})+u_{is}$: ``Publication bias'' is the slope $\gamma$ and ``Effect beyond bias'' the intercept $\varepsilon_{0}$ (each coefficient's standard error below it); the columns are OLS unweighted, FE inverse-variance weighted, RE random effects (DerSimonian--Laird), Study one median estimate per study, and Precision inverse-SE weighted. Standard errors are two-way clustered by study and by underlying database wherever a regression is estimated (OLS, FE, Precision, PEESE, and EK); RE reports the DerSimonian--Laird standard error and the study-level techniques are already one observation per study. \emph{Panel B}: PEESE \citep{stanley2014meta}, WAAP \citep{ioannidis2017power}, the top-10\% estimator, STEM \citep{furukawa2019publication}, and EK, the endogenous-kink meta-regression of \citet{bom2019kinked} whose kink point $a_1=|\varepsilon_0|/1.96$ is iterated to convergence. \emph{Panel C}: MAIVE \citep{irsova2023spurious} instruments the estimate variance with the inverse sample size (top 1\% of SEs dropped) and is reported only where the first-stage instrument is adequate ($F\ge10$; otherwise only the $F$ is shown); $p$-uniform* \citep{van2021correcting} works from study medians. ``Selection'' is a step-function selection model (a Vevea--Hedges / Andrews--Kasy-type maximum-likelihood correction \citep{vevea1995general,andrews2019identification} with a cut at the one-sided 2.5\% threshold). ``Observations'' is the number of estimates in the sample, all of which enter every test; the study-level techniques (Study, STEM, $p$-uniform*) then aggregate them to one value per study, and WAAP and the top-10\% estimator average a precision-selected subset of the same estimates. $^{***}$, $^{**}$, and $^{*}$ denote statistical significance at the 1\%, 5\%, and 10\% level.}\\
\endlastfoot
\toprule
\multicolumn{6}{@{}l}{\textbf{Part 1: Instrumented (IV/GMM)}}\\
\midrule
\multicolumn{6}{@{}l}{\emph{Panel A: Linear techniques}}\\
\addlinespace[2pt]
 & OLS & FE & RE & Study & Precision\\
\midrule
Publication bias & -1.037$^{***}$ & -3.819$^{***}$ & -1.462$^{***}$ & -1.722$^{***}$ & -1.841$^{***}$\\
\emph{\hspace{2mm}(Standard error)} & (0.241) & (0.695) & (0.109) & (0.361) & (0.375)\\
\addlinespace[2pt]
Effect beyond bias & -0.203$^{***}$ & -0.063$^{***}$ & -0.177$^{***}$ & -0.119$^{***}$ & -0.124$^{***}$\\
\emph{\hspace{2mm}(Constant)} & (0.039) & (0.023) & (0.009) & (0.039) & (0.036)\\
\addlinespace[2pt]
Observations & 549 & 549 & 549 & 549 & 549\\
\midrule
\multicolumn{6}{@{}l}{\emph{Panel B: Nonlinear techniques}}\\
\addlinespace[2pt]
 & Top 10\% & WAAP & PEESE & STEM & EK\\
\midrule
Effect beyond bias & -0.074$^{***}$ & -0.102$^{***}$ & -0.276$^{***}$ & -0.079$^{***}$ & -0.102$^{***}$\\
 & (0.001) & (0.001) & (0.030) & (0.022) & (0.028)\\
\addlinespace[2pt]
Observations & 549 & 549 & 549 & 549 & 549\\
\midrule
\multicolumn{6}{@{}l}{\emph{Panel C: Endogeneity- and selection-robust techniques}}\\
\addlinespace[2pt]
 & & & p-uniform* & MAIVE & Selection \\
\midrule
Effect beyond bias &  &  & -0.118$^{**}$ & -0.278$^{***}$ & -0.164$^{***}$\\
 &  &  & (0.056) & (0.042) & (0.020)\\
\addlinespace[2pt]
First-stage \emph{F} (MAIVE) &  &  &  & 17 & \\
Observations &  &  & 549 & 549 & 549\\
\addlinespace[2pt]
\addlinespace[2pt]
\toprule
\multicolumn{6}{@{}l}{\textbf{Part 2: Design-based (RCT/nat.\ exp./DID)}}\\
\midrule
\multicolumn{6}{@{}l}{\emph{Panel A: Linear techniques}}\\
\addlinespace[2pt]
 & OLS & FE & RE & Study & Precision\\
\midrule
Publication bias & 0.029 & 0.467 & -0.486$^{**}$ & -0.556 & -0.185\\
\emph{\hspace{2mm}(Standard error)} & (0.596) & (1.055) & (0.204) & (0.572) & (0.612)\\
\addlinespace[2pt]
Effect beyond bias & -0.095$^{*}$ & -0.084$^{**}$ & -0.046$^{***}$ & -0.117 & -0.066$^{**}$\\
\emph{\hspace{2mm}(Constant)} & (0.050) & (0.036) & (0.011) & (0.103) & (0.030)\\
\addlinespace[2pt]
Observations & 81 & 81 & 81 & 81 & 81\\
\midrule
\multicolumn{6}{@{}l}{\emph{Panel B: Nonlinear techniques}}\\
\addlinespace[2pt]
 & Top 10\% & WAAP & PEESE & STEM & EK\\
\midrule
Effect beyond bias & -0.087$^{***}$ & -0.080$^{***}$ & -0.114$^{**}$ & -0.108$^{***}$ & -0.077$^{***}$\\
 & (0.003) & (0.002) & (0.056) & (0.038) & (0.028)\\
\addlinespace[2pt]
Observations & 81 & 81 & 81 & 81 & 81\\
\midrule
\multicolumn{6}{@{}l}{\emph{Panel C: Endogeneity- and selection-robust techniques}}\\
\addlinespace[2pt]
 & & & p-uniform* & MAIVE & Selection \\
\midrule
Effect beyond bias &  &  & -0.116 & -0.115$^{***}$ & -0.066$^{***}$\\
 &  &  & (0.078) & (0.022) & (0.014)\\
\addlinespace[2pt]
First-stage \emph{F} (MAIVE) &  &  &  & 123 & \\
Observations &  &  & 81 & 81 & 81\\
\end{longtable}
\end{singlespace}
\clearpage
\begin{singlespace}\footnotesize\renewcommand{\arraystretch}{0.9}\setlength{\tabcolsep}{4pt}
\begin{longtable}{@{}>{\raggedright\arraybackslash}p{\dimexpr\linewidth-2.2cm*5-\tabcolsep*10\relax}*{5}{>{\centering\arraybackslash}p{2.2cm}}@{}}
\caption{Publication-bias battery: time-of-use and marginal-price subsamples}\label{tab:pbprice}\\
\endfirsthead
\multicolumn{6}{@{}l}{\emph{\autoref{tab:pbprice} continued}}\\
\endhead
\midrule\multicolumn{6}{r}{\scriptsize\emph{Continued on next page}}\\
\endfoot
\bottomrule
\multicolumn{6}{@{}>{\scriptsize}p{\linewidth}@{}}{\emph{Notes:} Publication-bias--corrected mean price elasticity from each estimator, on the 1\%-winsorized headline sample. \emph{Panel A} is the FAT-PET meta-regression $\varepsilon_{is}=\varepsilon_{0}+\gamma\,SE(\varepsilon_{is})+u_{is}$: ``Publication bias'' is the slope $\gamma$ and ``Effect beyond bias'' the intercept $\varepsilon_{0}$ (each coefficient's standard error below it); the columns are OLS unweighted, FE inverse-variance weighted, RE random effects (DerSimonian--Laird), Study one median estimate per study, and Precision inverse-SE weighted. Standard errors are two-way clustered by study and by underlying database wherever a regression is estimated (OLS, FE, Precision, PEESE, and EK); RE reports the DerSimonian--Laird standard error and the study-level techniques are already one observation per study. \emph{Panel B}: PEESE \citep{stanley2014meta}, WAAP \citep{ioannidis2017power}, the top-10\% estimator, STEM \citep{furukawa2019publication}, and EK, the endogenous-kink meta-regression of \citet{bom2019kinked} whose kink point $a_1=|\varepsilon_0|/1.96$ is iterated to convergence. \emph{Panel C}: MAIVE \citep{irsova2023spurious} instruments the estimate variance with the inverse sample size (top 1\% of SEs dropped) and is reported only where the first-stage instrument is adequate ($F\ge10$; otherwise only the $F$ is shown); $p$-uniform* \citep{van2021correcting} works from study medians. ``Selection'' is a step-function selection model (a Vevea--Hedges / Andrews--Kasy-type maximum-likelihood correction \citep{vevea1995general,andrews2019identification} with a cut at the one-sided 2.5\% threshold). ``Observations'' is the number of estimates in the sample, all of which enter every test; the study-level techniques (Study, STEM, $p$-uniform*) then aggregate them to one value per study, and WAAP and the top-10\% estimator average a precision-selected subset of the same estimates. $^{***}$, $^{**}$, and $^{*}$ denote statistical significance at the 1\%, 5\%, and 10\% level.}\\
\endlastfoot
\toprule
\multicolumn{6}{@{}l}{\textbf{Part 1: Time-of-use tariff}}\\
\midrule
\multicolumn{6}{@{}l}{\emph{Panel A: Linear techniques}}\\
\addlinespace[2pt]
 & OLS & FE & RE & Study & Precision\\
\midrule
Publication bias & -0.517$^{**}$ & -1.754$^{***}$ & -1.241$^{***}$ & -0.907$^{*}$ & -1.001$^{***}$\\
\emph{\hspace{2mm}(Standard error)} & (0.262) & (0.490) & (0.120) & (0.484) & (0.372)\\
\addlinespace[2pt]
Effect beyond bias & -0.091$^{**}$ & 0.001 & -0.022$^{***}$ & -0.147$^{*}$ & -0.018\\
\emph{\hspace{2mm}(Constant)} & (0.037) & (0.012) & (0.008) & (0.083) & (0.015)\\
\addlinespace[2pt]
Observations & 210 & 210 & 210 & 210 & 210\\
\midrule
\multicolumn{6}{@{}l}{\emph{Panel B: Nonlinear techniques}}\\
\addlinespace[2pt]
 & Top 10\% & WAAP & PEESE & STEM & EK\\
\midrule
Effect beyond bias & -0.005$^{***}$ & -0.011$^{***}$ & -0.154$^{***}$ & -0.051 & 0.001\\
 & (0.001) & (0.001) & (0.056) & (0.038) & (0.012)\\
\addlinespace[2pt]
Observations & 210 & 210 & 210 & 210 & 210\\
\midrule
\multicolumn{6}{@{}l}{\emph{Panel C: Endogeneity- and selection-robust techniques}}\\
\addlinespace[2pt]
 & & & p-uniform* & MAIVE & Selection \\
\midrule
Effect beyond bias &  &  & -0.228$^{***}$ & -0.090$^{***}$ & -0.075$^{***}$\\
 &  &  & (0.076) & (0.032) & (0.020)\\
\addlinespace[2pt]
First-stage \emph{F} (MAIVE) &  &  &  & 15 & \\
Observations &  &  & 210 & 210 & 210\\
\addlinespace[2pt]
\addlinespace[2pt]
\toprule
\multicolumn{6}{@{}l}{\textbf{Part 2: Marginal price}}\\
\midrule
\multicolumn{6}{@{}l}{\emph{Panel A: Linear techniques}}\\
\addlinespace[2pt]
 & OLS & FE & RE & Study & Precision\\
\midrule
Publication bias & -0.969$^{***}$ & -2.945$^{***}$ & -1.530$^{***}$ & -1.239$^{***}$ & -1.396$^{***}$\\
\emph{\hspace{2mm}(Standard error)} & (0.297) & (0.847) & (0.098) & (0.193) & (0.349)\\
\addlinespace[2pt]
Effect beyond bias & -0.148$^{***}$ & -0.027 & -0.114$^{***}$ & -0.135$^{***}$ & -0.080$^{**}$\\
\emph{\hspace{2mm}(Constant)} & (0.057) & (0.022) & (0.008) & (0.047) & (0.031)\\
\addlinespace[2pt]
Observations & 401 & 401 & 401 & 401 & 401\\
\midrule
\multicolumn{6}{@{}l}{\emph{Panel B: Nonlinear techniques}}\\
\addlinespace[2pt]
 & Top 10\% & WAAP & PEESE & STEM & EK\\
\midrule
Effect beyond bias & -0.041$^{***}$ & -0.049$^{***}$ & -0.230$^{***}$ & -0.076$^{*}$ & -0.049$^{**}$\\
 & (0.001) & (0.001) & (0.047) & (0.039) & (0.025)\\
\addlinespace[2pt]
Observations & 401 & 401 & 401 & 401 & 401\\
\midrule
\multicolumn{6}{@{}l}{\emph{Panel C: Endogeneity- and selection-robust techniques}}\\
\addlinespace[2pt]
 & & & p-uniform* & MAIVE & Selection \\
\midrule
Effect beyond bias &  &  & -0.124$^{**}$ &  & -0.155$^{***}$\\
 &  &  & (0.061) &  & (0.022)\\
\addlinespace[2pt]
First-stage \emph{F} (MAIVE) &  &  &  & 7 & \\
Observations &  &  & 401 & 401 & 401\\
\end{longtable}
\end{singlespace}
\clearpage

\FloatBarrier
\section{Heterogeneity: Bayesian model averaging}\label{app:bma}

\begin{figure}[p]\begin{center}
\caption{Model inclusion in Bayesian model averaging}\label{fig:bmaincl}
\begin{subfigure}{0.49\textwidth}\centering\includegraphics[width=\textwidth]{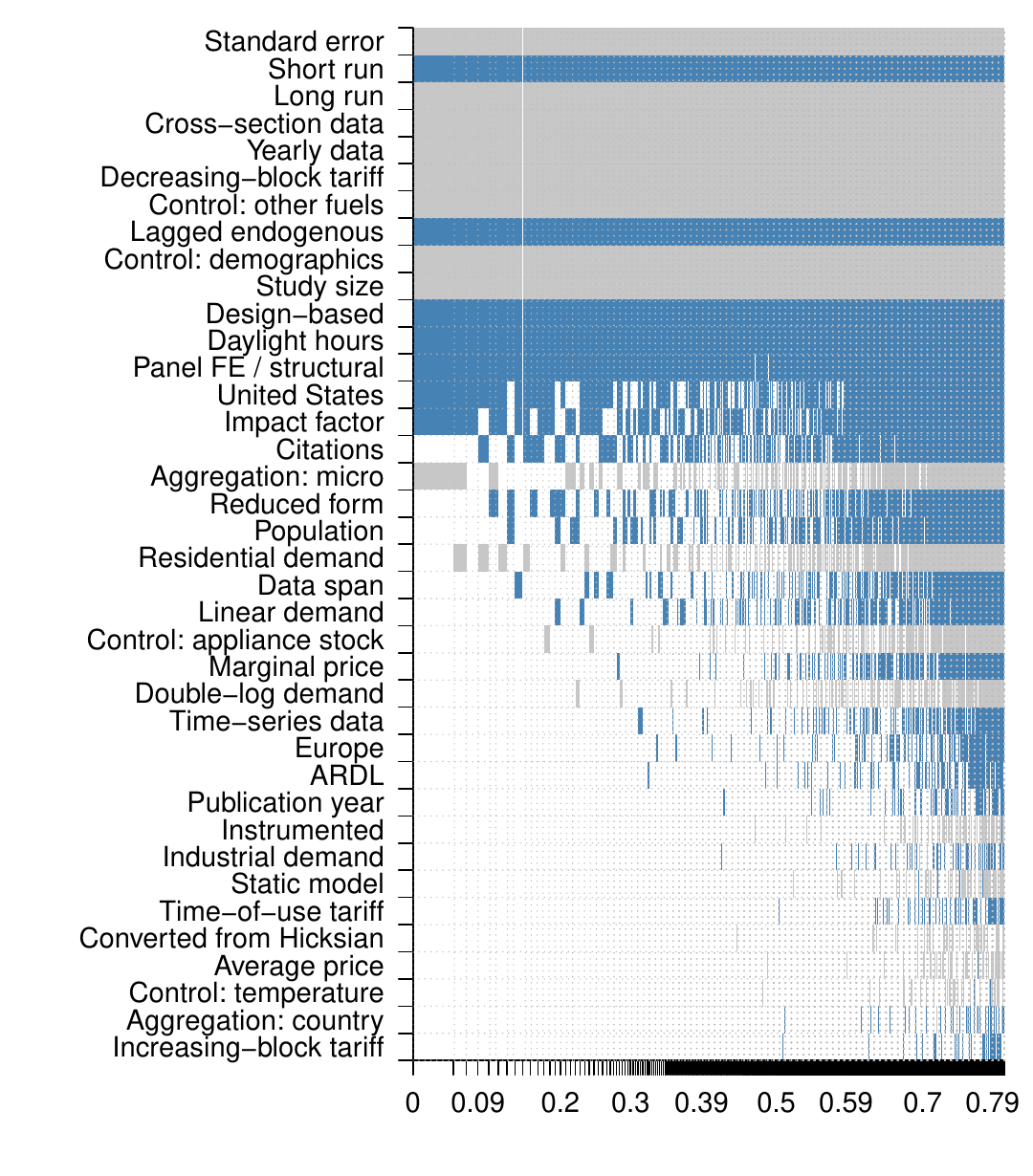}\caption{Full headline sample}\end{subfigure}\hfill
\begin{subfigure}{0.49\textwidth}\centering\includegraphics[width=\textwidth]{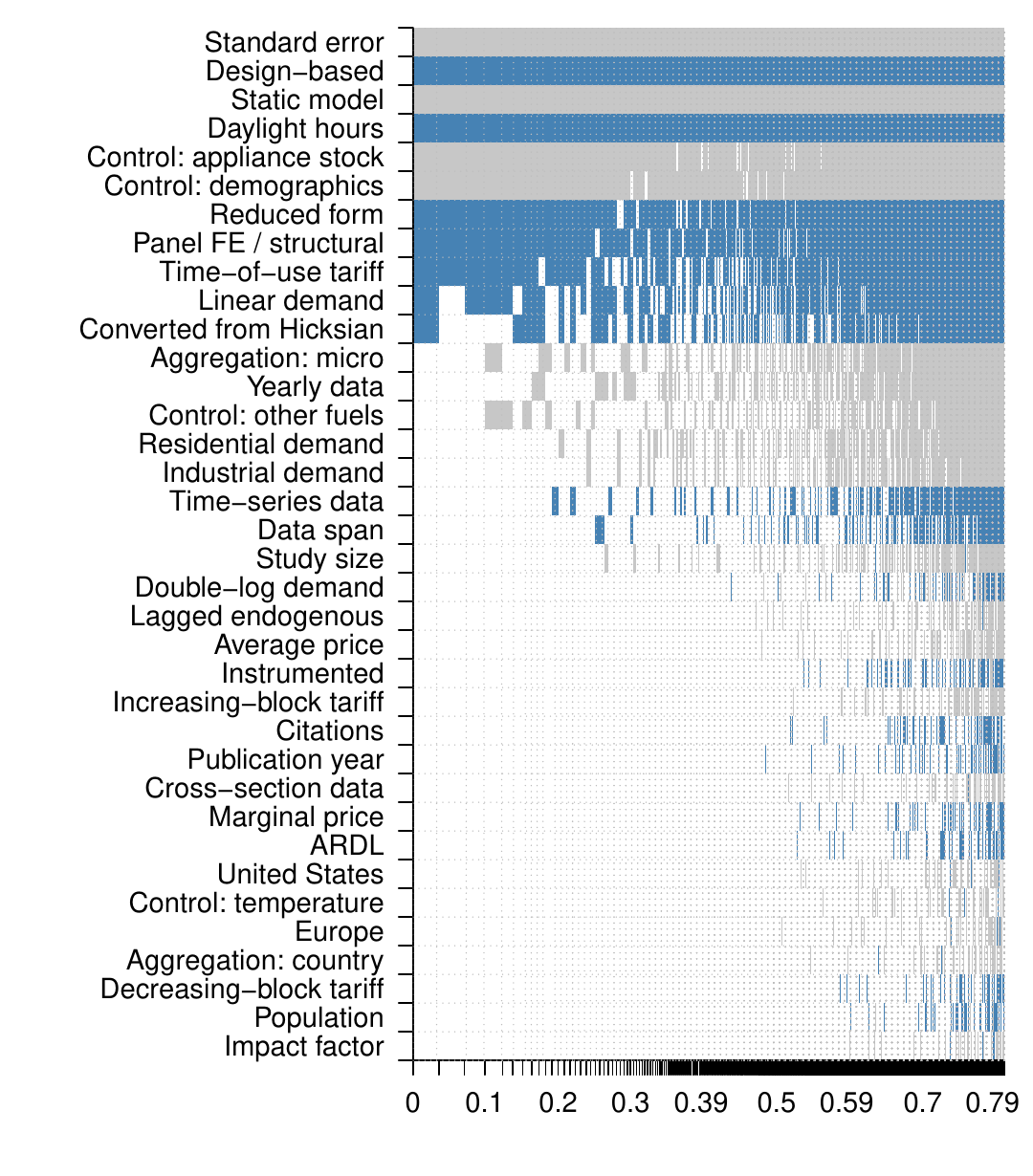}\caption{Short-run estimates}\end{subfigure}\\[0.2cm]
\begin{subfigure}{0.49\textwidth}\centering\includegraphics[width=\textwidth]{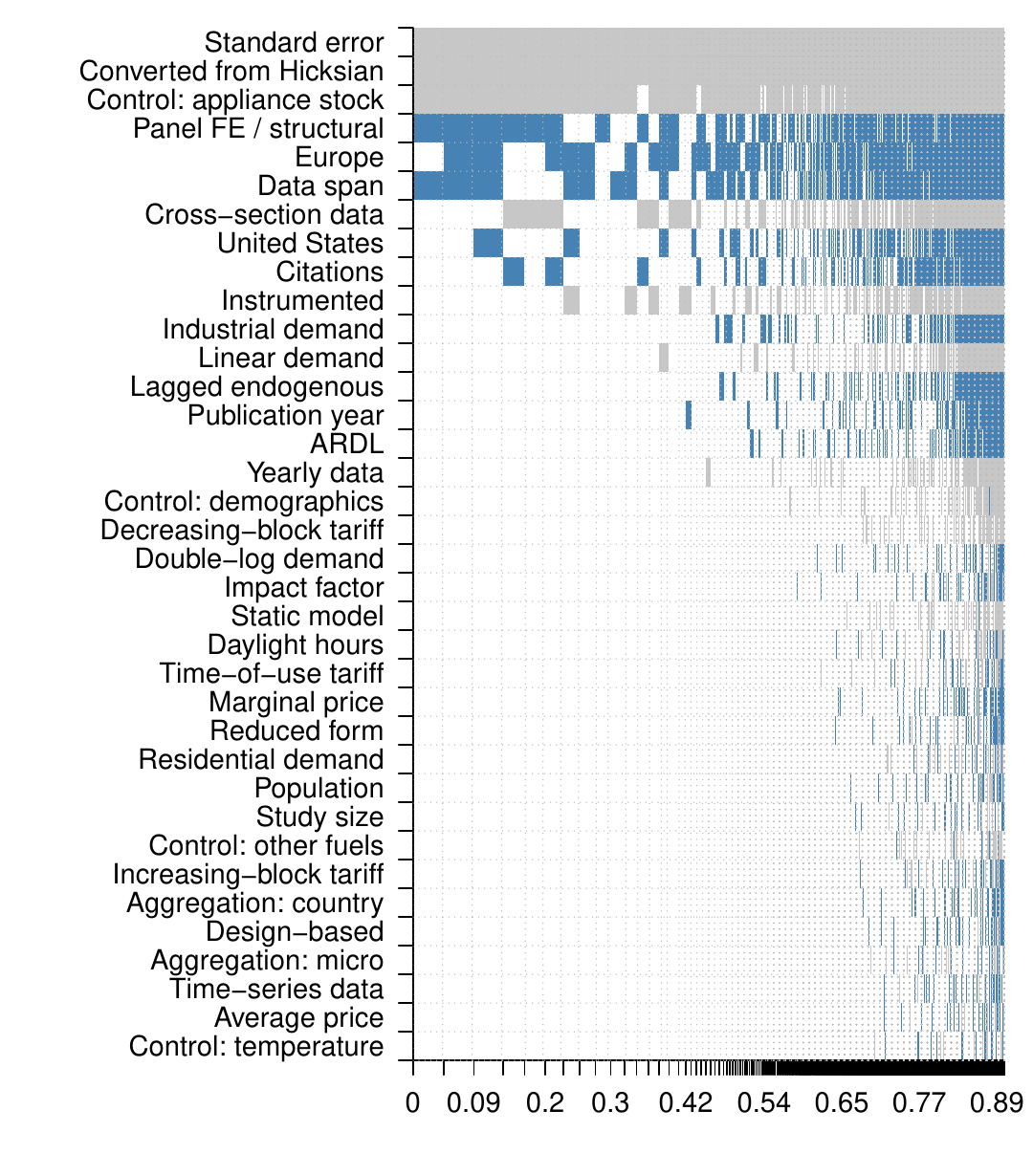}\caption{Long-run estimates}\end{subfigure}
\par\smallskip
\begin{minipage}{\textwidth}\footnotesize\textit{Notes:} Each column is a regression model; column width is proportional to the model's posterior probability, cumulated along the horizontal axis. Rows are regressors, sorted by posterior inclusion probability. Blue means the variable enters with a positive coefficient, grey negative, white excluded. The estimation uses the unit-information $g$-prior and the dilution model prior.\end{minipage}
\end{center}\end{figure}

\begin{figure}[p]\begin{center}
\caption{Regressor correlation matrix (collinearity diagnostic)}\label{fig:bmacorr}
\includegraphics[width=0.9\textwidth]{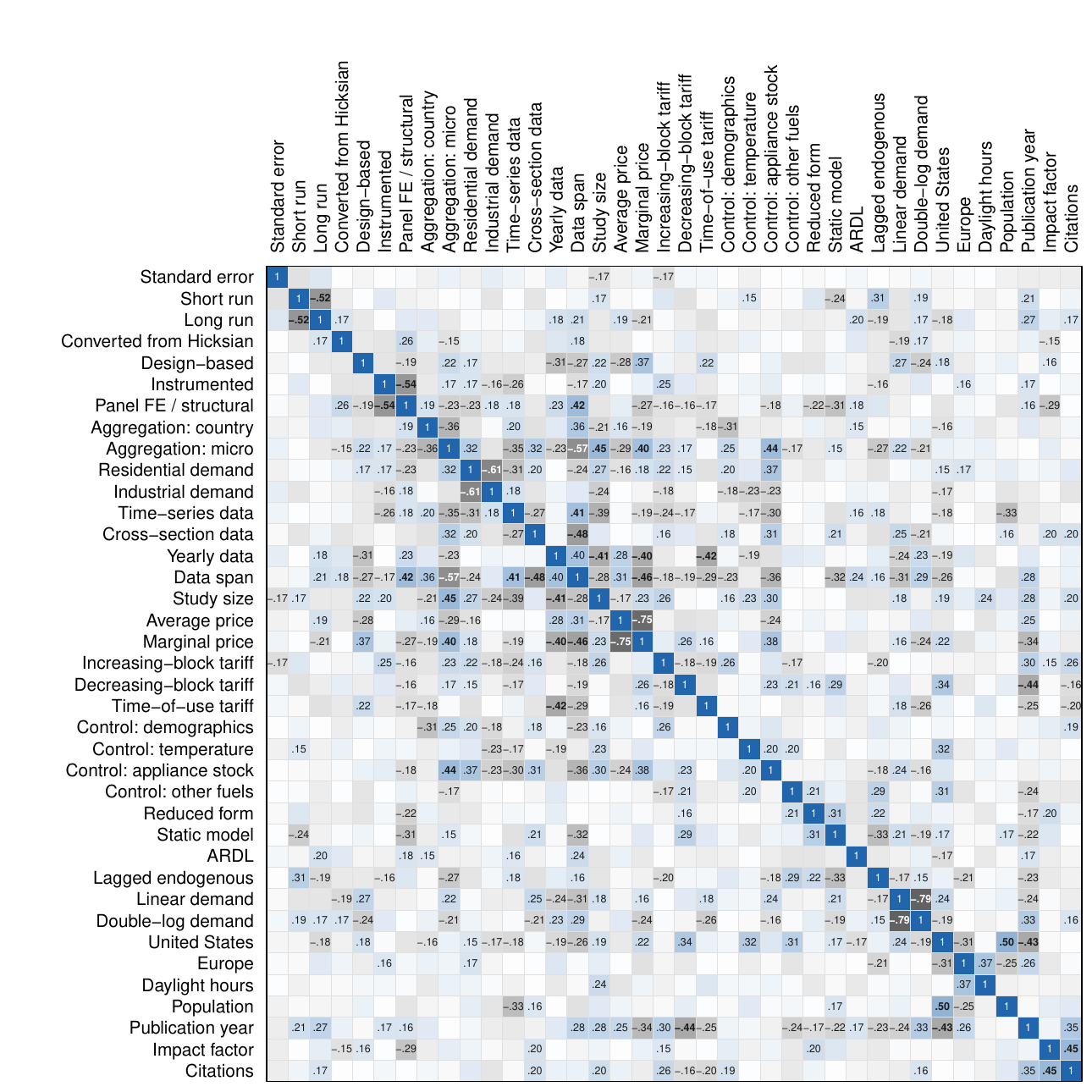}
\par\smallskip
\begin{minipage}{0.9\textwidth}\footnotesize\textit{Notes:} Pairwise correlations among the integrated-model regressors; cells with $|r|<0.15$ are left blank and $|r|\ge0.4$ are bold. Blue $=$ positive, grey $=$ negative. The visible clusters (among the identification, model-form and data-structure indicators) are why we average under the collinearity-penalizing dilution prior rather than reporting a single regression.\end{minipage}
\end{center}\end{figure}

The identification ladder and the two-clock test isolate the two dimensions the paper cares about most; the literature varies along many others, and this appendix maps that variation with Bayesian model averaging (BMA), computed with the \texttt{BMS} package \citep{zeugner2015bayesian}.\footnote{The replication package (data and code) for the full paper, including all appendices, is available at \url{https://meta-analysis.cz/electricity}.} The exercise is descriptive and deliberately secondary to the trend and ladder findings: it asks which of the coded study characteristics move the reported elasticity once all of them compete for inclusion, and it doubles as a robustness check on the publication-bias term, which here must survive against every moderator at once. Because the paper's spine is the adjustment horizon, we run the averaging on three samples (the full headline sample and, separately, its short-run and long-run subsamples) with variable definitions and summary statistics in \autoref{tab:vardefs} (full sample), \autoref{tab:vardefssr} (short run) and \autoref{tab:vardefslr} (long run). \autoref{tab:bma} reports the integrated averaging beside a two-way (study $\times$ database) clustered OLS check and a frequentist model average; \autoref{tab:bmasr} and \autoref{tab:bmalr} report the horizon-specific models; \autoref{tab:bmarobust} the sensitivity to the priors and \autoref{tab:bmadiag} the convergence diagnostics; \autoref{fig:bmaincl} shows the model-inclusion picture for the three samples and \autoref{fig:bmacorr} the regressor correlation matrix.

The response variable is the 1\%-winsorized price elasticity and the publication-bias term is its 1\%-winsorized standard error, as throughout the paper. The moderators, their reference categories, and the log transforms follow the variable list that ships with the final dataset. Because the identification, model-form, data-structure and price-measurement indicators are strongly collinear (a naive static reduced-form model on yearly average prices is one bundle, a dynamic instrumented panel another), we average under the \emph{dilution} prior of \citet{george2010dilution}, which down-weights models whose regressor correlation matrix is near-singular and so prices in the redundancy the single-regression approach ignores; the scale-laden moderators (study size, data span, publication year, impact factor, citations) enter in logs. Two-way clustering by study and by underlying database sharpens the frequentist check: many estimates share a data source across studies, and clustering on the database as well as the study nets out that second layer of dependence. The two-way correction changes little, because shared datasets are rare in this corpus: the 366 headline studies draw on 344 distinct underlying data sources, only 17 databases are shared by two or more studies and none by more than four, so study-level clustering already absorbs almost all of the cross-estimate dependence. Re-running the headline inferences with two-way (study $\times$ database) clustering leaves them materially unchanged: the short-run PET is $-0.16$ with study-clustered and two-way-clustered standard errors identical to three decimals ($0.025$), the design-versus-naive ladder contrast moves from $p=0.006$ to $p=0.007$, and the composition-adjusted data-clock trend stays null ($p=0.53$). The lone coefficient in which the second clustering dimension is visible is the pooled design-based term of the integrated model (\autoref{tab:bma}), borderline at $p=0.08$, a reflection of its larger moderator set and pooled horizons rather than of the clustering. Nine country-level covariate values (daylight and population) are median-imputed; the elasticity and its standard error are never imputed.

\clearpage
\begin{singlespace}\footnotesize
\begin{longtable}{@{}lp{8.9cm}rr@{}}
\caption{Definition and summary statistics of the model-averaging variables: full sample}\label{tab:vardefs}\\
\toprule
Variable & Description & Mean & SD\\
\midrule\endfirsthead
\multicolumn{4}{@{}l}{\emph{\autoref{tab:vardefs} continued}}\\
\toprule
Variable & Description & Mean & SD\\
\midrule\endhead
\bottomrule\multicolumn{4}{r}{\scriptsize\emph{Continued on next page}}\\\endfoot
\bottomrule
\multicolumn{4}{@{}>{\scriptsize}p{\linewidth}@{}}{\emph{Notes:} Mean and standard deviation over all 3{,}324 headline estimates (Marshallian-equivalent: directly reported or Slutsky-converted, usable standard error) from 366 studies. Binary indicators are reported as means (shares); the elasticity and its standard error are the 1\%-winsorized values used throughout. Reference categories are shown for completeness but drop from the averaging. The identification reference category (Naive) totals 821 estimates, of which 730 are labeled naive (tier 6) and 91 could not be classified; the averaging does not distinguish between the two. Scale-laden variables enter the averaging in logs, as noted in their descriptions. Country-level covariates are drawn from external sources: GDP per capita (current US\$) and population from the World Bank's \emph{World Development Indicators} \citep{worldbank2025wdi}; surface temperature (degrees Celsius) from the Berkeley Earth record \citep{berkeleyearth2015}; and the length of the longest day from the daylight calculator of \citet{moesen2010daylight}. Study-level bibliometrics are citation counts from Google Scholar and journal impact factors from the RePEc recursive ranking. The replication package's codebook maps every coded variable to its source.}\\
\endlastfoot
Price elasticity & The own-price elasticity of electricity demand (1\%-winsorized), the response variable. & -0.39 & 0.44 \\
Standard error & The standard error of the elasticity estimate (1\%-winsorized); its coefficient measures publication selection. & 0.15 & 0.20 \\
\midrule
\multicolumn{4}{@{}l}{\emph{Horizon}}\\
\hspace{0.4cm}Short run & =1 if the estimate is a short-run (within about one year) elasticity. & 0.50 & 0.50 \\
\hspace{0.4cm}Long run & =1 if the estimate is a long-run (beyond about five years; capital fully adjusted) elasticity. & 0.22 & 0.41 \\
\hspace{0.4cm}Intermediate run & =1 if the estimate is an intermediate-run (about 2--5 years) elasticity (reference category). & 0.29 & 0.45 \\
\midrule
\multicolumn{4}{@{}l}{\emph{Estimate type}}\\
\hspace{0.4cm}Converted from Hicksian & =1 if the estimate is a compensated (Hicksian) own-price elasticity converted to a Marshallian footing via the Slutsky relation $\varepsilon_M=\varepsilon_H-s\eta$; =0 for a directly reported Marshallian estimate. Controls for any residual difference between the converted and the directly reported evidence. & 0.22 & 0.42 \\
\midrule
\multicolumn{4}{@{}l}{\emph{Identification}}\\
\hspace{0.4cm}Design-based & =1 if a design-based strategy identifies the price response: a randomized or mandated pricing experiment, a natural experiment, or a difference-in-differences design (tiers 1--3). & 0.05 & 0.21 \\
\hspace{0.4cm}Instrumented & =1 if the price is instrumented (IV, two- or three-stage least squares, or GMM; tier 4). & 0.27 & 0.45 \\
\hspace{0.4cm}Panel FE / structural & =1 if identification rests on panel fixed effects or a structural demand system, without instrumentation (tier 5). & 0.43 & 0.50 \\
\hspace{0.4cm}Naive & =1 if a naive regression with no price-identification strategy is used (tier 6; the reference category, which also absorbs estimates whose strategy could not be classified). & 0.22 & 0.41 \\
\midrule
\multicolumn{4}{@{}l}{\emph{Estimate aggregation}}\\
\hspace{0.4cm}Aggregation: country & =1 if the data are aggregated to the country (or larger) level. & 0.26 & 0.44 \\
\hspace{0.4cm}Aggregation: micro & =1 if the data are micro (household, plant or firm-level). & 0.27 & 0.44 \\
\hspace{0.4cm}Aggregation: region & =1 if the data are aggregated to a sub-country region or city (reference category). & 0.30 & 0.46 \\
\midrule
\multicolumn{4}{@{}l}{\emph{Sector}}\\
\hspace{0.4cm}Residential demand & =1 if the estimate refers to residential electricity demand. & 0.62 & 0.49 \\
\hspace{0.4cm}Industrial demand & =1 if the estimate refers to industrial electricity demand. & 0.19 & 0.39 \\
\hspace{0.4cm}Commercial demand & =1 if the estimate refers to commercial or mixed demand (reference category). & 0.09 & 0.28 \\
\midrule
\multicolumn{4}{@{}l}{\emph{Data characteristics}}\\
\hspace{0.4cm}Time-series data & =1 if the study uses time-series data. & 0.34 & 0.47 \\
\hspace{0.4cm}Cross-section data & =1 if the study uses cross-sectional data. & 0.12 & 0.33 \\
\hspace{0.4cm}Panel data & =1 if the study uses panel data (reference category). & 0.54 & 0.50 \\
\hspace{0.4cm}Yearly data & =1 if the data are of yearly frequency. & 0.66 & 0.47 \\
\hspace{0.4cm}Sub-yearly data & =1 if the data are of monthly or quarterly frequency (reference category). & 0.28 & 0.45 \\
\hspace{0.4cm}Data span & The logarithm of one plus the number of years the dataset covers. & 2.35 & 1.00 \\
\hspace{0.4cm}Study size & The logarithm of the number of observations used in the estimation. & 6.13 & 2.54 \\
\midrule
\multicolumn{4}{@{}l}{\emph{Price and tariff}}\\
\hspace{0.4cm}Average price & =1 if the price is measured as average price (revenue over quantity). & 0.64 & 0.48 \\
\hspace{0.4cm}Marginal price & =1 if the price is measured as marginal price. & 0.24 & 0.43 \\
\hspace{0.4cm}Other price & =1 if another price concept is used (lump-sum, Shin, undefined; reference category). & 0.08 & 0.27 \\
\hspace{0.4cm}Increasing-block tariff & =1 if an increasing-block tariff is in force. & 0.19 & 0.39 \\
\hspace{0.4cm}Decreasing-block tariff & =1 if a decreasing-block tariff is in force. & 0.12 & 0.32 \\
\hspace{0.4cm}Time-of-use tariff & =1 if a time-of-use rate (different unit prices by time block) is used. & 0.14 & 0.34 \\
\hspace{0.4cm}Flat tariff & =1 if a flat tariff is in force (reference category). & 0.04 & 0.20 \\
\midrule
\multicolumn{4}{@{}l}{\emph{Demand controls}}\\
\hspace{0.4cm}Control: demographics & =1 if the demand equation controls for demographic variation. & 0.36 & 0.48 \\
\hspace{0.4cm}Control: temperature & =1 if the demand equation controls for temperature or degree-days. & 0.50 & 0.50 \\
\hspace{0.4cm}Control: appliance stock & =1 if the demand equation controls for the stock of electrical appliances. & 0.22 & 0.41 \\
\hspace{0.4cm}Control: other fuels & =1 if the demand equation controls for prices of substitute fuels. & 0.41 & 0.49 \\
\midrule
\multicolumn{4}{@{}l}{\emph{Model specification}}\\
\hspace{0.4cm}Reduced form & =1 if a single-equation reduced-form model is estimated. & 0.48 & 0.50 \\
\hspace{0.4cm}Structural form & =1 if a structural (multi-equation) model is estimated (reference category). & 0.50 & 0.50 \\
\hspace{0.4cm}Static model & =1 if the model is static (no lagged demand). & 0.32 & 0.47 \\
\hspace{0.4cm}Dynamic model & =1 if the model is dynamic, with lagged adjustment (reference category). & 0.68 & 0.47 \\
\hspace{0.4cm}ARDL & =1 if an autoregressive distributed-lag (partial-adjustment) model is used. & 0.04 & 0.20 \\
\hspace{0.4cm}Lagged endogenous & =1 if a lagged-endogenous model is used. & 0.19 & 0.40 \\
\hspace{0.4cm}Linear demand & =1 if the demand function is linear. & 0.13 & 0.34 \\
\hspace{0.4cm}Double-log demand & =1 if the demand function is double-logarithmic. & 0.80 & 0.40 \\
\hspace{0.4cm}Semi-log demand & =1 if the demand function is semi-logarithmic or Box-Cox (reference category). & 0.01 & 0.11 \\
\midrule
\multicolumn{4}{@{}l}{\emph{Geography and context}}\\
\hspace{0.4cm}United States & =1 if the estimate is for the United States. & 0.41 & 0.49 \\
\hspace{0.4cm}Europe & =1 if the estimate is for a European country. & 0.18 & 0.38 \\
\hspace{0.4cm}Other country & =1 if the estimate is for a country outside the US and Europe (reference category). & 0.42 & 0.49 \\
\hspace{0.4cm}Daylight hours & The average length of the longest day for the country under examination (source: \citealp{moesen2010daylight}). & 14.76 & 1.58 \\
\hspace{0.4cm}Population & The logarithm of the population of the country or entity (source: \citealp{worldbank2025wdi}). & 18.36 & 1.58 \\
\midrule
\multicolumn{4}{@{}l}{\emph{Publication characteristics}}\\
\hspace{0.4cm}Publication year & The logarithm of one plus the number of years since the earliest study in the sample. & 3.91 & 0.35 \\
\hspace{0.4cm}Impact factor & The logarithm of one plus the journal's RePEc recursive impact factor; 0 for unpublished work. & 0.26 & 0.55 \\
\hspace{0.4cm}Citations & The logarithm of one plus the study's mean annual Google Scholar citation count. & 0.87 & 0.67 \\
\end{longtable}
\end{singlespace}

\begin{singlespace}\footnotesize
\begin{longtable}{@{}lp{8.9cm}rr@{}}
\caption{Definition and summary statistics of the model-averaging variables: short-run estimates}\label{tab:vardefssr}\\
\toprule
Variable & Description & Mean & SD\\
\midrule\endfirsthead
\multicolumn{4}{@{}l}{\emph{\autoref{tab:vardefssr} continued}}\\
\toprule
Variable & Description & Mean & SD\\
\midrule\endhead
\bottomrule\multicolumn{4}{r}{\scriptsize\emph{Continued on next page}}\\\endfoot
\bottomrule
\multicolumn{4}{@{}>{\scriptsize}p{\linewidth}@{}}{\emph{Notes:} Mean and standard deviation over the 1{,}647 short-run headline estimates. The horizon indicators are constant on this subsample and drop from the short-run averaging. Binary indicators are reported as means (shares); the elasticity and its standard error are the 1\%-winsorized values.}\\
\endlastfoot
Price elasticity & The own-price elasticity of electricity demand (1\%-winsorized), the response variable. & -0.27 & 0.33 \\
Standard error & The standard error of the elasticity estimate (1\%-winsorized); its coefficient measures publication selection. & 0.12 & 0.17 \\
\midrule
\multicolumn{4}{@{}l}{\emph{Horizon}}\\
\hspace{0.4cm}Short run & =1 if the estimate is a short-run (within about one year) elasticity. & 1.00 & 0.00 \\
\hspace{0.4cm}Long run & =1 if the estimate is a long-run (beyond about five years; capital fully adjusted) elasticity. & 0.00 & 0.00 \\
\hspace{0.4cm}Intermediate run & =1 if the estimate is an intermediate-run (about 2--5 years) elasticity (reference category). & 0.00 & 0.00 \\
\midrule
\multicolumn{4}{@{}l}{\emph{Estimate type}}\\
\hspace{0.4cm}Converted from Hicksian & =1 if the estimate is a compensated (Hicksian) own-price elasticity converted to a Marshallian footing via the Slutsky relation $\varepsilon_M=\varepsilon_H-s\eta$; =0 for a directly reported Marshallian estimate. Controls for any residual difference between the converted and the directly reported evidence. & 0.18 & 0.39 \\
\midrule
\multicolumn{4}{@{}l}{\emph{Identification}}\\
\hspace{0.4cm}Design-based & =1 if a design-based strategy identifies the price response: a randomized or mandated pricing experiment, a natural experiment, or a difference-in-differences design (tiers 1--3). & 0.05 & 0.22 \\
\hspace{0.4cm}Instrumented & =1 if the price is instrumented (IV, two- or three-stage least squares, or GMM; tier 4). & 0.33 & 0.47 \\
\hspace{0.4cm}Panel FE / structural & =1 if identification rests on panel fixed effects or a structural demand system, without instrumentation (tier 5). & 0.40 & 0.49 \\
\hspace{0.4cm}Naive & =1 if a naive regression with no price-identification strategy is used (tier 6; the reference category, which also absorbs estimates whose strategy could not be classified). & 0.19 & 0.39 \\
\midrule
\multicolumn{4}{@{}l}{\emph{Estimate aggregation}}\\
\hspace{0.4cm}Aggregation: country & =1 if the data are aggregated to the country (or larger) level. & 0.24 & 0.43 \\
\hspace{0.4cm}Aggregation: micro & =1 if the data are micro (household, plant or firm-level). & 0.26 & 0.44 \\
\hspace{0.4cm}Aggregation: region & =1 if the data are aggregated to a sub-country region or city (reference category). & 0.34 & 0.47 \\
\midrule
\multicolumn{4}{@{}l}{\emph{Sector}}\\
\hspace{0.4cm}Residential demand & =1 if the estimate refers to residential electricity demand. & 0.64 & 0.48 \\
\hspace{0.4cm}Industrial demand & =1 if the estimate refers to industrial electricity demand. & 0.13 & 0.34 \\
\hspace{0.4cm}Commercial demand & =1 if the estimate refers to commercial or mixed demand (reference category). & 0.10 & 0.30 \\
\midrule
\multicolumn{4}{@{}l}{\emph{Data characteristics}}\\
\hspace{0.4cm}Time-series data & =1 if the study uses time-series data. & 0.34 & 0.47 \\
\hspace{0.4cm}Cross-section data & =1 if the study uses cross-sectional data. & 0.08 & 0.27 \\
\hspace{0.4cm}Panel data & =1 if the study uses panel data (reference category). & 0.59 & 0.49 \\
\hspace{0.4cm}Yearly data & =1 if the data are of yearly frequency. & 0.65 & 0.48 \\
\hspace{0.4cm}Sub-yearly data & =1 if the data are of monthly or quarterly frequency (reference category). & 0.27 & 0.44 \\
\hspace{0.4cm}Data span & The logarithm of one plus the number of years the dataset covers. & 2.40 & 0.98 \\
\hspace{0.4cm}Study size & The logarithm of the number of observations used in the estimation. & 6.55 & 2.56 \\
\pagebreak
\multicolumn{4}{@{}l}{\emph{Price and tariff}}\\
\hspace{0.4cm}Average price & =1 if the price is measured as average price (revenue over quantity). & 0.63 & 0.48 \\
\hspace{0.4cm}Marginal price & =1 if the price is measured as marginal price. & 0.24 & 0.43 \\
\hspace{0.4cm}Other price & =1 if another price concept is used (lump-sum, Shin, undefined; reference category). & 0.07 & 0.25 \\
\hspace{0.4cm}Increasing-block tariff & =1 if an increasing-block tariff is in force. & 0.25 & 0.43 \\
\hspace{0.4cm}Decreasing-block tariff & =1 if a decreasing-block tariff is in force. & 0.07 & 0.26 \\
\hspace{0.4cm}Time-of-use tariff & =1 if a time-of-use rate (different unit prices by time block) is used. & 0.13 & 0.33 \\
\hspace{0.4cm}Flat tariff & =1 if a flat tariff is in force (reference category). & 0.04 & 0.20 \\
\midrule
\multicolumn{4}{@{}l}{\emph{Demand controls}}\\
\hspace{0.4cm}Control: demographics & =1 if the demand equation controls for demographic variation. & 0.39 & 0.49 \\
\hspace{0.4cm}Control: temperature & =1 if the demand equation controls for temperature or degree-days. & 0.58 & 0.49 \\
\hspace{0.4cm}Control: appliance stock & =1 if the demand equation controls for the stock of electrical appliances. & 0.21 & 0.40 \\
\hspace{0.4cm}Control: other fuels & =1 if the demand equation controls for prices of substitute fuels. & 0.42 & 0.49 \\
\midrule
\multicolumn{4}{@{}l}{\emph{Model specification}}\\
\hspace{0.4cm}Reduced form & =1 if a single-equation reduced-form model is estimated. & 0.55 & 0.50 \\
\hspace{0.4cm}Structural form & =1 if a structural (multi-equation) model is estimated (reference category). & 0.43 & 0.50 \\
\hspace{0.4cm}Static model & =1 if the model is static (no lagged demand). & 0.21 & 0.40 \\
\hspace{0.4cm}Dynamic model & =1 if the model is dynamic, with lagged adjustment (reference category). & 0.79 & 0.41 \\
\hspace{0.4cm}ARDL & =1 if an autoregressive distributed-lag (partial-adjustment) model is used. & 0.03 & 0.18 \\
\hspace{0.4cm}Lagged endogenous & =1 if a lagged-endogenous model is used. & 0.32 & 0.47 \\
\hspace{0.4cm}Linear demand & =1 if the demand function is linear. & 0.09 & 0.29 \\
\hspace{0.4cm}Double-log demand & =1 if the demand function is double-logarithmic. & 0.88 & 0.33 \\
\hspace{0.4cm}Semi-log demand & =1 if the demand function is semi-logarithmic or Box-Cox (reference category). & 0.01 & 0.10 \\
\midrule
\multicolumn{4}{@{}l}{\emph{Geography and context}}\\
\hspace{0.4cm}United States & =1 if the estimate is for the United States. & 0.38 & 0.49 \\
\hspace{0.4cm}Europe & =1 if the estimate is for a European country. & 0.17 & 0.37 \\
\hspace{0.4cm}Other country & =1 if the estimate is for a country outside the US and Europe (reference category). & 0.47 & 0.50 \\
\hspace{0.4cm}Daylight hours & The average length of the longest day for the country under examination (source: \citealp{moesen2010daylight}). & 14.66 & 1.64 \\
\hspace{0.4cm}Population & The logarithm of the population of the country or entity (source: \citealp{worldbank2025wdi}). & 18.32 & 1.56 \\
\midrule
\multicolumn{4}{@{}l}{\emph{Publication characteristics}}\\
\hspace{0.4cm}Publication year & The logarithm of one plus the number of years since the earliest study in the sample. & 3.99 & 0.29 \\
\hspace{0.4cm}Impact factor & The logarithm of one plus the journal's RePEc recursive impact factor; 0 for unpublished work. & 0.26 & 0.56 \\
\hspace{0.4cm}Citations & The logarithm of one plus the study's mean annual Google Scholar citation count. & 0.91 & 0.59 \\
\end{longtable}
\end{singlespace}

\begin{singlespace}\footnotesize
\begin{longtable}{@{}lp{8.9cm}rr@{}}
\caption{Definition and summary statistics of the model-averaging variables: long-run estimates}\label{tab:vardefslr}\\
\toprule
Variable & Description & Mean & SD\\
\midrule\endfirsthead
\multicolumn{4}{@{}l}{\emph{\autoref{tab:vardefslr} continued}}\\
\toprule
Variable & Description & Mean & SD\\
\midrule\endhead
\bottomrule\multicolumn{4}{r}{\scriptsize\emph{Continued on next page}}\\\endfoot
\bottomrule
\multicolumn{4}{@{}>{\scriptsize}p{\linewidth}@{}}{\emph{Notes:} Mean and standard deviation over the 723 long-run headline estimates. The horizon indicators are constant on this subsample and drop from the long-run averaging. Binary indicators are reported as means (shares); the elasticity and its standard error are the 1\%-winsorized values.}\\
\endlastfoot
Price elasticity & The own-price elasticity of electricity demand (1\%-winsorized), the response variable. & -0.58 & 0.52 \\
Standard error & The standard error of the elasticity estimate (1\%-winsorized); its coefficient measures publication selection. & 0.21 & 0.26 \\
\midrule
\multicolumn{4}{@{}l}{\emph{Horizon}}\\
\hspace{0.4cm}Short run & =1 if the estimate is a short-run (within about one year) elasticity. & 0.00 & 0.00 \\
\hspace{0.4cm}Long run & =1 if the estimate is a long-run (beyond about five years; capital fully adjusted) elasticity. & 1.00 & 0.00 \\
\hspace{0.4cm}Intermediate run & =1 if the estimate is an intermediate-run (about 2--5 years) elasticity (reference category). & 0.00 & 0.00 \\
\midrule
\multicolumn{4}{@{}l}{\emph{Estimate type}}\\
\hspace{0.4cm}Converted from Hicksian & =1 if the estimate is a compensated (Hicksian) own-price elasticity converted to a Marshallian footing via the Slutsky relation $\varepsilon_M=\varepsilon_H-s\eta$; =0 for a directly reported Marshallian estimate. Controls for any residual difference between the converted and the directly reported evidence. & 0.36 & 0.48 \\
\midrule
\multicolumn{4}{@{}l}{\emph{Identification}}\\
\hspace{0.4cm}Design-based & =1 if a design-based strategy identifies the price response: a randomized or mandated pricing experiment, a natural experiment, or a difference-in-differences design (tiers 1--3). & 0.01 & 0.07 \\
\hspace{0.4cm}Instrumented & =1 if the price is instrumented (IV, two- or three-stage least squares, or GMM; tier 4). & 0.26 & 0.44 \\
\hspace{0.4cm}Panel FE / structural & =1 if identification rests on panel fixed effects or a structural demand system, without instrumentation (tier 5). & 0.55 & 0.50 \\
\hspace{0.4cm}Naive & =1 if a naive regression with no price-identification strategy is used (tier 6; the reference category, which also absorbs estimates whose strategy could not be classified). & 0.17 & 0.37 \\
\midrule
\multicolumn{4}{@{}l}{\emph{Estimate aggregation}}\\
\hspace{0.4cm}Aggregation: country & =1 if the data are aggregated to the country (or larger) level. & 0.37 & 0.48 \\
\hspace{0.4cm}Aggregation: micro & =1 if the data are micro (household, plant or firm-level). & 0.18 & 0.38 \\
\hspace{0.4cm}Aggregation: region & =1 if the data are aggregated to a sub-country region or city (reference category). & 0.38 & 0.49 \\
\midrule
\multicolumn{4}{@{}l}{\emph{Sector}}\\
\hspace{0.4cm}Residential demand & =1 if the estimate refers to residential electricity demand. & 0.64 & 0.48 \\
\hspace{0.4cm}Industrial demand & =1 if the estimate refers to industrial electricity demand. & 0.16 & 0.36 \\
\hspace{0.4cm}Commercial demand & =1 if the estimate refers to commercial or mixed demand (reference category). & 0.10 & 0.30 \\
\midrule
\multicolumn{4}{@{}l}{\emph{Data characteristics}}\\
\hspace{0.4cm}Time-series data & =1 if the study uses time-series data. & 0.36 & 0.48 \\
\hspace{0.4cm}Cross-section data & =1 if the study uses cross-sectional data. & 0.17 & 0.37 \\
\hspace{0.4cm}Panel data & =1 if the study uses panel data (reference category). & 0.47 & 0.50 \\
\hspace{0.4cm}Yearly data & =1 if the data are of yearly frequency. & 0.82 & 0.39 \\
\hspace{0.4cm}Sub-yearly data & =1 if the data are of monthly or quarterly frequency (reference category). & 0.12 & 0.32 \\
\hspace{0.4cm}Data span & The logarithm of one plus the number of years the dataset covers. & 2.74 & 0.95 \\
\hspace{0.4cm}Study size & The logarithm of the number of observations used in the estimation. & 5.62 & 2.10 \\
\pagebreak
\multicolumn{4}{@{}l}{\emph{Price and tariff}}\\
\hspace{0.4cm}Average price & =1 if the price is measured as average price (revenue over quantity). & 0.81 & 0.39 \\
\hspace{0.4cm}Marginal price & =1 if the price is measured as marginal price. & 0.07 & 0.26 \\
\hspace{0.4cm}Other price & =1 if another price concept is used (lump-sum, Shin, undefined; reference category). & 0.10 & 0.30 \\
\hspace{0.4cm}Increasing-block tariff & =1 if an increasing-block tariff is in force. & 0.21 & 0.41 \\
\hspace{0.4cm}Decreasing-block tariff & =1 if a decreasing-block tariff is in force. & 0.06 & 0.24 \\
\hspace{0.4cm}Time-of-use tariff & =1 if a time-of-use rate (different unit prices by time block) is used. & 0.04 & 0.20 \\
\hspace{0.4cm}Flat tariff & =1 if a flat tariff is in force (reference category). & 0.04 & 0.20 \\
\midrule
\multicolumn{4}{@{}l}{\emph{Demand controls}}\\
\hspace{0.4cm}Control: demographics & =1 if the demand equation controls for demographic variation. & 0.35 & 0.48 \\
\hspace{0.4cm}Control: temperature & =1 if the demand equation controls for temperature or degree-days. & 0.44 & 0.50 \\
\hspace{0.4cm}Control: appliance stock & =1 if the demand equation controls for the stock of electrical appliances. & 0.19 & 0.39 \\
\hspace{0.4cm}Control: other fuels & =1 if the demand equation controls for prices of substitute fuels. & 0.32 & 0.47 \\
\midrule
\multicolumn{4}{@{}l}{\emph{Model specification}}\\
\hspace{0.4cm}Reduced form & =1 if a single-equation reduced-form model is estimated. & 0.40 & 0.49 \\
\hspace{0.4cm}Structural form & =1 if a structural (multi-equation) model is estimated (reference category). & 0.57 & 0.50 \\
\hspace{0.4cm}Static model & =1 if the model is static (no lagged demand). & 0.25 & 0.43 \\
\hspace{0.4cm}Dynamic model & =1 if the model is dynamic, with lagged adjustment (reference category). & 0.74 & 0.44 \\
\hspace{0.4cm}ARDL & =1 if an autoregressive distributed-lag (partial-adjustment) model is used. & 0.12 & 0.32 \\
\hspace{0.4cm}Lagged endogenous & =1 if a lagged-endogenous model is used. & 0.05 & 0.21 \\
\hspace{0.4cm}Linear demand & =1 if the demand function is linear. & 0.05 & 0.21 \\
\hspace{0.4cm}Double-log demand & =1 if the demand function is double-logarithmic. & 0.93 & 0.26 \\
\hspace{0.4cm}Semi-log demand & =1 if the demand function is semi-logarithmic or Box-Cox (reference category). & 0.01 & 0.11 \\
\midrule
\multicolumn{4}{@{}l}{\emph{Geography and context}}\\
\hspace{0.4cm}United States & =1 if the estimate is for the United States. & 0.24 & 0.43 \\
\hspace{0.4cm}Europe & =1 if the estimate is for a European country. & 0.28 & 0.45 \\
\hspace{0.4cm}Other country & =1 if the estimate is for a country outside the US and Europe (reference category). & 0.49 & 0.50 \\
\hspace{0.4cm}Daylight hours & The average length of the longest day for the country under examination (source: \citealp{moesen2010daylight}). & 14.64 & 1.88 \\
\hspace{0.4cm}Population & The logarithm of the population of the country or entity (source: \citealp{worldbank2025wdi}). & 18.21 & 1.79 \\
\midrule
\multicolumn{4}{@{}l}{\emph{Publication characteristics}}\\
\hspace{0.4cm}Publication year & The logarithm of one plus the number of years since the earliest study in the sample. & 4.09 & 0.22 \\
\hspace{0.4cm}Impact factor & The logarithm of one plus the journal's RePEc recursive impact factor; 0 for unpublished work. & 0.40 & 0.78 \\
\hspace{0.4cm}Citations & The logarithm of one plus the study's mean annual Google Scholar citation count. & 1.08 & 0.99 \\
\end{longtable}
\end{singlespace}

\begin{table}[p]\centering\scriptsize\singlespace
\caption{Bayesian model averaging of the price elasticity: integrated headline sample}\label{tab:bma}
\begin{threeparttable}
\renewcommand{\arraystretch}{0.88}
\begin{tabular*}{\hsize}{@{\hskip\tabcolsep\extracolsep\fill}l*{10}{c}@{}}
\toprule
 & \multicolumn{3}{c}{Bayesian model averaging} & & \multicolumn{3}{c}{OLS, two-way clustered} & & \multicolumn{2}{c}{FMA}\\
 & \multicolumn{3}{c}{(UIP $g$-prior, dilution)} & & \multicolumn{3}{c}{(study $\times$ database)} & & \multicolumn{2}{c}{(smoothed AIC)}\\
\cmidrule(lr){2-4}\cmidrule(lr){6-8}\cmidrule(lr){10-11}
 & P.\ mean & P.\ SD & PIP & & Coef. & SE & $p$ & & Coef. & Wt.\\
\midrule
Constant & -0.480 & -- & 1.000 & & -0.937 & 0.257 & 0.000 & & -0.668 & 1.000 \\
Standard error & -0.830 & 0.034 & 1.000 & & -0.830 & 0.100 & 0.000 & & -0.834 & 1.000 \\
\midrule
\multicolumn{11}{@{}l}{\emph{Horizon}}\\
\hspace{0.4cm}Short run & 0.114 & 0.017 & 1.000 & & 0.096 & 0.034 & 0.005 & & 0.114 & 1.000 \\
\hspace{0.4cm}Long run & -0.103 & 0.021 & 1.000 & & -0.122 & 0.039 & 0.002 & & -0.101 & 1.000 \\
\midrule
\multicolumn{11}{@{}l}{\emph{Estimate type}}\\
\hspace{0.4cm}Converted from Hicksian & -0.000 & 0.003 & 0.014 & & -0.011 & 0.028 & 0.685 & & -0.000 & 0.012 \\
\midrule
\multicolumn{11}{@{}l}{\emph{Identification}}\\
\hspace{0.4cm}Design-based & 0.159 & 0.039 & 0.994 & & 0.122 & 0.069 & 0.080 & & 0.144 & 0.999 \\
\hspace{0.4cm}Instrumented & -0.001 & 0.006 & 0.019 & & -0.023 & 0.038 & 0.549 & & -0.000 & 0.003 \\
\hspace{0.4cm}Panel FE / structural & 0.075 & 0.019 & 0.989 & & 0.058 & 0.030 & 0.055 & & 0.072 & 1.000 \\
\midrule
\multicolumn{11}{@{}l}{\emph{Estimate aggregation}}\\
\hspace{0.4cm}Aggregation: country & 0.000 & 0.002 & 0.010 & & -0.009 & 0.028 & 0.739 & & 0.000 & 0.002 \\
\hspace{0.4cm}Aggregation: micro & -0.022 & 0.031 & 0.376 & & -0.039 & 0.040 & 0.340 & & -0.020 & 0.394 \\
\midrule
\multicolumn{11}{@{}l}{\emph{Sector}}\\
\hspace{0.4cm}Residential demand & -0.012 & 0.020 & 0.293 & & -0.028 & 0.028 & 0.310 & & -0.018 & 0.460 \\
\hspace{0.4cm}Industrial demand & 0.000 & 0.004 & 0.018 & & 0.002 & 0.036 & 0.961 & & 0.000 & 0.014 \\
\midrule
\multicolumn{11}{@{}l}{\emph{Data characteristics}}\\
\hspace{0.4cm}Time-series data & 0.003 & 0.010 & 0.074 & & 0.022 & 0.032 & 0.492 & & 0.004 & 0.135 \\
\hspace{0.4cm}Cross-section data & -0.166 & 0.029 & 1.000 & & -0.146 & 0.046 & 0.002 & & -0.154 & 1.000 \\
\hspace{0.4cm}Yearly data & -0.125 & 0.019 & 1.000 & & -0.108 & 0.033 & 0.001 & & -0.125 & 1.000 \\
\hspace{0.4cm}Data span & 0.005 & 0.012 & 0.200 & & 0.019 & 0.021 & 0.362 & & 0.019 & 0.664 \\
\hspace{0.4cm}Study size & -0.017 & 0.004 & 0.995 & & -0.018 & 0.006 & 0.005 & & -0.018 & 1.000 \\
\midrule
\multicolumn{11}{@{}l}{\emph{Price and tariff}}\\
\hspace{0.4cm}Average price & -0.000 & 0.002 & 0.013 & & 0.010 & 0.034 & 0.780 & & -0.000 & 0.007 \\
\hspace{0.4cm}Marginal price & 0.005 & 0.016 & 0.118 & & 0.087 & 0.055 & 0.116 & & 0.050 & 0.874 \\
\hspace{0.4cm}Increasing-block tariff & 0.000 & 0.002 & 0.009 & & 0.025 & 0.035 & 0.483 & & 0.000 & 0.002 \\
\hspace{0.4cm}Decreasing-block tariff & -0.174 & 0.024 & 1.000 & & -0.142 & 0.054 & 0.008 & & -0.178 & 1.000 \\
\hspace{0.4cm}Time-of-use tariff & 0.000 & 0.004 & 0.014 & & 0.038 & 0.046 & 0.405 & & 0.001 & 0.032 \\
\midrule
\multicolumn{11}{@{}l}{\emph{Demand controls}}\\
\hspace{0.4cm}Control: demographics & -0.067 & 0.016 & 0.998 & & -0.057 & 0.027 & 0.038 & & -0.058 & 1.000 \\
\hspace{0.4cm}Control: temperature & -0.000 & 0.002 & 0.010 & & -0.012 & 0.025 & 0.622 & & -0.000 & 0.002 \\
\hspace{0.4cm}Control: appliance stock & -0.005 & 0.015 & 0.118 & & -0.032 & 0.034 & 0.341 & & -0.025 & 0.546 \\
\hspace{0.4cm}Control: other fuels & -0.074 & 0.016 & 1.000 & & -0.062 & 0.028 & 0.029 & & -0.070 & 1.000 \\
\midrule
\multicolumn{11}{@{}l}{\emph{Model specification}}\\
\hspace{0.4cm}Reduced form & 0.014 & 0.021 & 0.338 & & 0.041 & 0.029 & 0.157 & & 0.024 & 0.661 \\
\hspace{0.4cm}Static model & -0.000 & 0.004 & 0.016 & & -0.041 & 0.035 & 0.245 & & -0.002 & 0.061 \\
\hspace{0.4cm}ARDL & 0.002 & 0.011 & 0.034 & & 0.050 & 0.043 & 0.251 & & 0.002 & 0.034 \\
\hspace{0.4cm}Lagged endogenous & 0.101 & 0.020 & 1.000 & & 0.083 & 0.036 & 0.020 & & 0.095 & 1.000 \\
\hspace{0.4cm}Linear demand & 0.010 & 0.023 & 0.188 & & 0.049 & 0.052 & 0.351 & & 0.041 & 0.701 \\
\hspace{0.4cm}Double-log demand & -0.004 & 0.014 & 0.094 & & -0.005 & 0.041 & 0.904 & & -0.003 & 0.067 \\
\midrule
\multicolumn{11}{@{}l}{\emph{Geography and context}}\\
\hspace{0.4cm}United States & 0.037 & 0.030 & 0.665 & & 0.054 & 0.034 & 0.110 & & 0.027 & 0.512 \\
\hspace{0.4cm}Europe & 0.002 & 0.010 & 0.046 & & 0.043 & 0.041 & 0.297 & & 0.009 & 0.184 \\
\hspace{0.4cm}Daylight hours & 0.019 & 0.005 & 0.992 & & 0.018 & 0.007 & 0.009 & & 0.019 & 0.999 \\
\hspace{0.4cm}Population & 0.004 & 0.007 & 0.313 & & 0.014 & 0.010 & 0.150 & & 0.011 & 0.740 \\
\midrule
\multicolumn{11}{@{}l}{\emph{Publication characteristics}}\\
\hspace{0.4cm}Publication year & 0.001 & 0.007 & 0.022 & & 0.065 & 0.050 & 0.194 & & 0.006 & 0.088 \\
\hspace{0.4cm}Impact factor & 0.034 & 0.027 & 0.665 & & 0.031 & 0.024 & 0.190 & & 0.028 & 0.653 \\
\hspace{0.4cm}Citations & 0.018 & 0.021 & 0.474 & & 0.019 & 0.021 & 0.352 & & 0.020 & 0.611 \\
\midrule
Observations & 3,324 & & & & 3,324 & & & & 3,324 & \\
Studies / databases & 366\,/\,344 & & & & & & & & & \\
\bottomrule
\end{tabular*}
\begin{tablenotes}[flushleft]\footnotesize\item \emph{Notes:} The response variable is the 1\%-winsorized own-price elasticity of electricity demand. \emph{Bayesian model averaging} (left) uses the unit-information $g$-prior with the dilution model prior, which penalizes collinearity by down-weighting models whose regressor correlation matrix is close to singular; P.\ mean, P.\ SD and PIP are the posterior mean, posterior standard deviation and posterior inclusion probability. \emph{OLS, two-way clustered} (middle) regresses the elasticity on the same moderators with standard errors clustered simultaneously by study and by underlying database. \emph{FMA} (right) is a frequentist model average with smoothed-AIC weights over the models visited by the sampler, reporting each moderator's model-averaged coefficient (\emph{Coef.}) and its smoothed-AIC inclusion weight (\emph{Wt.}, the frequentist analogue of the posterior inclusion probability); because those weights depend on which models the sampler visits over a very large model space, this column is a Monte Carlo estimate that is not exactly reproducible run to run (up to about $0.05$), unlike the reproducible Bayesian and OLS columns. The standard error of the estimate absorbs residual publication selection. Scale-laden moderators enter in logs; reference categories are omitted. Variables are defined in \autoref{tab:vardefs}.\end{tablenotes}
\end{threeparttable}
\end{table}

\begin{table}[p]\centering\footnotesize\singlespace
\caption{Bayesian model averaging of the short-run price elasticity}\label{tab:bmasr}
\begin{threeparttable}
\renewcommand{\arraystretch}{0.88}
\begin{tabular*}{\hsize}{@{\hskip\tabcolsep\extracolsep\fill}l*{10}{c}@{}}
\toprule
 & \multicolumn{3}{c}{Bayesian model averaging} & & \multicolumn{3}{c}{OLS, two-way clustered} & & \multicolumn{2}{c}{FMA}\\
 & \multicolumn{3}{c}{(UIP $g$-prior, dilution)} & & \multicolumn{3}{c}{(study $\times$ database)} & & \multicolumn{2}{c}{(smoothed AIC)}\\
\cmidrule(lr){2-4}\cmidrule(lr){6-8}\cmidrule(lr){10-11}
 & P.\ mean & P.\ SD & PIP & & Coef. & SE & $p$ & & Coef. & Wt.\\
\midrule
Constant & -0.449 & -- & 1.000 & & -0.766 & 0.367 & 0.037 & & -0.460 & 1.000 \\
Standard error & -0.846 & 0.045 & 1.000 & & -0.860 & 0.140 & 0.000 & & -0.833 & 1.000 \\
\midrule
\multicolumn{11}{@{}l}{\emph{Estimate type}}\\
\hspace{0.4cm}Converted from Hicksian & 0.033 & 0.038 & 0.483 & & 0.055 & 0.034 & 0.105 & & 0.062 & 0.960 \\
\midrule
\multicolumn{11}{@{}l}{\emph{Identification}}\\
\hspace{0.4cm}Design-based & 0.259 & 0.040 & 1.000 & & 0.258 & 0.095 & 0.006 & & 0.250 & 1.000 \\
\hspace{0.4cm}Instrumented & 0.001 & 0.005 & 0.020 & & 0.056 & 0.049 & 0.255 & & 0.007 & 0.186 \\
\hspace{0.4cm}Panel FE / structural & 0.066 & 0.029 & 0.901 & & 0.085 & 0.037 & 0.020 & & 0.065 & 0.992 \\
\midrule
\multicolumn{11}{@{}l}{\emph{Estimate aggregation}}\\
\hspace{0.4cm}Aggregation: country & -0.000 & 0.002 & 0.009 & & -0.038 & 0.031 & 0.215 & & -0.001 & 0.050 \\
\hspace{0.4cm}Aggregation: micro & -0.021 & 0.033 & 0.331 & & -0.025 & 0.043 & 0.559 & & -0.032 & 0.566 \\
\midrule
\multicolumn{11}{@{}l}{\emph{Sector}}\\
\hspace{0.4cm}Residential demand & -0.015 & 0.030 & 0.238 & & -0.039 & 0.034 & 0.246 & & -0.018 & 0.351 \\
\hspace{0.4cm}Industrial demand & -0.014 & 0.032 & 0.175 & & -0.070 & 0.046 & 0.129 & & -0.021 & 0.350 \\
\midrule
\multicolumn{11}{@{}l}{\emph{Data characteristics}}\\
\hspace{0.4cm}Time-series data & 0.006 & 0.016 & 0.154 & & 0.029 & 0.035 & 0.409 & & 0.010 & 0.294 \\
\hspace{0.4cm}Cross-section data & -0.000 & 0.005 & 0.012 & & -0.006 & 0.071 & 0.936 & & -0.000 & 0.016 \\
\hspace{0.4cm}Yearly data & -0.019 & 0.031 & 0.312 & & -0.067 & 0.040 & 0.092 & & -0.048 & 0.854 \\
\hspace{0.4cm}Data span & 0.003 & 0.010 & 0.094 & & 0.019 & 0.024 & 0.427 & & 0.009 & 0.350 \\
\hspace{0.4cm}Study size & -0.001 & 0.003 & 0.093 & & -0.011 & 0.007 & 0.138 & & -0.002 & 0.235 \\
\midrule
\multicolumn{11}{@{}l}{\emph{Price and tariff}}\\
\hspace{0.4cm}Average price & -0.001 & 0.004 & 0.022 & & -0.035 & 0.035 & 0.315 & & -0.001 & 0.034 \\
\hspace{0.4cm}Marginal price & 0.000 & 0.003 & 0.012 & & 0.006 & 0.059 & 0.921 & & 0.000 & 0.019 \\
\hspace{0.4cm}Increasing-block tariff & -0.001 & 0.005 & 0.019 & & -0.035 & 0.033 & 0.288 & & -0.001 & 0.026 \\
\hspace{0.4cm}Decreasing-block tariff & 0.000 & 0.003 & 0.009 & & 0.018 & 0.049 & 0.716 & & 0.000 & 0.011 \\
\hspace{0.4cm}Time-of-use tariff & 0.065 & 0.045 & 0.733 & & 0.057 & 0.063 & 0.367 & & 0.066 & 0.858 \\
\midrule
\multicolumn{11}{@{}l}{\emph{Demand controls}}\\
\hspace{0.4cm}Control: demographics & -0.064 & 0.023 & 0.948 & & -0.052 & 0.034 & 0.122 & & -0.054 & 0.981 \\
\hspace{0.4cm}Control: temperature & -0.000 & 0.002 & 0.010 & & -0.014 & 0.027 & 0.589 & & -0.001 & 0.028 \\
\hspace{0.4cm}Control: appliance stock & -0.083 & 0.029 & 0.950 & & -0.068 & 0.043 & 0.111 & & -0.085 & 1.000 \\
\hspace{0.4cm}Control: other fuels & -0.010 & 0.021 & 0.232 & & -0.019 & 0.032 & 0.566 & & -0.027 & 0.675 \\
\midrule
\multicolumn{11}{@{}l}{\emph{Model specification}}\\
\hspace{0.4cm}Reduced form & 0.059 & 0.025 & 0.912 & & 0.084 & 0.030 & 0.005 & & 0.074 & 1.000 \\
\hspace{0.4cm}Static model & -0.133 & 0.023 & 1.000 & & -0.123 & 0.047 & 0.009 & & -0.133 & 1.000 \\
\hspace{0.4cm}ARDL & 0.000 & 0.006 & 0.012 & & 0.015 & 0.038 & 0.701 & & 0.001 & 0.019 \\
\hspace{0.4cm}Lagged endogenous & -0.001 & 0.006 & 0.023 & & -0.024 & 0.041 & 0.551 & & -0.000 & 0.021 \\
\hspace{0.4cm}Linear demand & 0.056 & 0.049 & 0.627 & & 0.145 & 0.098 & 0.136 & & 0.093 & 0.989 \\
\hspace{0.4cm}Double-log demand & 0.001 & 0.011 & 0.030 & & 0.056 & 0.065 & 0.390 & & 0.003 & 0.055 \\
\midrule
\multicolumn{11}{@{}l}{\emph{Geography and context}}\\
\hspace{0.4cm}United States & -0.000 & 0.002 & 0.011 & & -0.005 & 0.044 & 0.908 & & -0.000 & 0.012 \\
\hspace{0.4cm}Europe & -0.000 & 0.003 & 0.011 & & -0.026 & 0.047 & 0.577 & & -0.000 & 0.016 \\
\hspace{0.4cm}Daylight hours & 0.020 & 0.005 & 0.996 & & 0.024 & 0.006 & 0.000 & & 0.020 & 1.000 \\
\hspace{0.4cm}Population & 0.000 & 0.001 & 0.008 & & 0.009 & 0.011 & 0.424 & & 0.000 & 0.016 \\
\midrule
\multicolumn{11}{@{}l}{\emph{Publication characteristics}}\\
\hspace{0.4cm}Publication year & 0.000 & 0.005 & 0.015 & & 0.030 & 0.084 & 0.723 & & 0.001 & 0.029 \\
\hspace{0.4cm}Impact factor & -0.000 & 0.001 & 0.007 & & -0.009 & 0.021 & 0.677 & & -0.000 & 0.012 \\
\hspace{0.4cm}Citations & 0.000 & 0.002 & 0.015 & & 0.010 & 0.036 & 0.781 & & 0.001 & 0.037 \\
\midrule
Observations & 1,647 & & & & 1,647 & & & & 1,647 & \\
Studies / databases & 226\,/\,218 & & & & & & & & & \\
\bottomrule
\end{tabular*}
\begin{tablenotes}[flushleft]\footnotesize\item \emph{Notes:} As in \autoref{tab:bma}, estimated on the 1{,}647 short-run headline estimates; the horizon indicators are constant here and drop out. Variables are defined in \autoref{tab:vardefssr}.\end{tablenotes}
\end{threeparttable}
\end{table}

\begin{table}[p]\centering\footnotesize\singlespace
\caption{Bayesian model averaging of the long-run price elasticity}\label{tab:bmalr}
\begin{threeparttable}
\renewcommand{\arraystretch}{0.88}
\begin{tabular*}{\hsize}{@{\hskip\tabcolsep\extracolsep\fill}l*{10}{c}@{}}
\toprule
 & \multicolumn{3}{c}{Bayesian model averaging} & & \multicolumn{3}{c}{OLS, two-way clustered} & & \multicolumn{2}{c}{FMA}\\
 & \multicolumn{3}{c}{(UIP $g$-prior, dilution)} & & \multicolumn{3}{c}{(study $\times$ database)} & & \multicolumn{2}{c}{(smoothed AIC)}\\
\cmidrule(lr){2-4}\cmidrule(lr){6-8}\cmidrule(lr){10-11}
 & P.\ mean & P.\ SD & PIP & & Coef. & SE & $p$ & & Coef. & Wt.\\
\midrule
Constant & -0.515 & -- & 1.000 & & -1.525 & 0.839 & 0.069 & & -1.056 & 1.000 \\
Standard error & -1.072 & 0.069 & 1.000 & & -1.085 & 0.138 & 0.000 & & -1.067 & 1.000 \\
\midrule
\multicolumn{11}{@{}l}{\emph{Estimate type}}\\
\hspace{0.4cm}Converted from Hicksian & -0.186 & 0.043 & 0.995 & & -0.135 & 0.067 & 0.045 & & -0.202 & 1.000 \\
\midrule
\multicolumn{11}{@{}l}{\emph{Identification}}\\
\hspace{0.4cm}Design-based & 0.001 & 0.020 & 0.007 & & -0.260 & 0.191 & 0.174 & & 0.000 & 0.004 \\
\hspace{0.4cm}Instrumented & -0.028 & 0.057 & 0.221 & & -0.102 & 0.077 & 0.185 & & -0.050 & 0.387 \\
\hspace{0.4cm}Panel FE / structural & 0.087 & 0.071 & 0.651 & & 0.079 & 0.082 & 0.337 & & 0.082 & 0.629 \\
\midrule
\multicolumn{11}{@{}l}{\emph{Estimate aggregation}}\\
\hspace{0.4cm}Aggregation: country & 0.000 & 0.004 & 0.008 & & 0.022 & 0.063 & 0.729 & & 0.000 & 0.008 \\
\hspace{0.4cm}Aggregation: micro & -0.000 & 0.006 & 0.008 & & 0.068 & 0.105 & 0.518 & & 0.000 & 0.003 \\
\midrule
\multicolumn{11}{@{}l}{\emph{Sector}}\\
\hspace{0.4cm}Residential demand & -0.000 & 0.005 & 0.009 & & 0.044 & 0.066 & 0.499 & & -0.000 & 0.005 \\
\hspace{0.4cm}Industrial demand & 0.012 & 0.037 & 0.113 & & 0.114 & 0.076 & 0.133 & & 0.041 & 0.433 \\
\midrule
\multicolumn{11}{@{}l}{\emph{Data characteristics}}\\
\hspace{0.4cm}Time-series data & 0.000 & 0.004 & 0.007 & & -0.002 & 0.075 & 0.983 & & 0.000 & 0.007 \\
\hspace{0.4cm}Cross-section data & -0.085 & 0.120 & 0.376 & & -0.064 & 0.114 & 0.572 & & -0.042 & 0.240 \\
\hspace{0.4cm}Yearly data & -0.005 & 0.023 & 0.047 & & -0.071 & 0.100 & 0.477 & & -0.012 & 0.142 \\
\hspace{0.4cm}Data span & 0.047 & 0.043 & 0.582 & & 0.062 & 0.046 & 0.178 & & 0.077 & 0.910 \\
\hspace{0.4cm}Study size & 0.000 & 0.001 & 0.009 & & -0.012 & 0.019 & 0.521 & & -0.001 & 0.066 \\
\midrule
\multicolumn{11}{@{}l}{\emph{Price and tariff}}\\
\hspace{0.4cm}Average price & 0.000 & 0.004 & 0.007 & & 0.099 & 0.110 & 0.366 & & 0.000 & 0.006 \\
\hspace{0.4cm}Marginal price & 0.001 & 0.011 & 0.012 & & 0.247 & 0.173 & 0.154 & & 0.021 & 0.160 \\
\hspace{0.4cm}Increasing-block tariff & 0.000 & 0.005 & 0.008 & & -0.054 & 0.084 & 0.519 & & 0.000 & 0.006 \\
\hspace{0.4cm}Decreasing-block tariff & -0.002 & 0.018 & 0.016 & & -0.028 & 0.149 & 0.850 & & -0.001 & 0.012 \\
\hspace{0.4cm}Time-of-use tariff & -0.001 & 0.016 & 0.011 & & 0.010 & 0.183 & 0.957 & & 0.000 & 0.007 \\
\midrule
\multicolumn{11}{@{}l}{\emph{Demand controls}}\\
\hspace{0.4cm}Control: demographics & -0.001 & 0.012 & 0.021 & & -0.056 & 0.075 & 0.455 & & -0.003 & 0.052 \\
\hspace{0.4cm}Control: temperature & 0.000 & 0.003 & 0.007 & & -0.084 & 0.059 & 0.156 & & -0.000 & 0.009 \\
\hspace{0.4cm}Control: appliance stock & -0.174 & 0.073 & 0.912 & & -0.151 & 0.059 & 0.011 & & -0.157 & 0.995 \\
\hspace{0.4cm}Control: other fuels & -0.000 & 0.005 & 0.009 & & -0.051 & 0.068 & 0.457 & & -0.000 & 0.011 \\
\midrule
\multicolumn{11}{@{}l}{\emph{Model specification}}\\
\hspace{0.4cm}Reduced form & 0.000 & 0.006 & 0.010 & & 0.071 & 0.080 & 0.373 & & 0.001 & 0.013 \\
\hspace{0.4cm}Static model & -0.001 & 0.008 & 0.013 & & -0.069 & 0.090 & 0.447 & & -0.000 & 0.006 \\
\hspace{0.4cm}ARDL & 0.007 & 0.032 & 0.064 & & 0.110 & 0.066 & 0.096 & & 0.070 & 0.536 \\
\hspace{0.4cm}Lagged endogenous & 0.015 & 0.055 & 0.086 & & 0.235 & 0.085 & 0.006 & & 0.208 & 0.939 \\
\hspace{0.4cm}Linear demand & -0.016 & 0.057 & 0.088 & & -0.446 & 0.179 & 0.013 & & -0.158 & 0.804 \\
\hspace{0.4cm}Double-log demand & 0.001 & 0.013 & 0.015 & & -0.200 & 0.100 & 0.047 & & 0.001 & 0.017 \\
\midrule
\multicolumn{11}{@{}l}{\emph{Geography and context}}\\
\hspace{0.4cm}United States & 0.035 & 0.062 & 0.270 & & 0.313 & 0.096 & 0.001 & & 0.169 & 0.997 \\
\hspace{0.4cm}Europe & 0.085 & 0.074 & 0.632 & & 0.206 & 0.078 & 0.008 & & 0.177 & 1.000 \\
\hspace{0.4cm}Daylight hours & 0.000 & 0.001 & 0.011 & & -0.012 & 0.016 & 0.454 & & -0.000 & 0.015 \\
\hspace{0.4cm}Population & 0.000 & 0.001 & 0.008 & & -0.007 & 0.020 & 0.704 & & -0.000 & 0.028 \\
\midrule
\multicolumn{11}{@{}l}{\emph{Publication characteristics}}\\
\hspace{0.4cm}Publication year & 0.013 & 0.053 & 0.067 & & 0.342 & 0.197 & 0.083 & & 0.110 & 0.451 \\
\hspace{0.4cm}Impact factor & 0.000 & 0.005 & 0.013 & & -0.016 & 0.048 & 0.745 & & 0.000 & 0.004 \\
\hspace{0.4cm}Citations & 0.015 & 0.030 & 0.232 & & 0.065 & 0.033 & 0.048 & & 0.028 & 0.525 \\
\midrule
Observations & 723 & & & & 723 & & & & 723 & \\
Studies / databases & 151\,/\,143 & & & & & & & & & \\
\bottomrule
\end{tabular*}
\begin{tablenotes}[flushleft]\footnotesize\item \emph{Notes:} As in \autoref{tab:bma}, estimated on the 723 long-run headline estimates; the horizon indicators are constant here and drop out. Variables are defined in \autoref{tab:vardefslr}.\end{tablenotes}
\end{threeparttable}
\end{table}

\begin{table}[p]\centering\footnotesize\singlespace
\caption{Model averaging under alternative priors (integrated sample)}\label{tab:bmarobust}
\begin{threeparttable}
\begin{tabular*}{\hsize}{@{\hskip\tabcolsep\extracolsep\fill}l*{8}{c}@{}}
\toprule
Variable & \multicolumn{2}{c}{A} & \multicolumn{2}{c}{B} & \multicolumn{2}{c}{C} & \multicolumn{2}{c}{D} \\
\cmidrule(lr){2-3} \cmidrule(lr){4-5} \cmidrule(lr){6-7} \cmidrule(lr){8-9}
 & P.\ mean & PIP & P.\ mean & PIP & P.\ mean & PIP & P.\ mean & PIP \\
\midrule
Standard error & -0.830 & 1.000 & -0.830 & 1.000 & -0.830 & 1.000 & -0.829 & 1.000 \\
Short run & 0.114 & 1.000 & 0.114 & 1.000 & 0.114 & 1.000 & 0.114 & 1.000 \\
Long run & -0.103 & 1.000 & -0.102 & 1.000 & -0.103 & 1.000 & -0.102 & 1.000 \\
Cross-section data & -0.166 & 1.000 & -0.166 & 1.000 & -0.166 & 1.000 & -0.161 & 1.000 \\
Yearly data & -0.125 & 1.000 & -0.125 & 1.000 & -0.125 & 1.000 & -0.124 & 1.000 \\
Decreasing-block tariff & -0.174 & 1.000 & -0.174 & 1.000 & -0.174 & 1.000 & -0.174 & 1.000 \\
Lagged endogenous & 0.101 & 1.000 & 0.100 & 1.000 & 0.101 & 1.000 & 0.098 & 1.000 \\
Control: other fuels & -0.074 & 1.000 & -0.074 & 1.000 & -0.074 & 1.000 & -0.073 & 1.000 \\
Control: demographics & -0.067 & 0.998 & -0.067 & 0.998 & -0.067 & 0.998 & -0.063 & 1.000 \\
Study size & -0.017 & 0.995 & -0.017 & 0.997 & -0.017 & 0.995 & -0.017 & 1.000 \\
Design-based & 0.159 & 0.994 & 0.159 & 0.994 & 0.159 & 0.994 & 0.154 & 0.996 \\
Daylight hours & 0.019 & 0.992 & 0.019 & 0.993 & 0.019 & 0.992 & 0.019 & 0.997 \\
Panel FE / structural & 0.075 & 0.989 & 0.075 & 0.990 & 0.075 & 0.989 & 0.074 & 0.998 \\
Impact factor & 0.034 & 0.665 & 0.034 & 0.668 & 0.034 & 0.665 & 0.032 & 0.689 \\
United States & 0.037 & 0.664 & 0.038 & 0.670 & 0.037 & 0.664 & 0.036 & 0.647 \\
Citations & 0.018 & 0.474 & 0.019 & 0.488 & 0.018 & 0.474 & 0.019 & 0.556 \\
Aggregation: micro & -0.022 & 0.376 & -0.022 & 0.386 & -0.022 & 0.376 & -0.023 & 0.433 \\
Reduced form & 0.014 & 0.338 & 0.015 & 0.366 & 0.014 & 0.338 & 0.020 & 0.509 \\
Population & 0.004 & 0.313 & 0.005 & 0.331 & 0.004 & 0.313 & 0.007 & 0.474 \\
Residential demand & -0.012 & 0.293 & -0.013 & 0.315 & -0.012 & 0.293 & -0.016 & 0.420 \\
Data span & 0.005 & 0.200 & 0.006 & 0.220 & 0.005 & 0.201 & 0.010 & 0.385 \\
Linear demand & 0.010 & 0.188 & 0.011 & 0.207 & 0.010 & 0.188 & 0.021 & 0.379 \\
Marginal price & 0.005 & 0.118 & 0.006 & 0.138 & 0.005 & 0.118 & 0.021 & 0.414 \\
Control: appliance stock & -0.005 & 0.118 & -0.005 & 0.130 & -0.005 & 0.118 & -0.012 & 0.268 \\
Double-log demand & -0.004 & 0.093 & -0.004 & 0.104 & -0.004 & 0.094 & -0.005 & 0.136 \\
Time-series data & 0.003 & 0.074 & 0.003 & 0.081 & 0.003 & 0.074 & 0.004 & 0.118 \\
Europe & 0.002 & 0.046 & 0.002 & 0.052 & 0.002 & 0.046 & 0.005 & 0.117 \\
ARDL & 0.002 & 0.034 & 0.002 & 0.039 & 0.002 & 0.034 & 0.003 & 0.063 \\
Publication year & 0.001 & 0.021 & 0.001 & 0.025 & 0.001 & 0.022 & 0.003 & 0.062 \\
Instrumented & -0.001 & 0.019 & -0.001 & 0.020 & -0.001 & 0.019 & -0.000 & 0.015 \\
Industrial demand & 0.000 & 0.018 & 0.000 & 0.020 & 0.000 & 0.018 & 0.000 & 0.024 \\
Static model & -0.000 & 0.016 & -0.000 & 0.018 & -0.000 & 0.016 & -0.001 & 0.043 \\
Time-of-use tariff & 0.000 & 0.014 & 0.000 & 0.016 & 0.000 & 0.014 & 0.001 & 0.026 \\
Converted from Hicksian & -0.000 & 0.013 & -0.000 & 0.015 & -0.000 & 0.013 & -0.000 & 0.019 \\
Average price & -0.000 & 0.012 & -0.000 & 0.015 & -0.000 & 0.013 & -0.000 & 0.021 \\
Control: temperature & -0.000 & 0.010 & -0.000 & 0.012 & -0.000 & 0.010 & -0.000 & 0.012 \\
Aggregation: country & 0.000 & 0.010 & 0.000 & 0.011 & 0.000 & 0.010 & 0.000 & 0.010 \\
Increasing-block tariff & 0.000 & 0.009 & 0.000 & 0.010 & 0.000 & 0.009 & 0.000 & 0.011 \\
\bottomrule
\end{tabular*}
\begin{tablenotes}[flushleft]\footnotesize\item \emph{Notes:} Posterior mean and inclusion probability (PIP) for the integrated model average under four prior specifications, ordered by baseline PIP. (A) unit-information $g$-prior, dilution model prior (baseline); (B) unit-information $g$-prior, uniform model prior; (C) BRIC $g$-prior, random (beta-binomial) model prior; (D) Hannan--Quinn $g$-prior, dilution model prior. The key moderators keep their sign and inclusion across priors.\end{tablenotes}
\end{threeparttable}
\end{table}

\begin{table}[h]\centering\small\singlespace
\caption{Model-averaging diagnostics}\label{tab:bmadiag}
\begin{threeparttable}
\begin{tabular}{lccccc}
\toprule
Model & Obs. & E[model size] & Models visited & Corr PMP & Shrinkage \\
\midrule
Integrated & 3,324 & 17.11 & 146,615 & 1.00 & 1.00 \\
Short run & 1,647 & 11.43 & 148,177 & 1.00 & 1.00 \\
Long run & 723 & 6.54 & 117,429 & 1.00 & 1.00 \\
\bottomrule
\end{tabular}
\begin{tablenotes}[flushleft]\footnotesize\item \emph{Notes:} Birth--death MCMC ($1.5\times10^{5}$ draws after $5\times10^{4}$ burn-in) for each averaging. ``E[model size]'' is the sum of inclusion probabilities; ``Corr PMP'' is the correlation between analytical and MCMC posterior model probabilities (one indicates convergence); ``Shrinkage'' is $g/(1+g)$.\end{tablenotes}
\end{threeparttable}
\end{table}

\clearpage
\emph{What the averaging shows.} Four features reinforce the body. First, publication selection is the most robust regularity in the corpus: the standard-error term enters every model in every subsample (PIP $=1.00$) with a negative posterior mean that grows with the horizon ($-0.83$ integrated, $-0.85$ short run, $-1.07$ long run), the funnel asymmetry of \autoref{tab:battery} recovered as a model-averaged coefficient that survives against every moderator at once. Second, and most telling for the flexibility premise, the best-identified and most technology-rich settings are, conditional on everything else, the \emph{least} price-responsive: in the short-run averaging the design-based tier ($+0.26$, PIP $=1.00$, and OLS-significant) is robustly included and pulls the elasticity toward zero, while the time-of-use indicator ($+0.07$, PIP $=0.73$, its OLS coefficient not distinguishable from zero) is weakly included and only sign-consistent, the model-averaging counterpart of the identification ladder and of the new-regime and time-of-use findings (\autoref{fig:newregime}, \autoref{sec:time}). Third, the vintage dimension itself carries \emph{negligible} weight: in the integrated model the publication-year moderator is almost never selected and the data span only rarely (PIP $=0.02$ and $0.20$), both entering with posterior means near zero, so once composition is averaged over there is no heterogeneity along the calendar axis a rising-flexibility world would light up, the averaging counterpart of the flat \emph{publication} clock; the data mid-year itself is not among the model-averaging moderators, so this appendix speaks to the reporting clock rather than directly to the behavioral one. Fourth, the heterogeneity that \emph{is} robust is compositional and methodological: data structure and frequency (cross-section $-0.17$, yearly $-0.12$, both PIP $=1.00$), study size, the dynamic-adjustment indicators (lagged endogenous $+0.10$ integrated, static model $-0.13$ short run), decreasing-block tariffs ($-0.17$) and a demographics control, exactly the study characteristics the ladder and horizon analyses hold fixed. The frequentist model average and the two-way clustered OLS agree in sign with the posterior means on every reliably included moderator, the core integrated-sample terms are stable across the four prior specifications (\autoref{tab:bmarobust}), while the horizon-specific positives (the short-run design-based and time-of-use coefficients) are checked only for sign in that integrated sweep, and the chains have converged (posterior-model-probability correlation $1.00$; \autoref{tab:bmadiag}), with the regressor correlations (\autoref{fig:bmacorr}) justifying the dilution prior.

\FloatBarrier
\subsection{Best-practice prediction, country by country}\label{app:bestpractice}

The model average also lets us read off a \emph{best-practice} elasticity for each country: the response an ideally-designed study would report there. We predict from the integrated two-way-clustered regression of \autoref{tab:bma}, evaluated not at the average study but at a best-practice profile $x^{\star}$. Following the variable list, (i)~the standard error is set to zero, stripping out the publication-selection component; (ii)~the journal impact factor and citation count are set to their 95th percentiles, a well-cited outlet without letting a single citation outlier drive the prediction; (iii)~the country-level moderators (the US and Europe indicators, daylight hours and log population) are set to \emph{each country's own level}; (iv)~the identification tier is set to best practice, with the design-based indicator on and the instrumented and panel-FE/structural indicators off (naive is the reference); (v)~the demand equation is fully controlled (demographics, temperature, appliance stock, and substitute-fuel prices) and estimated on micro, household-level data; and (vi)~every remaining moderator (model specification and functional form) is held at its sample mean, so the prediction describes a fully-specified, design-based, de-biased study conducted in that country. We report the prediction at both horizons: the short-run profile sets the short-run indicator on, the long-run profile the long-run indicator, with the other horizon dummies off.

\begin{figure}[h]
\centering
\caption{Best-practice price elasticity by country, short and long run}\label{fig:bpcountry}
\begin{threeparttable}
\includegraphics[width=.82\textwidth]{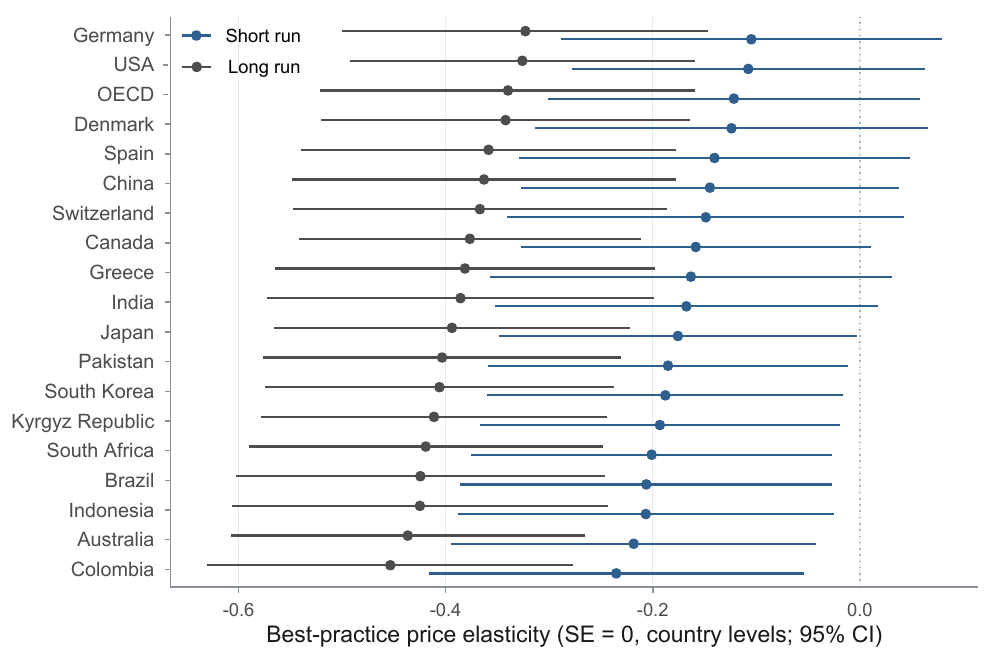}
\begin{tablenotes}[flush]
\footnotesize
\item \textit{Notes:} Best-practice prediction from the integrated regression of \autoref{tab:bma} (standard error set to zero, impact factor and citations at their 95th percentiles, country-level moderators at each country's own level, identification at best practice (design-based tier on; instrumented and structural tiers off, naive reference), demand controls on and micro-level data, remaining moderators at sample means), with 95\% intervals from the two-way (study $\times$ database) clustered covariance. Countries with at least thirty headline estimates, ranked by the short-run prediction.
\end{tablenotes}
\end{threeparttable}
\end{figure}

The best-practice elasticity is the linear prediction $\mathrm{Mean}^{\star}=x^{\star\prime}\widehat\beta$, and because it is a linear combination of the estimated coefficients its sampling variance is $x^{\star\prime}\widehat V\,x^{\star}$, where $\widehat V$ is the two-way (study $\times$ database) clustered covariance; we report the Wald interval $\mathrm{Mean}^{\star}\pm1.96\sqrt{x^{\star\prime}\widehat V\,x^{\star}}$. \autoref{fig:bpcountry} shows the two predictions for the countries with at least thirty headline estimates, ranked by the short-run value; \autoref{tab:bpcountry} lists every country with at least five. The message mirrors the pooled result: the best-practice short-run elasticity is small and negative everywhere (typically around $-0.18$), in line with the corrected short-run consensus of the body, though its confidence interval still spans zero in fewer than half the countries; the long-run prediction is roughly twice as large and significantly negative in most countries, the horizon gradient reproducing country by country. Because only the four country-level moderators vary across predictions while all method, sector, and horizon slopes are held common, the compressed cross-country spread is imposed by the pooled model rather than demonstrated as genuine homogeneity. The cross-country spread is modest relative to the within-country uncertainty, so the audit's central finding travels: no country's best-practice \emph{prediction} supports the large, rising responsiveness the flexibility premise assumes. For countries without an in-country design-based study, this value is a model extrapolation that sets the design-based indicator on, not an observed record.

\begin{singlespace}\footnotesize
\setlength\LTleft{\fill}\setlength\LTright{\fill}
\begin{longtable}{@{}l r@{\hskip 0.7em}>{\centering\arraybackslash}p{2.2cm}@{\hskip 2.4em} r@{\hskip 0.7em}>{\centering\arraybackslash}p{2.2cm}@{}}
\caption{Best-practice price elasticity by country}\label{tab:bpcountry}\\
\toprule
 & \multicolumn{2}{c}{Short run} & \multicolumn{2}{c}{Long run}\\
\cmidrule(lr){2-3}\cmidrule(lr){4-5}
Country & $\mathrm{Mean}^{\star}$ & 95\% CI & $\mathrm{Mean}^{\star}$ & 95\% CI\\
\midrule\endfirsthead
\multicolumn{5}{@{}l}{\emph{\autoref{tab:bpcountry} continued}}\\
\toprule
 & \multicolumn{2}{c}{Short run} & \multicolumn{2}{c}{Long run}\\
\cmidrule(lr){2-3}\cmidrule(lr){4-5}
Country & $\mathrm{Mean}^{\star}$ & 95\% CI & $\mathrm{Mean}^{\star}$ & 95\% CI\\
\midrule\endhead
\bottomrule\multicolumn{5}{r}{\scriptsize\emph{Continued on next page}}\\\endfoot
\bottomrule
\multicolumn{5}{@{}>{\scriptsize}p{10.6cm}@{}}{\emph{Notes:} $\mathrm{Mean}^{\star}$ is the best-practice predicted elasticity, the fitted value at a country's best-practice profile $x^\star$; the table reports short- and long-run $\mathrm{Mean}^{\star}$ with 95\% two-way (study $\times$ database) clustered intervals, for the countries contributing at least five headline estimates, ordered alphabetically; a few rows are country aggregates (EU, OECD, Middle income) rather than individual countries. Country-level moderators are set to each country's own level; all other moderators follow the best-practice profile of \autoref{fig:bpcountry}.}\\
\endlastfoot
Argentina & $-0.201$ & $[-0.374,\,-0.028]$ & $-0.419$ & $[-0.588,\,-0.250]$ \\
Australia & $-0.218$ & $[-0.395,\,-0.042]$ & $-0.437$ & $[-0.608,\,-0.266]$ \\
Austria & $-0.148$ & $[-0.339,\,0.044]$ & $-0.366$ & $[-0.546,\,-0.185]$ \\
Azerbaijan & $-0.469$ & $[-0.757,\,-0.182]$ & $-0.687$ & $[-0.970,\,-0.405]$ \\
Bangladesh & $-0.189$ & $[-0.365,\,-0.013]$ & $-0.407$ & $[-0.583,\,-0.232]$ \\
Brazil & $-0.206$ & $[-0.386,\,-0.026]$ & $-0.424$ & $[-0.603,\,-0.246]$ \\
Canada & $-0.158$ & $[-0.328,\,0.011]$ & $-0.377$ & $[-0.542,\,-0.212]$ \\
Chile & $-0.215$ & $[-0.390,\,-0.039]$ & $-0.433$ & $[-0.603,\,-0.263]$ \\
China & $-0.145$ & $[-0.327,\,0.038]$ & $-0.363$ & $[-0.548,\,-0.178]$ \\
Colombia & $-0.235$ & $[-0.416,\,-0.054]$ & $-0.453$ & $[-0.630,\,-0.277]$ \\
Costa Rica & $-0.265$ & $[-0.455,\,-0.076]$ & $-0.483$ & $[-0.664,\,-0.302]$ \\
Denmark & $-0.124$ & $[-0.313,\,0.066]$ & $-0.342$ & $[-0.520,\,-0.164]$ \\
Egypt & $-0.198$ & $[-0.372,\,-0.024]$ & $-0.416$ & $[-0.588,\,-0.245]$ \\
EU & $-0.091$ & $[-0.278,\,0.096]$ & $-0.309$ & $[-0.493,\,-0.126]$ \\
Finland & $-0.064$ & $[-0.254,\,0.125]$ & $-0.282$ & $[-0.461,\,-0.104]$ \\
France & $-0.123$ & $[-0.309,\,0.063]$ & $-0.341$ & $[-0.520,\,-0.162]$ \\
Germany & $-0.105$ & $[-0.289,\,0.079]$ & $-0.323$ & $[-0.500,\,-0.146]$ \\
Ghana & $-0.237$ & $[-0.418,\,-0.056]$ & $-0.455$ & $[-0.631,\,-0.280]$ \\
Greece & $-0.163$ & $[-0.357,\,0.031]$ & $-0.381$ & $[-0.565,\,-0.198]$ \\
Hong Kong & $-0.236$ & $[-0.416,\,-0.056]$ & $-0.454$ & $[-0.628,\,-0.280]$ \\
India & $-0.167$ & $[-0.352,\,0.017]$ & $-0.386$ & $[-0.573,\,-0.199]$ \\
Indonesia & $-0.207$ & $[-0.389,\,-0.025]$ & $-0.425$ & $[-0.606,\,-0.244]$ \\
Iran & $-0.188$ & $[-0.362,\,-0.015]$ & $-0.406$ & $[-0.577,\,-0.236]$ \\
Ireland & $-0.141$ & $[-0.334,\,0.052]$ & $-0.359$ & $[-0.540,\,-0.178]$ \\
Israel & $-0.226$ & $[-0.404,\,-0.047]$ & $-0.444$ & $[-0.616,\,-0.272]$ \\
Italy & $-0.131$ & $[-0.318,\,0.057]$ & $-0.349$ & $[-0.529,\,-0.169]$ \\
Jamaica & $-0.255$ & $[-0.442,\,-0.068]$ & $-0.473$ & $[-0.652,\,-0.295]$ \\
Japan & $-0.176$ & $[-0.349,\,-0.003]$ & $-0.394$ & $[-0.565,\,-0.222]$ \\
Kazakhstan & $-0.180$ & $[-0.351,\,-0.009]$ & $-0.398$ & $[-0.564,\,-0.232]$ \\
Kuwait & $-0.245$ & $[-0.431,\,-0.059]$ & $-0.463$ & $[-0.641,\,-0.286]$ \\
Kyrgyz Republic & $-0.193$ & $[-0.367,\,-0.019]$ & $-0.411$ & $[-0.579,\,-0.244]$ \\
Malaysia & $-0.242$ & $[-0.424,\,-0.060]$ & $-0.460$ & $[-0.637,\,-0.283]$ \\
Mexico & $-0.200$ & $[-0.376,\,-0.025]$ & $-0.418$ & $[-0.592,\,-0.245]$ \\
Middle income & $-0.076$ & $[-0.274,\,0.123]$ & $-0.294$ & $[-0.492,\,-0.095]$ \\
Mozambique & $-0.458$ & $[-0.743,\,-0.172]$ & $-0.676$ & $[-0.957,\,-0.395]$ \\
Netherlands & $-0.122$ & $[-0.308,\,0.064]$ & $-0.340$ & $[-0.517,\,-0.164]$ \\
New Zealand & $-0.218$ & $[-0.397,\,-0.039]$ & $-0.436$ & $[-0.607,\,-0.265]$ \\
Niger & $-0.247$ & $[-0.430,\,-0.063]$ & $-0.465$ & $[-0.641,\,-0.289]$ \\
Nigeria & $-0.217$ & $[-0.396,\,-0.038]$ & $-0.435$ & $[-0.611,\,-0.259]$ \\
Norway & $-0.089$ & $[-0.278,\,0.100]$ & $-0.307$ & $[-0.485,\,-0.130]$ \\
OECD & $-0.122$ & $[-0.301,\,0.058]$ & $-0.340$ & $[-0.521,\,-0.159]$ \\
Pakistan & $-0.185$ & $[-0.360,\,-0.011]$ & $-0.403$ & $[-0.576,\,-0.231]$ \\
Peru & $-0.231$ & $[-0.411,\,-0.052]$ & $-0.449$ & $[-0.625,\,-0.274]$ \\
Philippines & $-0.216$ & $[-0.395,\,-0.038]$ & $-0.435$ & $[-0.610,\,-0.259]$ \\
Portugal & $-0.161$ & $[-0.355,\,0.032]$ & $-0.380$ & $[-0.563,\,-0.196]$ \\
Saudi Arabia & $-0.222$ & $[-0.399,\,-0.045]$ & $-0.441$ & $[-0.612,\,-0.269]$ \\
South Africa & $-0.201$ & $[-0.375,\,-0.027]$ & $-0.419$ & $[-0.590,\,-0.248]$ \\
South Korea & $-0.188$ & $[-0.360,\,-0.016]$ & $-0.406$ & $[-0.575,\,-0.237]$ \\
Spain & $-0.140$ & $[-0.330,\,0.049]$ & $-0.359$ & $[-0.540,\,-0.177]$ \\
Sri Lanka & $-0.239$ & $[-0.420,\,-0.058]$ & $-0.457$ & $[-0.633,\,-0.281]$ \\
Sweden & $-0.078$ & $[-0.263,\,0.107]$ & $-0.296$ & $[-0.471,\,-0.121]$ \\
Switzerland & $-0.149$ & $[-0.340,\,0.043]$ & $-0.367$ & $[-0.548,\,-0.186]$ \\
Taiwan & $-0.217$ & $[-0.394,\,-0.041]$ & $-0.436$ & $[-0.607,\,-0.264]$ \\
Tunisia & $-0.215$ & $[-0.391,\,-0.039]$ & $-0.433$ & $[-0.603,\,-0.263]$ \\
Turkey & $-0.138$ & $[-0.328,\,0.051]$ & $-0.357$ & $[-0.538,\,-0.175]$ \\
Ukraine & $-0.150$ & $[-0.342,\,0.041]$ & $-0.369$ & $[-0.552,\,-0.185]$ \\
United Kingdom & $-0.099$ & $[-0.282,\,0.084]$ & $-0.317$ & $[-0.493,\,-0.142]$ \\
USA & $-0.108$ & $[-0.278,\,0.063]$ & $-0.326$ & $[-0.492,\,-0.159]$ \\
Yemen & $-0.237$ & $[-0.417,\,-0.056]$ & $-0.455$ & $[-0.629,\,-0.280]$ \\
\end{longtable}
\end{singlespace}

\end{appendices}
\end{document}